\documentclass[10pt,letterpaper]{article}

\usepackage{natbib}
\usepackage{hyperref}

\usepackage[latin1]{inputenc}							
\usepackage{amsmath}									
\usepackage{empheq}

\usepackage{graphicx}

\usepackage{amsfonts}
\newcommand{\field}[1]{\mathbb{#1}}

\usepackage{xcolor}
\definecolor{bl}{rgb}{0.0,0.2,0.6}

\usepackage{sectsty}
\usepackage[compact]{titlesec} 
\allsectionsfont{\color{bl}\scshape\selectfont}

\makeatletter							
\def\printtitle{%
    {\color{bl} \centering \huge \sc \textbf{\@title}\par}}		
\makeatother							

\title{Skewness and kurtosis unbiased by \\Gaussian uncertainties}

\makeatletter							
\def\printauthor{%
    {\centering \@author}}				
\makeatother							

\author{%
\vspace{8pt}
\large Lorenzo Rimoldini \\
\vspace{5pt}
\small Observatoire astronomique de l'Universit\'e de Gen\`eve, chemin des Maillettes 51, CH-1290 Versoix, Switzerland \\
\small ISDC Data Centre for Astrophysics, Universit\'e de Gen\`eve, chemin d'Ecogia 16, CH-1290 Versoix, Switzerland \\
\vspace{3pt}
\small \texttt{lorenzo@rimoldini.info}\\
	\vspace{12pt}
Draft version: April 29, 2013\\
	\vspace{8pt}
	}

\usepackage{fancyhdr}
\usepackage{lastpage}	
\usepackage{abstract}			
\abslabeldelim{\quad}						%
\begin{document}

\printtitle 

\printauthor

\begin{abstract}
Noise is an unavoidable part of most measurements which can hinder a correct interpretation of the data. Uncertainties propagate in the data analysis and can lead to biased results even in basic descriptive statistics such as the central moments and cumulants. Expressions of  noise-unbiased estimates of central moments and cumulants up to the fourth order are presented under the assumption of independent Gaussian uncertainties, for weighted and unweighted statistics. These results are expected to be relevant for applications of the skewness and kurtosis estimators such as outlier detections, normality tests and in automated classification procedures. The comparison of  estimators corrected and not corrected for noise biases is illustrated with simulations as a function of signal-to-noise ratio, employing different sample sizes and weighting schemes.
\end{abstract}

\section{Introduction}

Measurements generally provide an approximate description of real phenomena, because data acquisition compounds many processes which contribute, to a different degree, to instrumental errors (e.g., related to sensitivity or systematic biases) and uncertainties of statistical nature.
While instrumental effects are addressed before data analysis, statistical uncertainties propagate in subsequent processing 
and can affect both precision 
and accuracy 
of  results,
especially at low signal-to-noise ($S/N$) ratios.
Correcting for biases generated by noise    
can help the characterization and interpretation of weak signals,  
and in some cases improve a significant fraction of all data
(e.g., the number of astronomical sources increases dramatically near the faint detection threshold, since there are many more sources far away than nearby).

In this paper, noise-unbiased estimates of central moments and cumulants up to the fourth order,  which are often employed to characterize the shape of the distribution of data, are derived analytically.
Some of the advantages of these estimators include the ease of computation and the ability to encapsulate important features in a few numbers. 
Skewness and kurtosis measure the degree of asymmetry and peakedness or weight of the tails of the distribution, respectively, and they are useful for the detection of outliers, the assessment of departures from normality of the data \citep{Dagostino}, the classification of  light variations of astronomical sources \citep{RimoldiniWeighted} and many other applications.
Various estimators of skewness and kurtosis are available in the literature \citep[e.g.,][]{Moors,Hosking,Groeneveld,Bowley}, some of which aim at mitigating the sensitivity to outliers of the conventional formulations.
On the other hand, robust measures might miss important features of signals, especially when these are characterized by outliers (as in astronomical time series where stellar bursts or eclipses from binary systems represent rare events in the light curve) and weighting might help distinguish true outliers from spurious data (employing additional information such as the accuracy of each measurement), so the traditional forms of weighted central moments and cumulants are employed in this work.

\newpage
Moments are usually computed on random variables. Herein, their application is extended to data generated from deterministic functions  and randomized by the uneven sampling of a finite number of measurements and by their uncertainties, 
whereas the corresponding `population' statistics are defined in the limit of an infinite regular sampling with no random or systematic errors.
This scenario is common in astronomical time series, where measurements are typically non-regular due to observational constraints, they are unavoidably affected by noise, and sometimes also not very numerous: all of these aspects introduce some level of randomness in the characterization of the underlying signal of a star. 

While the effects of sampling and sample size on time series are studied in \citet{RimoldiniWeighted,RimoldiniUnbiased}, this work addresses the bias, precision and accuracy of estimators when measurements are affected (mostly) by Gaussian uncertainties. 
Bias is defined as the difference between expectation and population values and thus expresses a systematic deviation from the true value.
Precision is described by the dispersion of measurements, while accuracy is related to the distance of an estimator from the true value and thus combines the bias and precision concepts (e.g., accuracy can be measured by  the mean square error, defined by the sum of bias and uncertainty in quadrature).

Noise-unbiased expressions are provided for the variance, skewness and kurtosis (central moments and cumulants), weighted and unweighted, assuming Gaussian uncertainties and independent measurements. 
The dependence of noise-unbiased estimators on S/N is illustrated with simulations employing different sample sizes and two weighting schemes: the common inverse-squared uncertainties and interpolation-based weights as described in \citet{RimoldiniWeighted}. 
The latter demonstrated a significant improvement in the precision of weighted estimators at the high $S/N$ end.

This paper is organized as follows. The notation employed throughout is defined in Sec.~\ref{sec:notation}, followed by the description of the method to estimate Gaussian-noise unbiased moments in Sec.~\ref{sec:method}.
Noise-unbiased estimates of moments and cumulants (biased and unbiased by sample-size) are presented in Sections~\ref{sec:sample} and \ref{sec:unbiased}, in both weighted and unweighted formulations, and the special case of error-weighted estimators is presented in Sec.~\ref{sec:special}. The noise-unbiased estimators are compared with the uncorrected (noise-biased) counterparts with simulated signals as a function of $S/N$ ratio in Sec.~\ref{sec:simulation}, including weighted and unweighted schemes and two different sample sizes. Conclusions are drawn in Sec.~\ref{sec:concl}, followed by detailed derivations of the noise-unbiased estimators in App.~\ref{app:derivations}.

\section{Notation}
\label{sec:notation}
For a set of $n$ measurements $\mathbf{x}=(x_1,x_2,...,x_n)$, the following quantities are defined.
\begin{itemize}
\item[(i)]
Population central moments $\mu_r=\langle (\mathbf{x}-\mu)^r\rangle$ with mean $\mu=\langle\mathbf{x}\rangle$, where $\langle .\rangle$ denotes expectation, and cumulants $\kappa_2=\mu_2$, $\kappa_3=\mu_3$, $\kappa_4=\mu_4-3\mu_2^2$  \citep[e.g.,][]{Kendall}.
\item[(ii)]
The sum of the $p$-th power of weights is defined as $V_p=\sum_{i=1}^n w_i^p$.
\item[(iii)]
The  mean $\bar{\theta}$ of a generic set of $n$ elements $\theta_i$  associated with weights $w_i$ is $\bar{\theta}= \sum_{i=1}^{n}w_i \theta_i/V_1$.
\item[(iv)]
Sample central moments $m_r=\sum_{i=1}^n w_i (x_i-\bar{x})^r/V_1$ and corresponding cumulants $k_r$.
\item[(v)]
Sample-size unbiased estimates of central moments  $M_i$ and cumulants $K_i$, i.e., $\langle M_i\rangle=\mu_i$ and $\langle K_i\rangle=\kappa_i$.
\item[(vi)]
The standardized skewness and kurtosis are defined as $g_1=k_3/k_2^{3/2}$,~ $g_2=k_4/k_2^2$,~ $G_1=K_3/K_2^{3/2}$, and
$G_2=K_4/K_2^2$, with population values $\gamma_1=\kappa_3/\kappa_2^{3/2}$ and $\gamma_2=\kappa_4/\kappa_2^2$. 
$G_1$ and $G_2$ satisfy consistency (for $n\rightarrow\infty$) but are not unbiased in general \citep[e.g., see][for exceptions]{Heijmans}.
\item[(vii)]
Noise-unbiased estimates of central moments and cumulants are denoted by an asterisk superscript.
\item[(viii)]
No systematic errors are considered herein and random errors are simply referred to as errors or uncertainties.
\item[(ix)]
Statistics weighted by the inverse-squared uncertainties are called `error-weighted' for brevity and interpolation-based weights computed in phase \citep{RimoldiniWeighted} are named `phase weights'.
\end{itemize}

\section{Method}
\label{sec:method}
The goal is to derive an estimator $T ^*\!(\mathbf{x},\mbox{{\boldmath $\epsilon$}})$ as a function of observables (measurements $\mathbf{x}$ with corresponding uncertainties {\boldmath $\epsilon$}) which is unbiased by the noise in the data, i.e., such that the expectation $\langle  T ^*\!(\mathbf{x},\mbox{{\boldmath $\epsilon$}})\rangle$ equals the estimator $T (\mbox{{\boldmath $\xi$}})$ in terms of the true (unknown) values {\boldmath $\xi$} aimed at by the measurements.

The noise-unbiased estimator $T ^*\!(\mathbf{x},\mbox{{\boldmath $\epsilon$}})$ is obtained with the following procedure and assumptions.
If $n$ independent measurements $\mathbf{x}$ are associated with independent Gaussian uncertainties {\boldmath $\epsilon$}, the expected value $\langle T (\mathbf{x})\rangle$ of the estimator $ T (\mathbf{x})$ is evaluated from measurements $\mathbf{x}$ 
and the joint probability density  $p(\mathbf{x}|\mbox{{\boldmath $\xi$}},\mbox{{\boldmath $\epsilon$}})$, for given true values {\boldmath $\xi$} and measurement uncertainties {\boldmath $\epsilon$}: 
\begin{equation}
 \langle T (\mathbf{x})\rangle =\int_{\field{R}^n} T (\mathbf{x}')\, p(\mathbf{x}'|\mbox{{\boldmath $\xi$}},\mbox{{\boldmath $\epsilon$}})\, {\mathrm d}^n\!x',
\end{equation}
where
\begin{equation}
p(\mathbf{x}'|\mbox{{\boldmath $\xi$}},\mbox{{\boldmath $\epsilon$}})=\prod_{i=1}^n \frac{1}{\sqrt{2 \pi}\, \epsilon_i} \exp\left[-\frac{(x'_i-\xi_i)^2}{2\epsilon_i^2}\right].
\label{eq:pdf}
\end{equation}
As shown in App.~\ref{app:derivations}, the expectation  $\langle T (\mathbf{x})\rangle$ of the estimators considered herein can be decomposed as
\begin{equation}
\langle T (\mathbf{x})\rangle = T (\mbox{{\boldmath $\xi$}})+f(\mbox{{\boldmath $\xi$}},\mbox{{\boldmath $\epsilon$}}).
\end{equation}
Thus, the noise-free estimator $ T (\mbox{{\boldmath $\xi$}})=\langle T (\mathbf{x})\rangle-f(\mbox{{\boldmath $\xi$}},\mbox{{\boldmath $\epsilon$}})$ can be estimated in terms of measurements $\mathbf{x}$ and uncertainties {\boldmath $\epsilon$} by the noise-unbiased estimator $ T ^*\!(\mathbf{x},\mbox{{\boldmath $\epsilon$}})= T (\mathbf{x})- f^*\!(\mathbf{x},\mbox{{\boldmath $\epsilon$}})$, 
where  $\langle f^*\!(\mathbf{x},\mbox{{\boldmath $\epsilon$}})\rangle =f(\mbox{{\boldmath $\xi$}},\mbox{{\boldmath $\epsilon$}})$ and, by definition, $\langle  T ^*\!(\mathbf{x},\mbox{{\boldmath $\epsilon$}})\rangle= T (\mbox{{\boldmath $\xi$}})$.
The $f^*\!(\mathbf{x},\mbox{{\boldmath $\epsilon$}})$ term is derived first by computing $f(\mbox{{\boldmath $\xi$}},\mbox{{\boldmath $\epsilon$}})=\langle T (\mathbf{x})\rangle- T (\mbox{{\boldmath $\xi$}})$ and then by replacing terms depending on {\boldmath $\xi$} in $f(\mbox{{\boldmath $\xi$}},\mbox{{\boldmath $\epsilon$}})$ with terms as a function of $\mathbf{x}$ which satisfy the requirement $\langle f^*\!(\mathbf{x},\mbox{{\boldmath $\epsilon$}})\rangle =f(\mbox{{\boldmath $\xi$}},\mbox{{\boldmath $\epsilon$}})$ (see App.~\ref{app:derivations}).
A property often used in the following sections is that a noise-unbiased linear combination of $N$ estimators  is equivalent to the linear combination of noise-unbiased estimators:
\begin{equation}
\left[\sum_{i=1}^N c_i \, T _i(\mathbf{x})\right]^*=\sum_{i=1}^N c_i\, T _i^*\!(\mathbf{x}),
\end{equation}
where the coefficients $c_i$ are independent of the measurements $\mathbf{x}$.

\section{Gaussian-noise unbiased sample moments and cumulants}
\label{sec:sample}
Weighted sample central moments unbiased by Gaussian uncertainties, such as  the variance $m^*_2$, skewness $m^*_3$,  kurtosis $m^*_4$ and the respective cumulants about the weighted mean $\bar{x}^*=\bar{x}$ are derived assuming independent measurements $x_i$, uncertainties $\epsilon_i$ and weights $w_i$, as described in full detail in App.~\ref{app:derivations}.
They are defined as follows:
\begin{align}
&m^*_2 = m_2-\frac{1}{V_1}\sum_{i=1}^{n}w_i  \epsilon_i^2\left(1-\frac{w_i}{V_1} \right) = k^*_2 \\
&m^*_3 =m_3-\frac{3}{V_1}\sum_{i=1}^{n}w_i \epsilon_i^2  \left(x_i-\bar{x}\right)\left(1-\frac{2 w_i}{V_1} \right)= k^*_3 \\
&m^*_4  =m_4- \frac{6}{V_1}\sum_{i=1}^{n}w_i  \epsilon_i^2 \left[\left(x_i-\bar{x}\right)^2\left(1-\frac{2w_i}{V_1} \right)-\frac{\epsilon_i^2}{2}\left(1-\frac{2w_i}{V_1}\right)^2
+\frac{m^*_2 w_i}{V_1}\right] -\frac{3}{V_1^4}\left(\sum_{i=1}^{n} w_i^2\epsilon_i^2\right)^2\\
&(m_2^2)^* =\left(m^*_2\right)^2 -\frac{4}{V_1^2}\sum_{i=1}^{n} w_i^2\epsilon_i^2 \left[ \left(x_i-\bar{x}\right)^2 -\frac{\epsilon_i^2}{2}\left(1-\frac{2w_i}{V_1} \right)\right]
+\frac{2}{V_1^4}\left(\sum_{i=1}^{n} w_i^2\epsilon_i^2\right)^2\\
&k^*_4  = m^*_4-3\,(m_2^2)^* .
\end{align}
By definition, the above expressions satisfy
\begin{equation}
\langle m^*_r\rangle=\frac{1}{V_1}\sum_{i=1}^{n}w_i \left(\xi_i-\mbox{$\bar{\xi}$}\,\right)^r.
\end{equation}

The unweighted forms can be obtained by substituting $w_i=1$ (for all $i$) and $V_p=n$ (for all $p$) in all terms, leading to: 
\begin{align}
&m^*_2 =m_2 -  \frac{n-1}{n^2}\sum_{i=1}^{n} \epsilon_i^2  = k^*_2 \\
&m^*_3 =m_3  - \frac{3(n-2)}{n^2}\sum_{i=1}^{n} \epsilon_i^2\left(x_i-\bar{x}\right)= k^*_3 \\ 
&m^*_4  =m_4  -\frac{6(n-2)}{n^2}\sum_{i=1}^{n}\epsilon_i^2\left(x_i-\bar{x}\right)^2
-\frac{6\,m_2^*}{n^2}\sum_{i=1}^n \epsilon_i^2
+\frac{3(n-2)^2}{n^3}\sum_{i=1}^n \epsilon_i^4
-\frac{3}{n^4}\left(\sum_{i=1}^n \epsilon_i^2 \right)^2 \\
&(m_2^2)^* = (m_2^*)^2
-\frac{4}{n^2}\sum_{i=1}^{n}\epsilon_i^2\left(x_i-\bar{x}\right)^2
+\frac{2(n-2)}{n^3}\sum_{i=1}^n \epsilon_i^4
+\frac{2}{n^4}\left(\sum_{i=1}^n \epsilon_i^2 \right)^2 \\
&k^*_4  = m^*_4-3\,(m_2^2)^* .
\end{align}

\section{Gaussian-noise and sample-size unbiased moments and cumulants}
\label{sec:unbiased}
The estimates of weighted central moments which are unbiased by both sample-size and Gaussian uncertainties, such as the variance $M^*_2$, skewness $M^*_3$,  kurtosis $M^*_4$ and the respective cumulants, are defined in terms of the noise-unbiased sample estimators as follows:
\begin{align}
M^*_2 =\,& \frac{V_1^2}{V_1^2-V_2}\,m^*_2 = K^*_2 \\
M^*_3 =\,&   \frac{V_1^3}{V_1^3-3V_1V_2+2V_3}\,m^*_3 = K^*_3 \\
M^*_4  =\,&   \frac{V_1^2(V_1^4-3V_1^2V_2+2V_1V_3+3V_2^2-3V_4)}{(V_1^2-V_2)(V_1^4-6V_1^2V_2+8V_1V_3+3V_2^2-6V_4)}\,m^*_4 \,+\notag\\
\,&  -\frac{3V_1^2(2V_1^2V_2-2V_1V_3-3V_2^2+3V_4)}{(V_1^2-V_2)(V_1^4-6V_1^2V_2+8V_1V_3+3V_2^2-6V_4)}\,(m_2^2)^* \\
K^*_4  =\,&   \frac{V_1^2(V_1^4-4V_1V_3+3V_2^2)}{(V_1^2-V_2)(V_1^4-6V_1^2V_2+8V_1V_3+3V_2^2-6V_4)}\,m^*_4 \,+\notag\\
\,&  -\frac{3V_1^2(V_1^4-2V_1^2V_2+4V_1V_3-3V_2^2)}{(V_1^2-V_2)(V_1^4-6V_1^2V_2+8V_1V_3+3V_2^2-6V_4)}\,(m_2^2)^*. 
\end{align}
The derivation of the sample-size unbiased weighted estimators is described in \citet{RimoldiniUnbiased}. The corresponding unweighted forms can be achieved by direct substitution $V_p=n$ for all $p$, 
 leading to:
\begin{align}
M^*_2 =\,& \frac{n}{n-1}\,m^*_2 = M_2 -  \frac{1}{n}\sum_{i=1}^{n} \epsilon_i^2 = K^*_2 \\
M^*_3 =\,&   \frac{n^2}{(n-1)(n-2)}\,m^*_3 =M_3  - \frac{3}{n-1}\sum_{i=1}^{n} \epsilon_i^2\left(x_i-\bar{x}\right)= K^*_3 \\
M^*_4  =\,&   \frac{n(n^2-2n+3)}{(n-1)(n-2)(n-3)}\,m^*_4 -\frac{3n(2n-3)}{(n-1)(n-2)(n-3)}\,(m_2^2)^* \\
K^*_4  =\,&   \frac{n^2(n+1)}{(n-1)(n-2)(n-3)}\,m^*_4 -\frac{3n^2}{(n-2)(n-3)}\,(m_2^2)^*.
\end{align}

\section{Special cases}
\label{sec:special}
If weights are related to measurement errors as $w_i=1/\epsilon_i^2$, the noise-unbiased weighted sample moments and cumulants reduce to the following expressions:
\begin{align}
&m^*_2 = m_2-\frac{n-1}{V_1} = k^*_2 \\
&m^*_3 =m_3- \frac{3}{V_1}\sum_{i=1}^{n} \left(x_i-\bar{x}\right)= k^*_3 \\
&m^*_4  =m_4-\frac{6}{V_1}\sum_{i=1}^{n} \left[\left(x_i-\bar{x}\right)^2-\frac{\epsilon_i^2}{2}\right]+\frac{6m^*_2}{V_1}-\frac{3}{V_1^2}\\
&(m_2^2)^* = (m^*_2)^2-\frac{2(m^*_2+m_2)}{V_1}\\
&k^*_4  = m^*_4-3\,(m_2^2)^*.
\end{align}

In the case of constant errors, i.e., $\epsilon_i=\epsilon_0$ for all $i$, some of the unweighted estimators are equivalent or similar to their noise-unbiased counterparts:
\begin{align}
\mbox{Skewness:}&~~~k^*_3=k_3 \mbox{~~and~~} K^*_3=K_3 \mbox{~~~(also $m^*_3=m_3$~~and~~$M^*_3=M_3$)}, \\
\mbox{Kurtosis:}&~~~k^*_4\approx k_4 \mbox{~~and~~} K^*_4=K_4, 
\end{align}
where the approximation $k^*_4\approx k_4$ holds for large values of $n$ or $S/N$ ratios since
\begin{equation}
\frac{k^*_4- k_4}{k_2^2}=\frac{6\, \epsilon_0^2\left(k^*_2  +k_2\right)}{n\, k_2^2}\approx\frac{6\left[1+2\,(S/N)^2\right]}{n\left[1+(S/N)^2 \right]^2},
\end{equation}
considering that, for constant errors, $\sum_i \epsilon_i^2/n=\epsilon_0^2$ and $(S/N)^2\approx k_2^*/\epsilon_0^2\approx k_2/\epsilon_0^2-1$.
For sample cumulants up to the fourth order, only the variance depends strongly on noise. However, this is an important estimator because it is often involved in  definitions of standardized skewness ($g_1$ and $G_1$) and kurtosis ($g_2$ and $G_2$) as follows:
\begin{align}
g_1=k_3/k_2^{3/2}, &~~~G_1=K_3/K_2^{3/2}, \label{eq:g1} \\
g_2=k_4/k_2^2,~~ &~~~G_2=K_4/K_2^2. \label{eq:g2}
\end{align}
For consistency with the above definitions, the noise-unbiased equivalents are defined as
\begin{align}
g^*_1=k^*_3/(k^*_2)^{3/2}, &~~~G^*_1=K^*_3/(K^*_2)^{3/2}, \label{eq:gi1}\\
g^*_2=k^*_4/(k^*_2)^2,~~ &~~~G^*_2=K^*_4/(K^*_2)^2, \label{eq:gi2}
\end{align}
although the truly noise-unbiased expressions should have been computed on the ratios in Eqs~(\ref{eq:g1})--(\ref{eq:g2}).
The application of Eqs~(\ref{eq:gi1})--(\ref{eq:gi2}) should generally be  restricted to larger samples (e.g., $n>50$) with $S/N$ ratios greater than a few, in order to avoid non-positive values of $k^*_2$ or $K^*_2$.

\section{Estimators as a function of signal-to-noise ratio}
\label{sec:simulation}
Noise-biased and unbiased estimators are compared as a function of  signal-to-noise ratio $S/N$ with simulated data and different weighting schemes for specific signals, sampling and error laws.
The  values of the population moments of the continuous simulated periodic `true' signal $\xi(\phi)$ are computed averaging in phase $\phi$ as follows:
\begin{align}
\mu_r=\frac{1}{2\pi}\int_0^{2\pi}\left[\xi(\phi)-\mu \right]^r {\mathrm d}\phi, ~~~~~~\mbox{where}~~~\mu=\frac{1}{2\pi}\int_0^{2\pi}\xi(\phi)\,{\mathrm d}\phi.
\end{align}

 \subsection{Simulation}
 \label{sec:}
Simulated signals are described by a sinusoidal function to the fourth power, which has a non-zero skewness and thus makes it possible to evaluate the precision and accuracy of the skewness standardized by the estimated variance without simply reflecting the accuracy of the variance.
The  $S/N$ level is evaluated by the ratio of the standard deviation $\sqrt{\mu_2}$ of the true signal $\xi(\phi)$ and the root of the mean of squared measurement uncertainties $\epsilon_i$ (assumed independent of the signal).
The signal $\xi(\phi)$ is sampled  $n=100$ and 1000 times at phases $\phi_i$ randomly drawn from a uniform distribution, while the $S/N$ ratio varies from 1 to 1000 and determines the uncertainties $\epsilon_i$ of measurements $x_i$ as follows:
\begin{empheq}[left=\empheqlbrace]{align}
&\xi(\phi)=A\sin^4\phi \label{eq:simuStart}\\
&x_i\sim {\cal N}(\xi_i,\epsilon_i^2) ~~~~~~~~~\mbox{for~~}\xi_i=\xi(\phi_i) \mbox {~~and~~}\phi_i\sim {\cal U}(0,2\pi) \\
&\epsilon_i^2=\left(1+\rho_i \right)\,\mu_2 \,/\, (S/N)^2 ~~~~~~~~~~~\mbox{for~~}\rho_i\sim{\cal U}(-0.8,0.8), \label{eq:simuCoreEnd}
\end{empheq}
where the $i$-th measurement $x_i$ is drawn from a normal distribution ${\cal N}(\xi_i,\epsilon_i^2)$ of mean $\xi_i$ and variance $\epsilon_i^2$.
The latter is defined in terms of a variable $\rho_i$  randomly drawn from a uniform distribution ${\cal U}(-0.8,0.8)$ so that measurement uncertainties vary by up to a factor of 3 for a given $\mu_2$ and $S/N$ ratio.
Simulations were repeated $10^4$ times for each $S/N$ ratio (for $n=100$ and 1000).

\begin{figure}
\centering
\includegraphics[width=8cm]{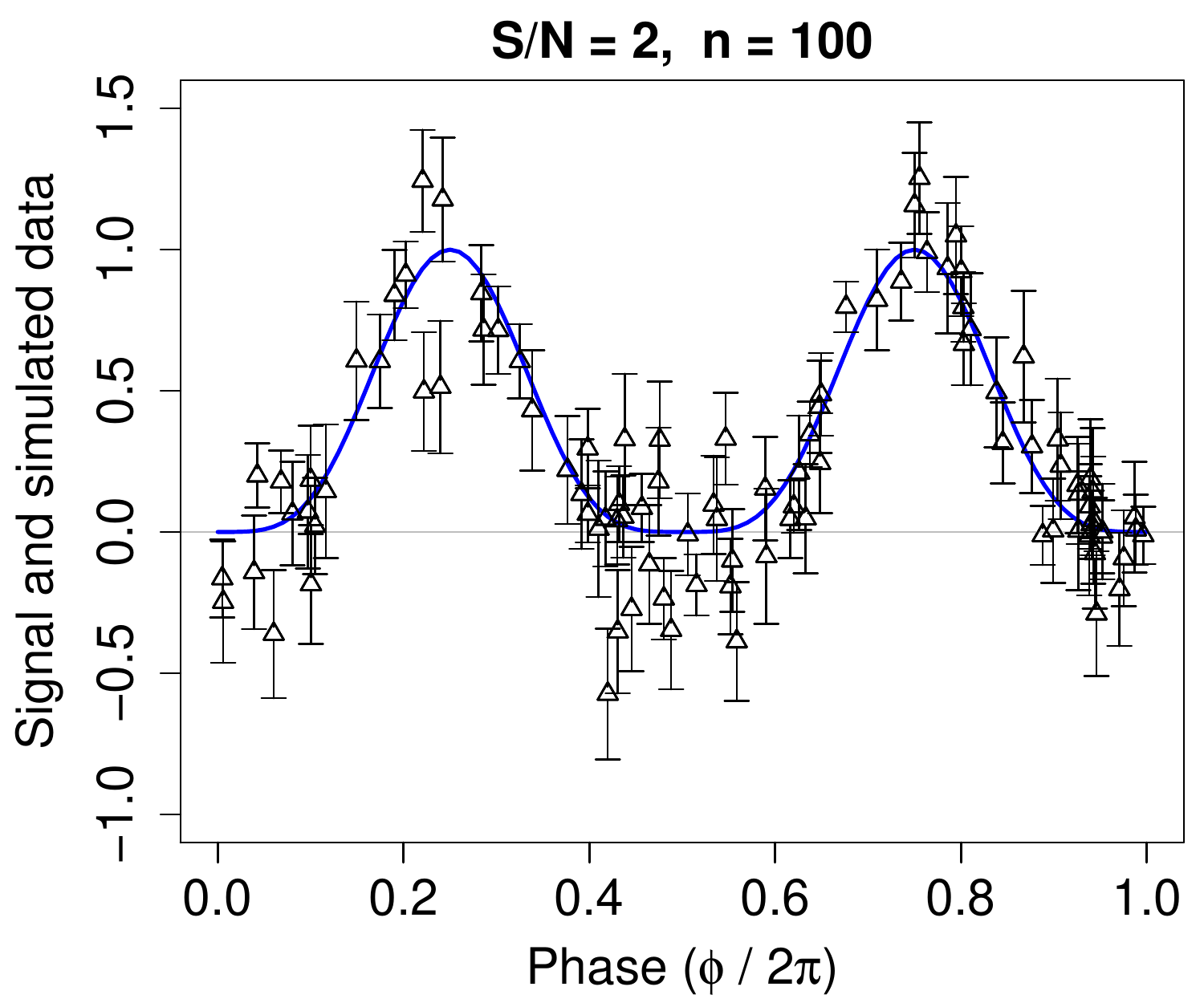} \\
\caption{A simulated signal of the form of $\sin^4\phi$ (blue curve) is irregularly sampled by 100 measurements (denoted by triangles) with $S/N=2$.}
\label{fig:simu}
\end{figure}

The dependence of weighted estimators on sample size and the corresponding unbiased expressions were presented in \citet{RimoldiniUnbiased}. Herein, only large sample sizes are employed so that sample-size biases are negligible with respect to the ones resulting from small $S/N$ ratios. 
A sample signal and simulated data are illustrated in Fig.~\ref{fig:simu} for $n=100$  and $S/N=2$. 
The reference population values of the mean, variance, skewness and kurtosis of the simulated signal are listed in table~1 of \citet{RimoldiniUnbiased}.

Error weights are defined by $w_i=1/\epsilon_i^2$, while mixed error-phase weights  follow \citet{RimoldiniWeighted}, assuming phase-sorted data:
\begin{empheq}[left=\empheqlbrace]{align}
&w_i= h(S/N|a,b)\,\frac{w'_i}{\sum_{j=1}^n w'_j}+ \left[1-h(S/N|a,b)\right]\frac{\epsilon_i^{-2}}{\sum_{j=1}^n \epsilon_j^{-2}}  ~~~~~~~\forall i\in(1,n)\label{eq:SNweights}\\
&w'_i=\phi_{i+1}-\phi_{i-1}~~~~~~~~~~~~~~~~~~~~~~~~~~\forall i\in(2,n-1) \label{eq:w_gap_ia}\\
&w'_1=\phi_{2}-\phi_{n}+2\pi \label{eq:w_gap_1a}\\
&w'_n=\phi_{1}-\phi_{n-1}+2\pi \label{eq:w_gap_na}\\
&h(S/N|a,b)=\frac{1}{1+e^{-(S/N-a)/b}}~~~~~~~~\mbox{for}~~a,b>0. \label{eq:simuEnd}
\end{empheq}
Weighting effectively decreases the sample size, since more importance is given to some data at the expense of other ones and results depend mostly on fewer `relevant' measurements (e.g., weighting by the inverse-squared uncertainties can worsen precision at high $S/N$ levels). 
Weighted procedures are desirable when the  dispersion and bias of estimators from an effectively reduced sample size are smaller than the improvements in precision and accuracy
(e.g., weighting by inverse-squared uncertainties can improve both precision and accuracy at low $S/N$ ratios). 
Also, weighting might exploit correlations in the data to improve precision, as it is shown employing phase weights \citep{RimoldiniWeighted}. Since correlated data do not satisfy the assumptions of the expressions derived herein, their application might return biased results. However, small biases could be justified if  improvements in precision are significant and, depending on the extent of the application, larger biases could be mitigated with mixed weighting schemes, such as the one described by Eqs~(\ref{eq:SNweights})--(\ref{eq:simuEnd}).

Estimators derived herein assume a single weighting scheme and combinations of estimators (like the variance and the mean in the standardized skewness and kurtosis) are expected to apply the same weights to terms associated with the same measurements.
The function $h(S/N|a,b)$ constitutes just an example to achieve a mixed weighting scheme: tuning parameters $a,b$ offer the possibility to control the transition from error-weighted to phase-weighted estimators (in the limits of low and high $S/N$, respectively) and thus
reach a compromise solution between precision and accuracy for all values of $S/N$,  according to the specific estimators,  signals, sampling, errors,  sample sizes and their distributions in the data. 

\subsection{Results}
The results of simulations are illustrated for sample estimators, since the conclusions in \citet{RimoldiniUnbiased} suggested that phase-weighted sample estimators can be more accurate and precise than the sample-size unbiased counterparts in most cases, especially for large sample sizes as considered herein. 

\begin{figure}
\begin{center}
~~~~~~~~{\bf\fbox{\parbox{0.15\textwidth}{\centering Mean \\ $(n=100,1000)$}}}\\
\end{center}
\begin{minipage}{0.5\columnwidth}
\centering
~~~~~~Unweighted \\
\includegraphics[width=\columnwidth]{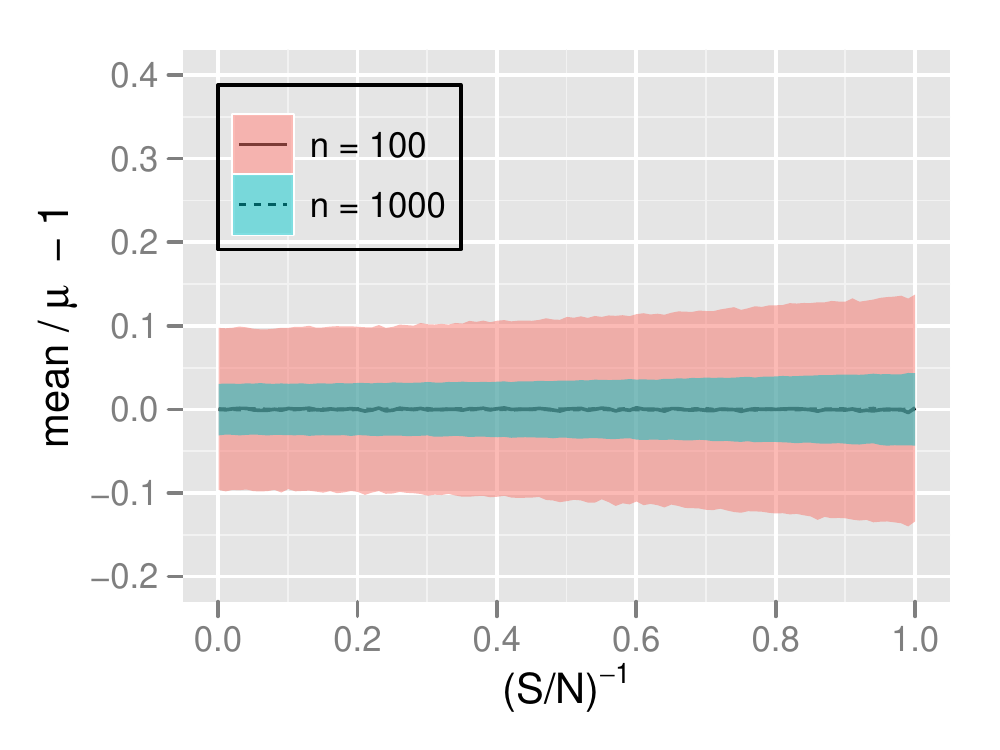} \\
~~~~~~~~Phase Weighted ($a,b\rightarrow 0$)\\
\includegraphics[width=\columnwidth]{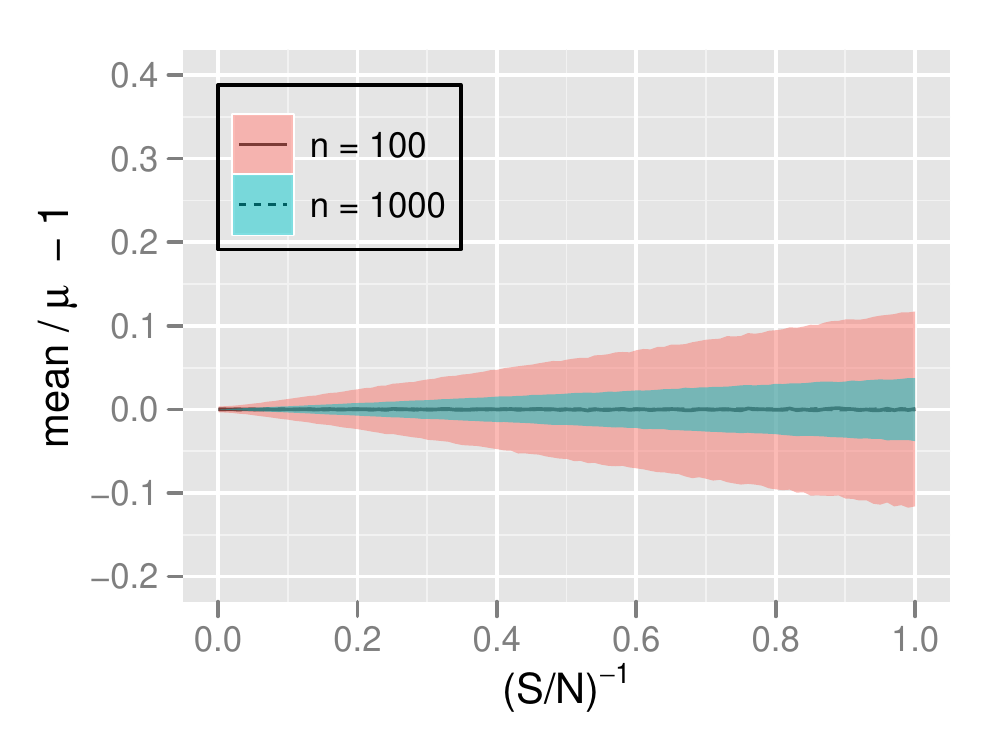}
\end{minipage}
\begin{minipage}{0.5\columnwidth}
\centering
~~~~~~~~Error Weighted\\
\includegraphics[width=\columnwidth]{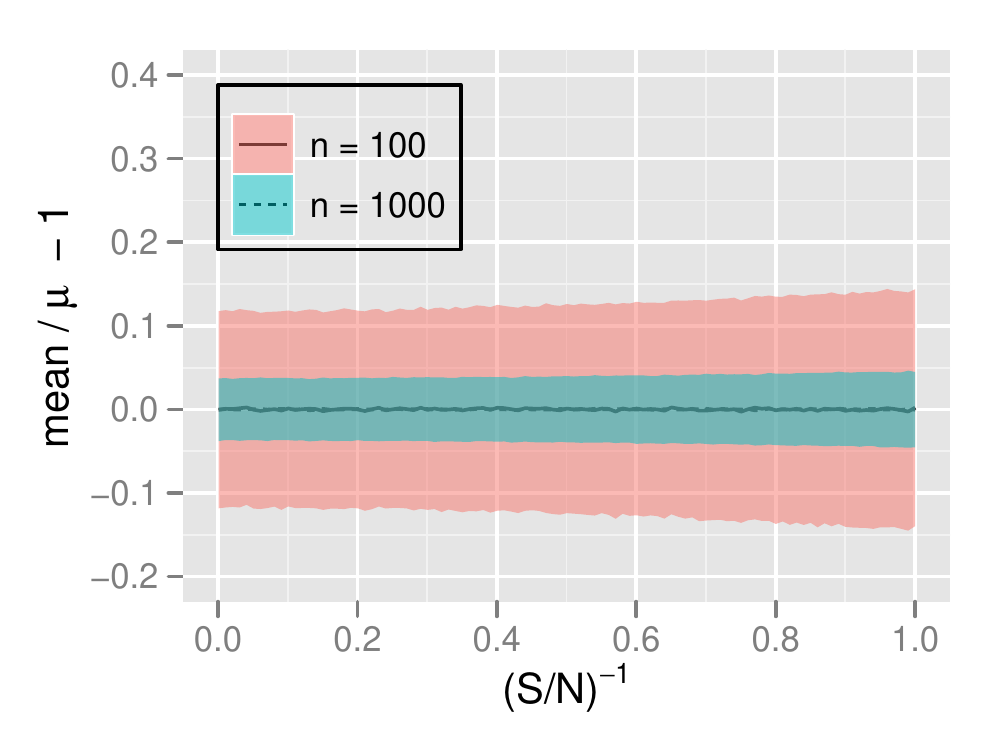}\\
~~~~~~~~Error-Phase Weighted ($a=2,b=0.3$)\\
\includegraphics[width=\columnwidth]{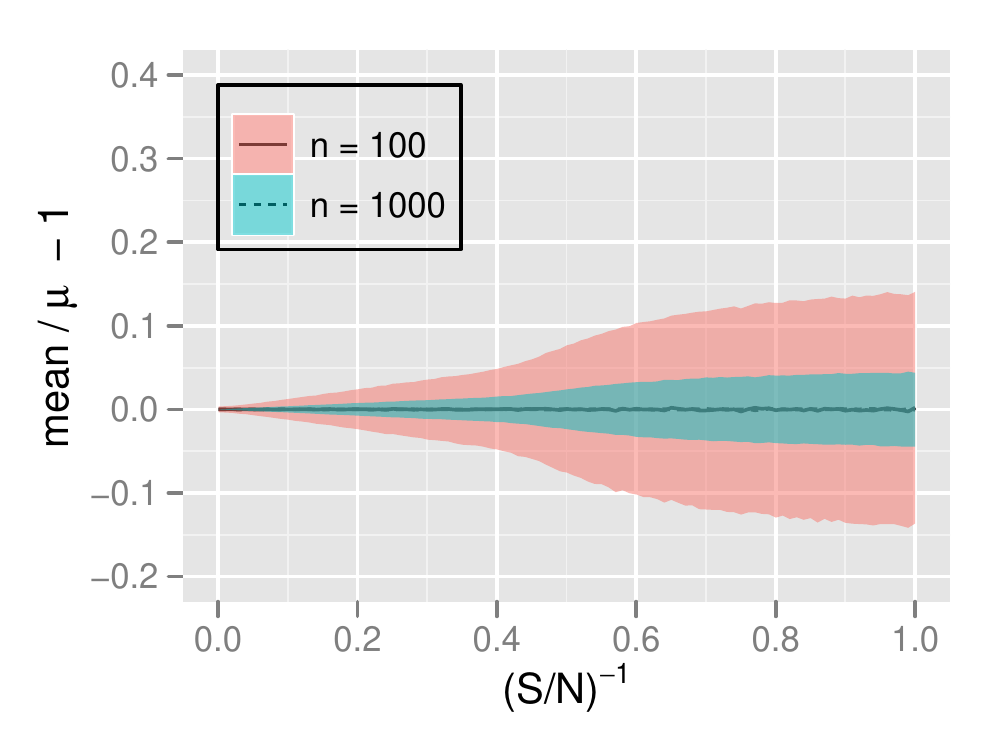}
\end{minipage}
\caption{Sample mean for $S/N>1$ and $n=100,1000$: unweighted on the top-left hand side, weighted by the inverse of squared measurement errors on the top-right hand side, and weighted by phases and errors, according to Eq.~(\ref{eq:SNweights}), with different parameter values, as specified above the lower panels. Shaded areas encompass one standard deviation from the average of the distribution of the mean employing simulations defined by Eqs~(\ref{eq:simuStart})--(\ref{eq:simuCoreEnd}). }
\label{fig:M1}
\end{figure}

Figure~\ref{fig:M1} illustrates the sample mean in the various scenarios considered in the simulations: sample sizes  of $n=100$ and 1000, unweighted and with different weighting schemes (error-weighted, phase-weighted and combined error-phase weighted). While accuracy is the same in all cases, the best precision of the mean is achieved employing phase weights (including the low $S/N$ end, unlike  other estimators).

\begin{figure}
\begin{center}
~~~~~~~~{\bf\fbox{\parbox{0.15\textwidth}{\centering Variance \\ $(n=100)$}}}\\
\end{center}
\begin{minipage}{0.5\columnwidth}
\centering
~~~~~~Unweighted \\
\includegraphics[width=\columnwidth]{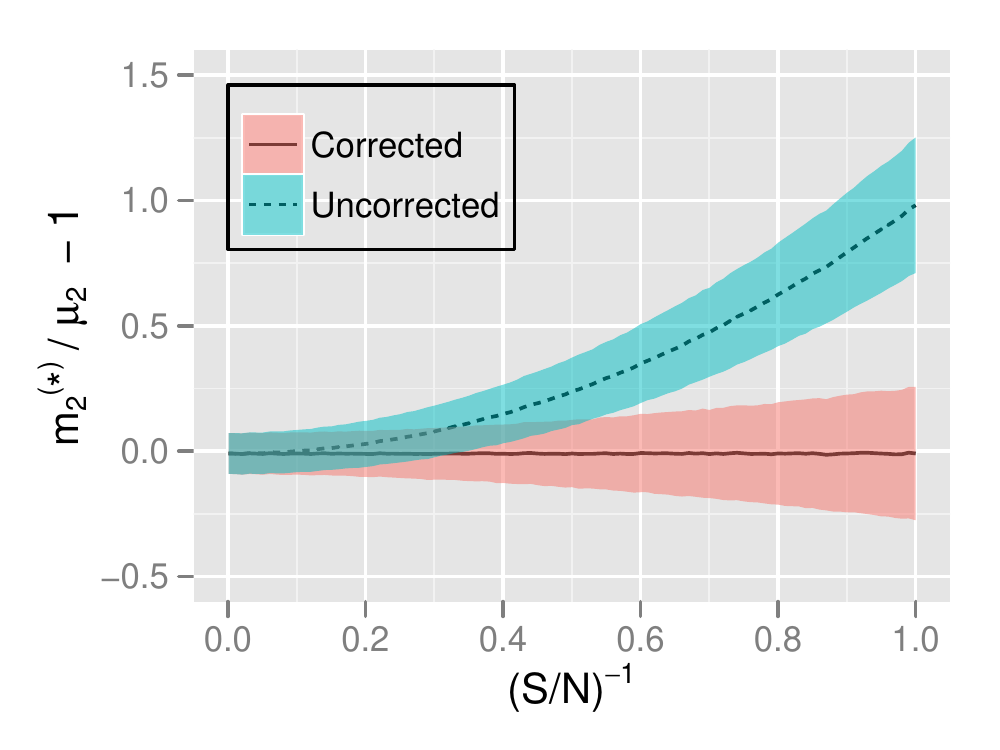} \\
~~~~~~~~Phase Weighted ($a,b\rightarrow 0$)\\
\includegraphics[width=\columnwidth]{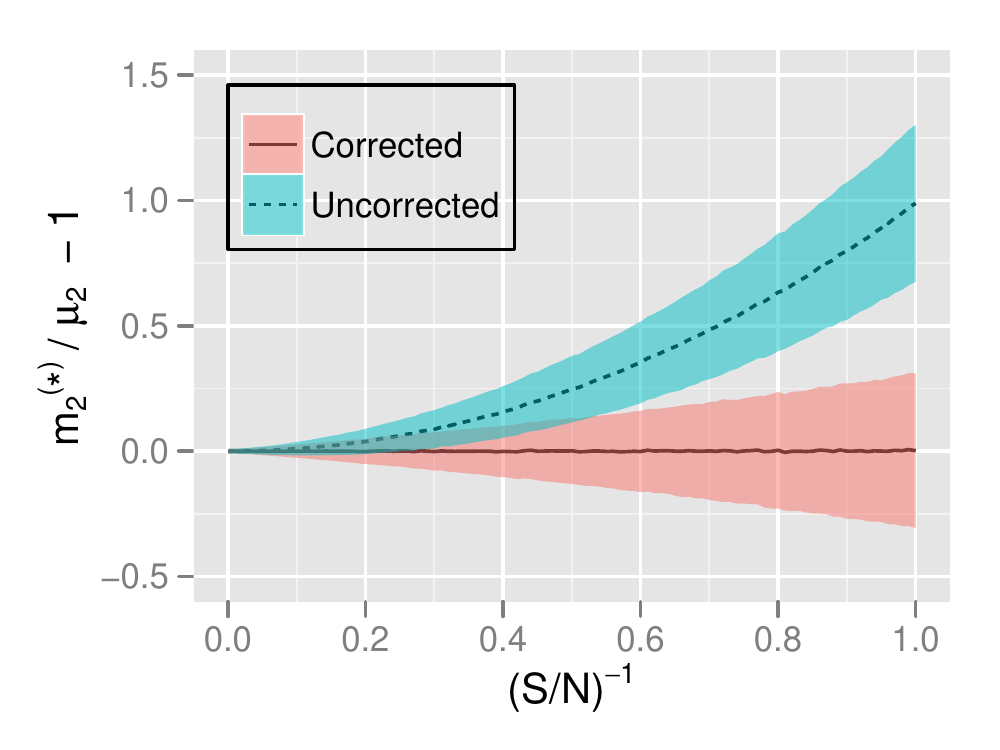}
\end{minipage}
\begin{minipage}{0.5\columnwidth}
\centering
~~~~~~~~Error Weighted\\
\includegraphics[width=\columnwidth]{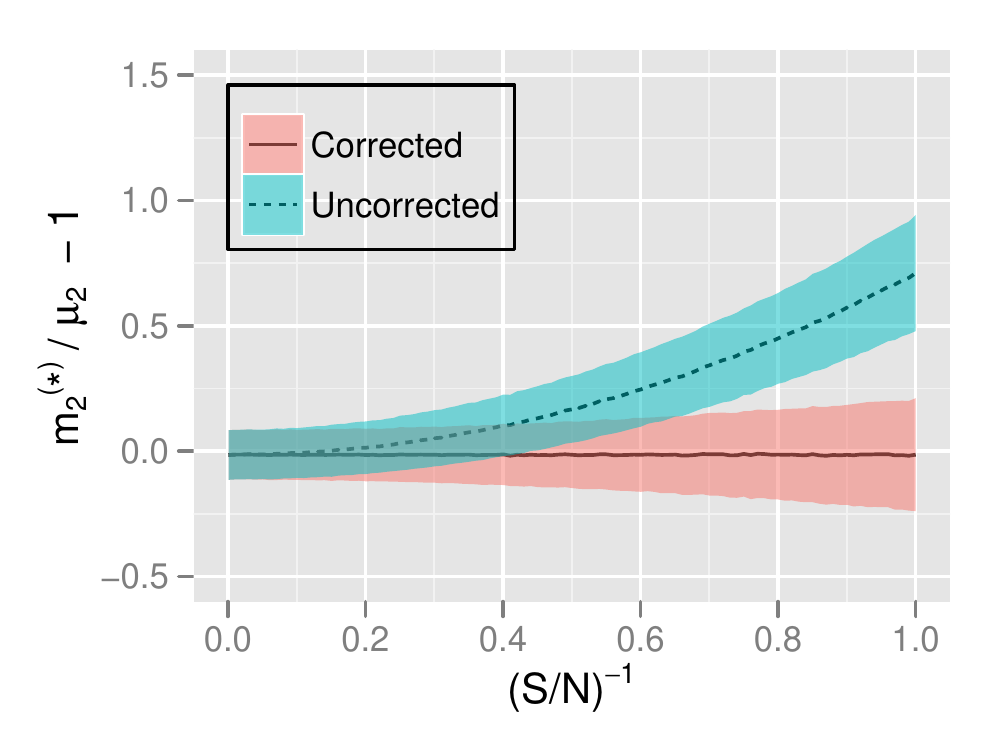}\\
~~~~~~~~Error-Phase Weighted ($a=2,b=0.3$)\\
\includegraphics[width=\columnwidth]{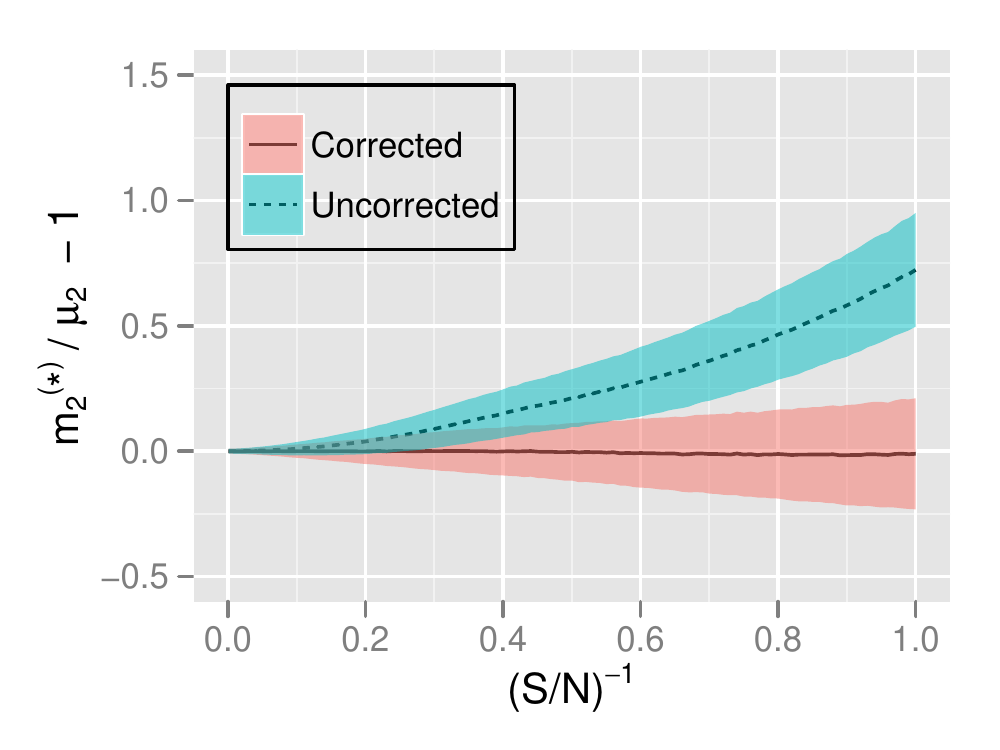}
\end{minipage}
\caption{Noise-biased  ({\it `uncorrected'}\,) versus noise-unbiased ({\it `corrected'}\,) sample variance for $S/N>1$ and $n=100$: unweighted on the top-left hand side, weighted by the inverse of squared measurement errors on the top-right hand side, and weighted by phases and errors, according to Eq.~(\ref{eq:SNweights}), with different parameter values, as specified above the lower panels.  Shaded areas encompass one standard deviation from the mean of the distribution of the variance employing simulations defined by Eqs~(\ref{eq:simuStart})--(\ref{eq:simuCoreEnd}). }
\label{fig:M2_100}
\end{figure}

\begin{figure}
\begin{center}
~~~~~~~~{\bf\fbox{\parbox{0.15\textwidth}{\centering Variance \\ $(n=1000)$}}}\\
\end{center}
\begin{minipage}{0.5\columnwidth}
\centering
~~~~~~Unweighted \\
\includegraphics[width=\columnwidth]{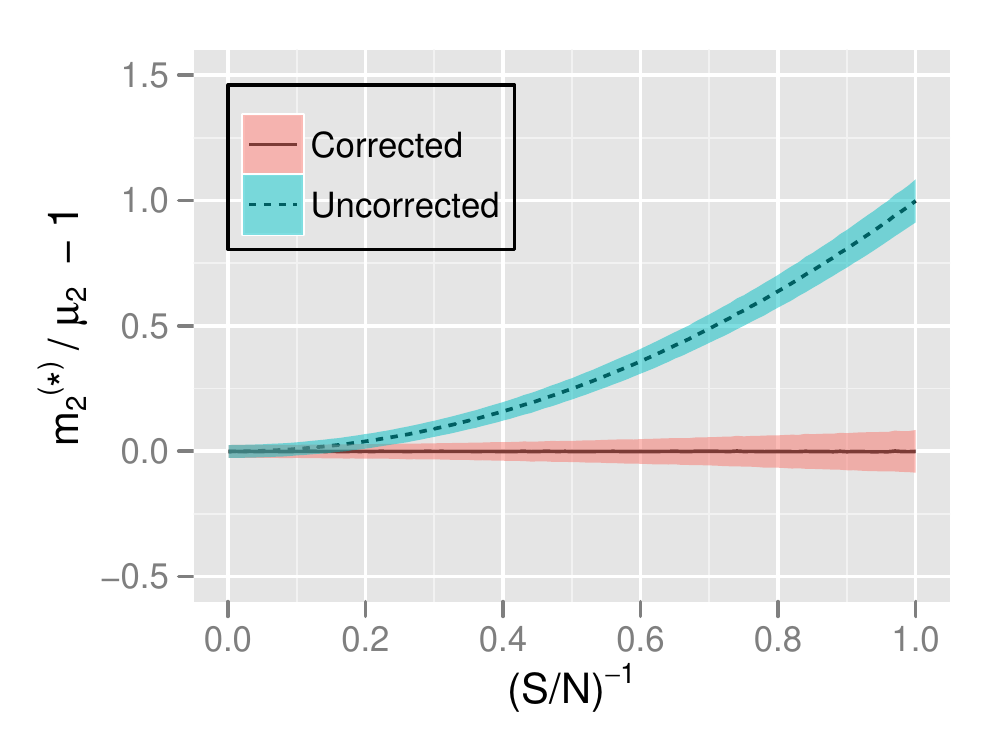}\\
~~~~~~~~Phase Weighted ($a,b\rightarrow 0$)\\
\includegraphics[width=\columnwidth]{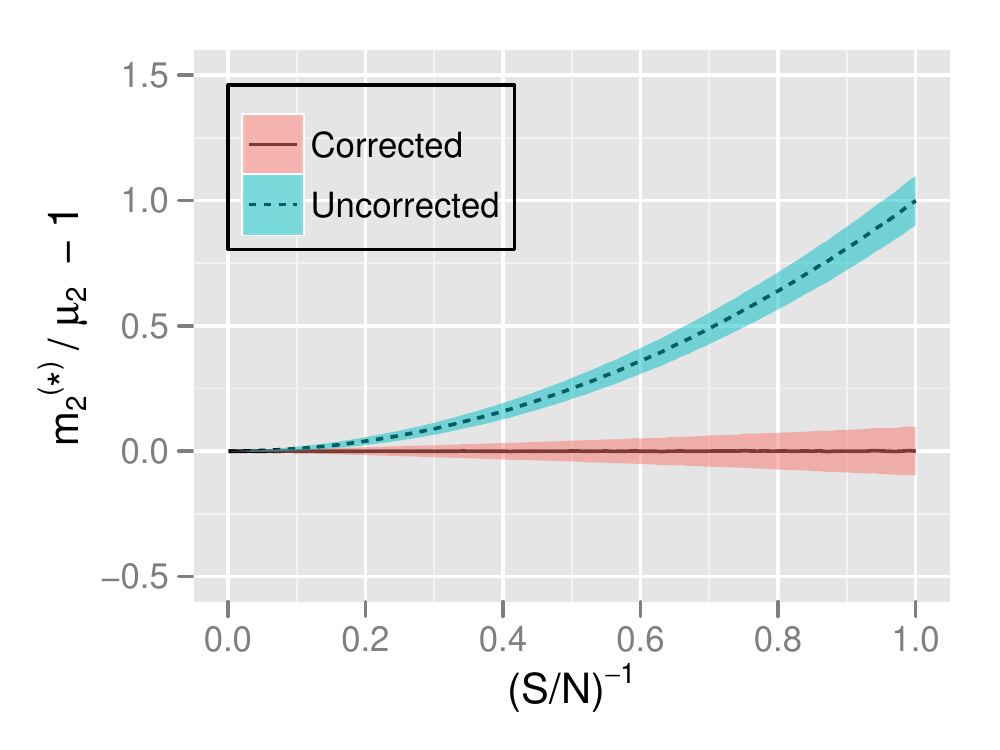}
\end{minipage}
\begin{minipage}{0.5\columnwidth}
\centering
~~~~~~~~Error Weighted\\
\includegraphics[width=\columnwidth]{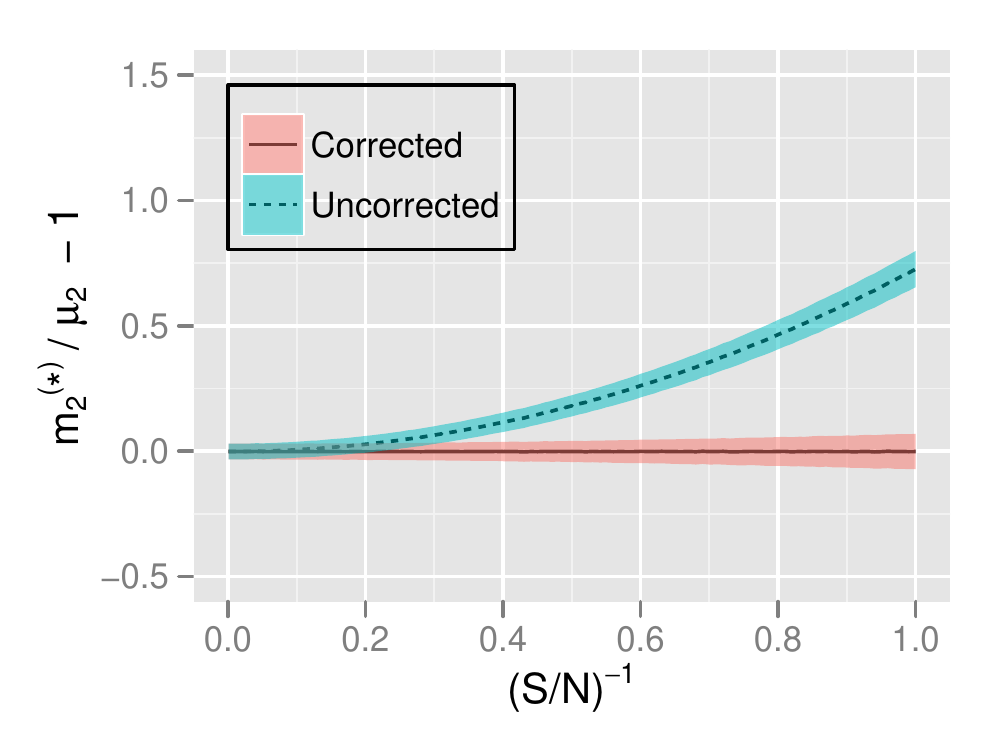}\\
~~~~~~~~Error-Phase Weighted ($a=2,b=0.3$)\\
\includegraphics[width=\columnwidth]{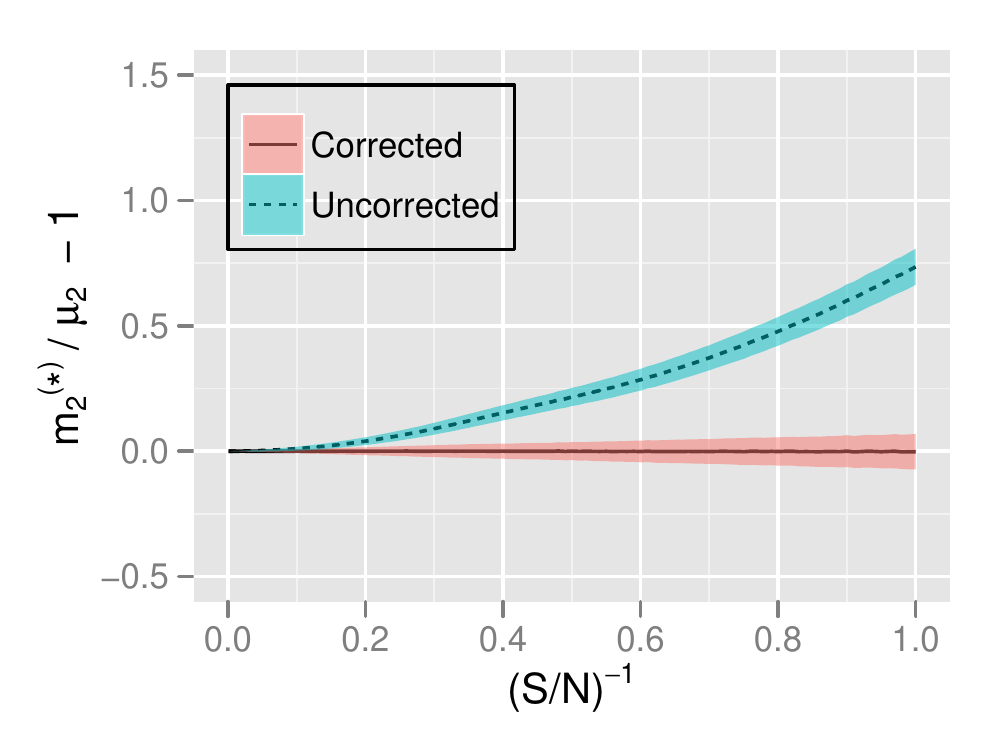}
\end{minipage}
\caption{Noise-biased  ({\it `uncorrected'}\,) versus noise-unbiased ({\it `corrected'}\,) sample variance for $S/N>1$ and $n=1000$: unweighted on the top-left hand side, weighted by the inverse of squared measurement errors on the top-right hand side, and weighted by phases and errors, according to Eq.~(\ref{eq:SNweights}), with different parameter values, as specified above the lower panels.  Shaded areas encompass one standard deviation from the mean of the distribution of the variance employing simulations defined by Eqs~(\ref{eq:simuStart})--(\ref{eq:simuCoreEnd}). }
\label{fig:M2_1000}
\end{figure}

\begin{figure}
\begin{center}
~~~~~~~~{\bf\fbox{\parbox{0.15\textwidth}{\centering Skewness \\ $(n=100)$}}}\\
\end{center}
\begin{minipage}{0.5\columnwidth}
\centering
~~~~~~Unweighted \\
\includegraphics[width=\columnwidth]{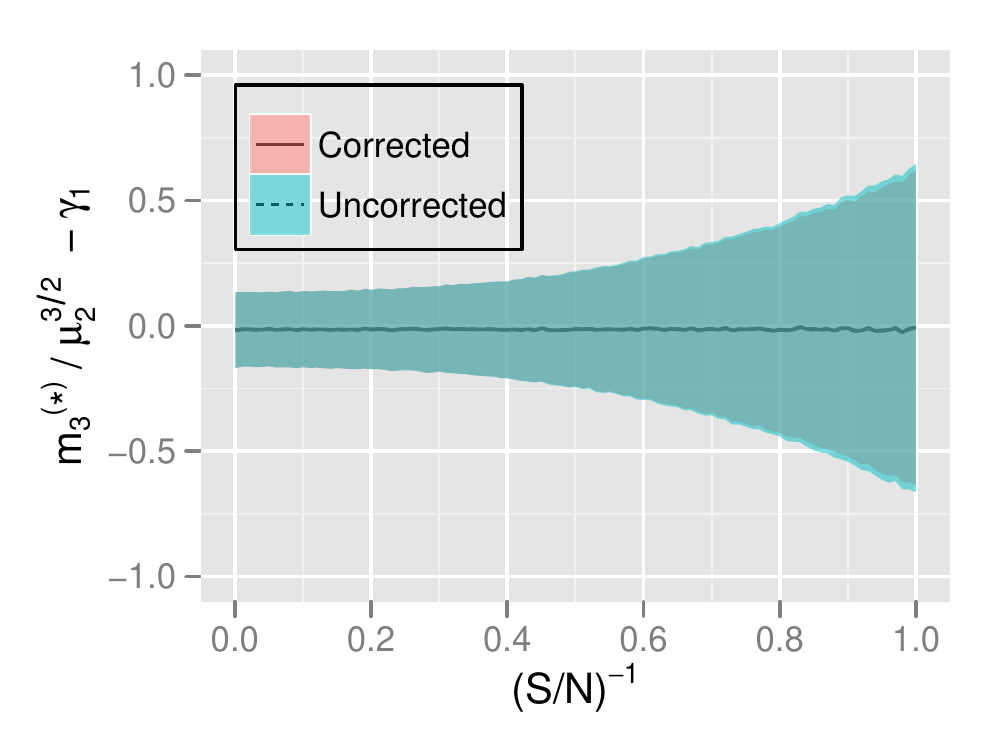}\\
~~~~~~~~Error Weighted\\
\includegraphics[width=\columnwidth]{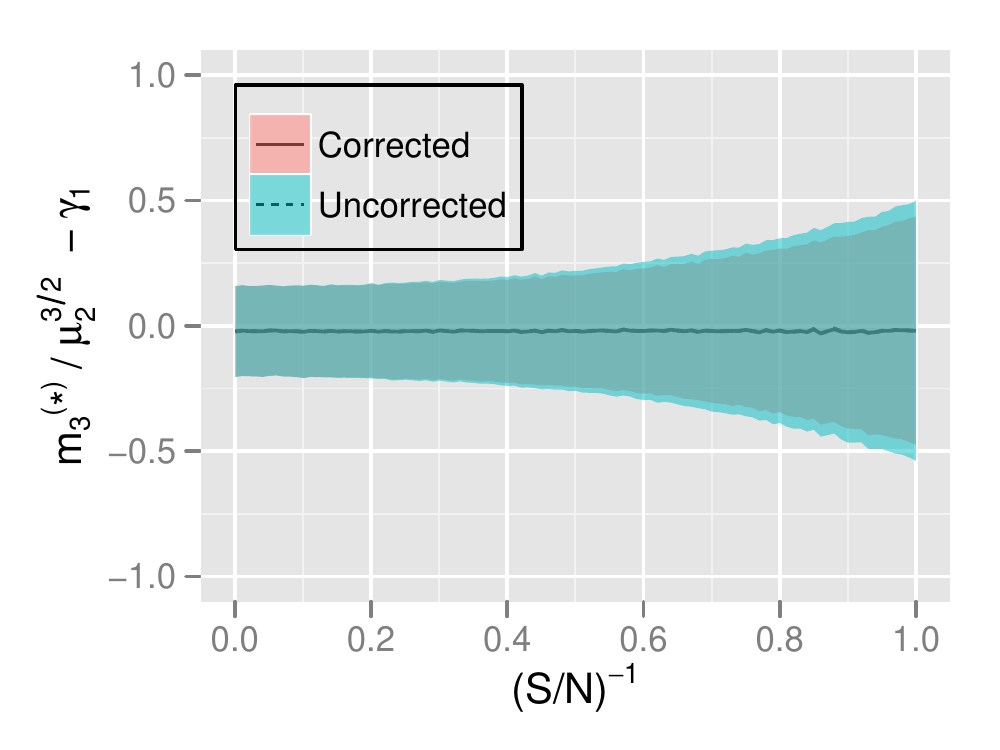}
\end{minipage}
\begin{minipage}{0.5\columnwidth}
\centering
~~~~~~Unweighted \\
\includegraphics[width=\columnwidth]{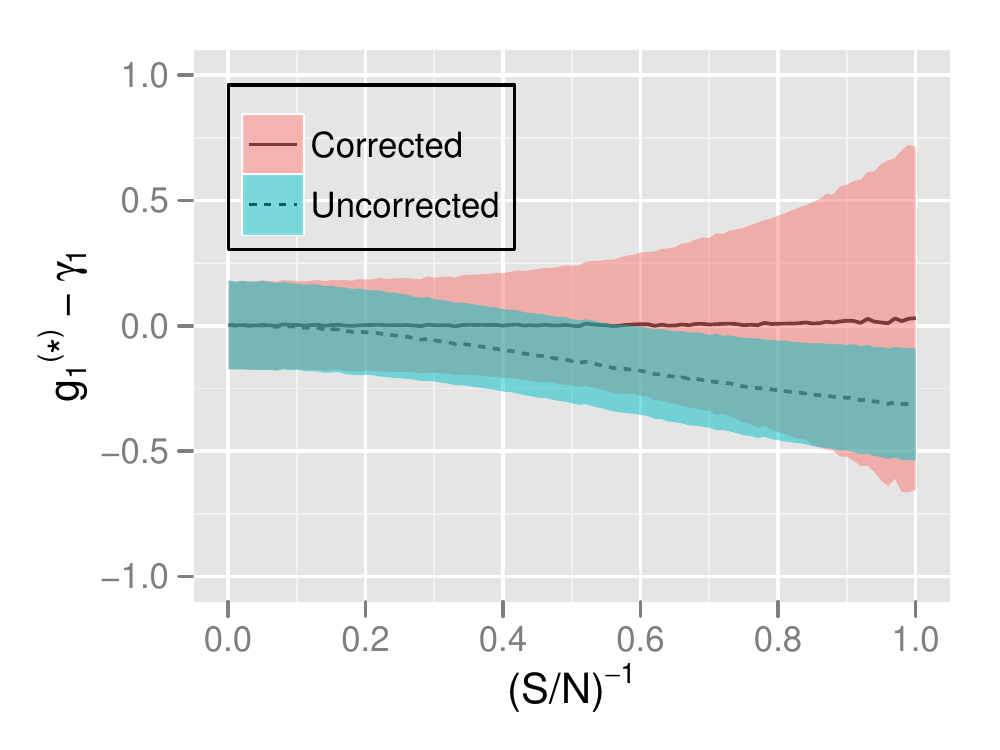}\\
~~~~~~~~Error Weighted\\
\includegraphics[width=\columnwidth]{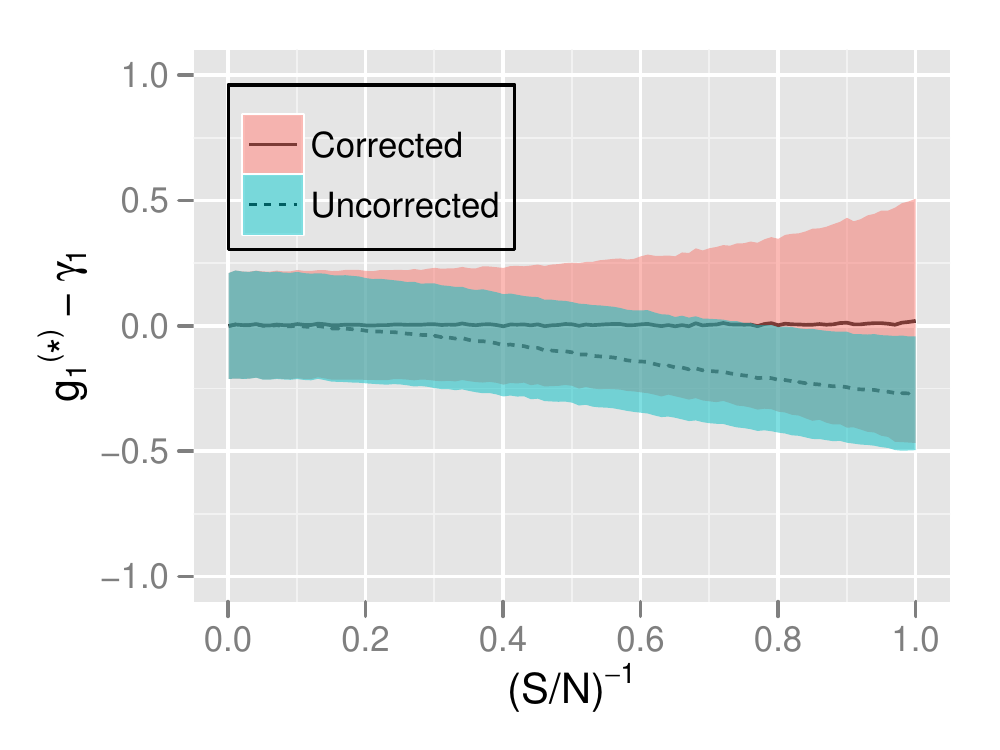}
\end{minipage}
\caption{Noise-biased  ({\it `uncorrected'}\,) versus noise-unbiased ({\it `corrected'}\,) sample skewness for $S/N>1$ and $n=100$: unweighted in the upper panels and weighted by the inverse of squared measurement errors in the lower panels. Shaded areas encompass one standard deviation from the mean of the distribution of the skewness employing simulations defined by Eqs~(\ref{eq:simuStart})--(\ref{eq:simuCoreEnd}).  
}
\label{fig:M3_100}
\end{figure}

\begin{figure}
\begin{center}
~~~~~~~~{\bf\fbox{\parbox{0.15\textwidth}{\centering Skewness \\ $(n=100)$}}}\\
\end{center}
\begin{minipage}{0.5\columnwidth}
\centering
~~~~~~~~Phase Weighted ($a,b\rightarrow 0$)\\
\includegraphics[width=\columnwidth]{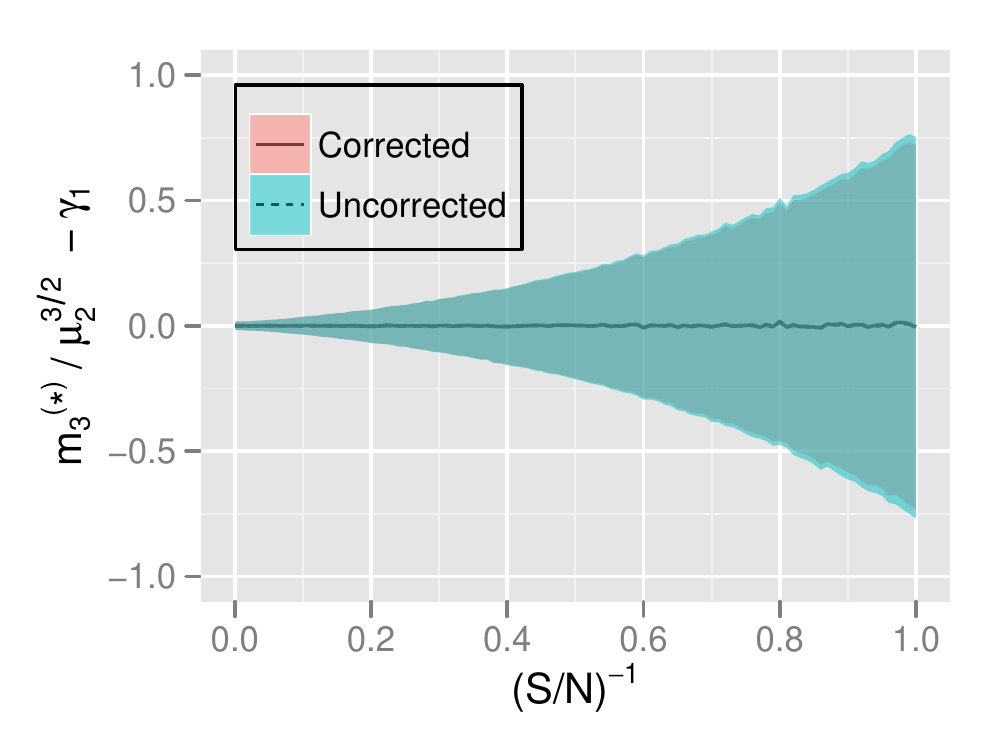}\\
~~~~~~~~Error-Phase Weighted ($a=2,b=0.3$)\\
\includegraphics[width=\columnwidth]{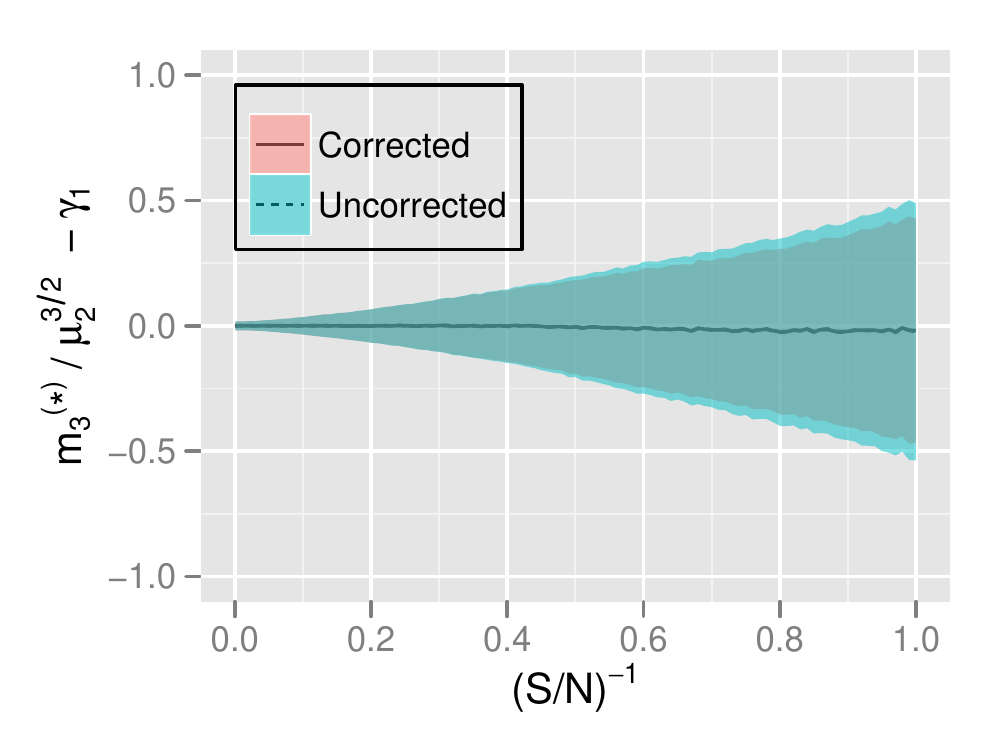}
\end{minipage}
\begin{minipage}{0.5\columnwidth}
\centering
~~~~~~~~Phase Weighted ($a,b\rightarrow 0$)\\
\includegraphics[width=\columnwidth]{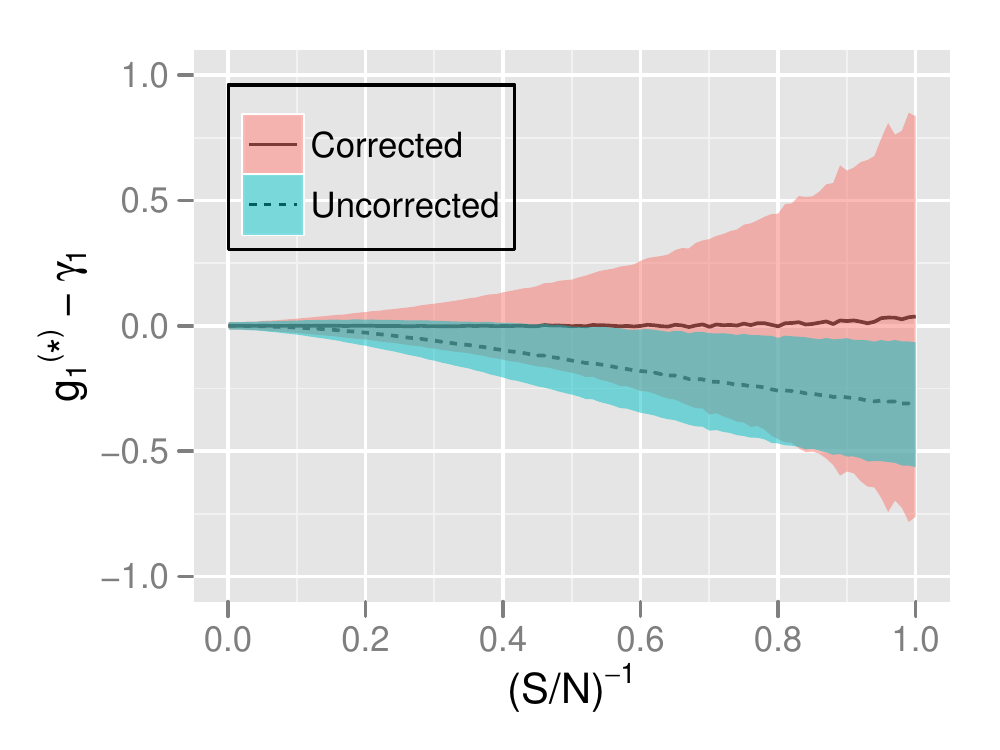}\\
~~~~~~~~Error-Phase Weighted ($a=2,b=0.3$)\\
\includegraphics[width=\columnwidth]{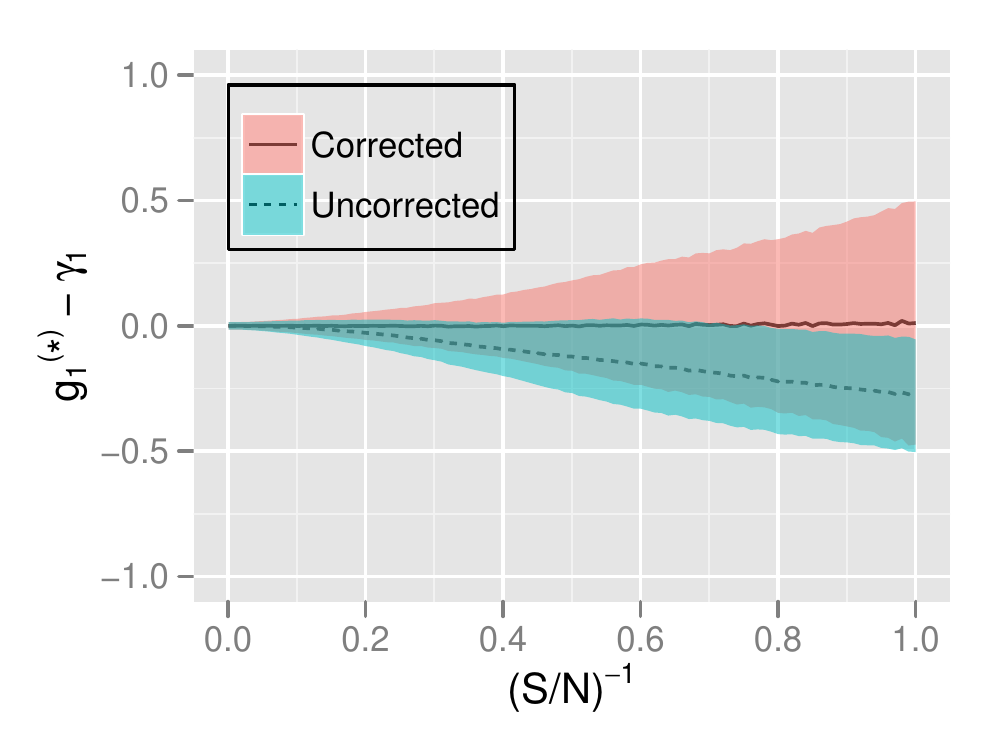}
\end{minipage}
\caption{Noise-biased  ({\it `uncorrected'}\,) versus noise-unbiased ({\it `corrected'}\,) sample skewness for $S/N>1$ and $n=100$, weighted by phases and errors, according to Eq.~(\ref{eq:SNweights}), with different parameter values, as specified above each panel.   
Shaded areas encompass one standard deviation from the mean of the distribution of the skewness employing simulations defined by Eqs~(\ref{eq:simuStart})--(\ref{eq:simuCoreEnd}). }
\label{fig:M3_100ph}
\end{figure}

\begin{figure}
\begin{center}
~~~~~~~~{\bf\fbox{\parbox{0.15\textwidth}{\centering Skewness \\ $(n=1000)$}}}\\
\end{center}
\begin{minipage}{0.5\columnwidth}
\centering
~~~~~~Unweighted \\
\includegraphics[width=\columnwidth]{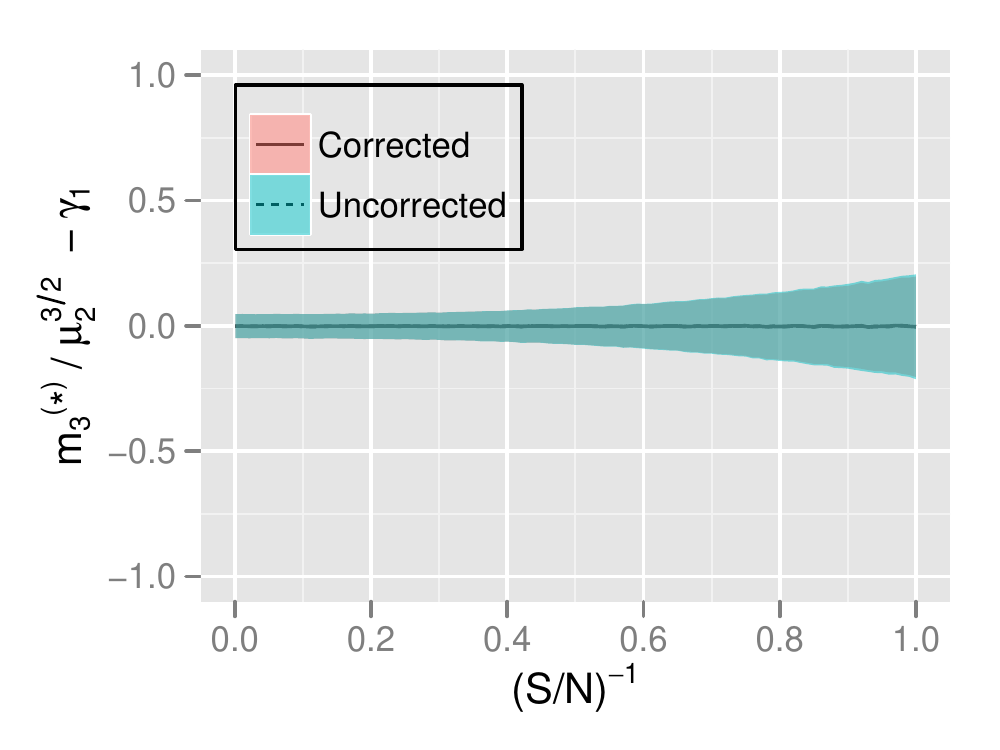}\\
~~~~~~~~Error Weighted\\
\includegraphics[width=\columnwidth]{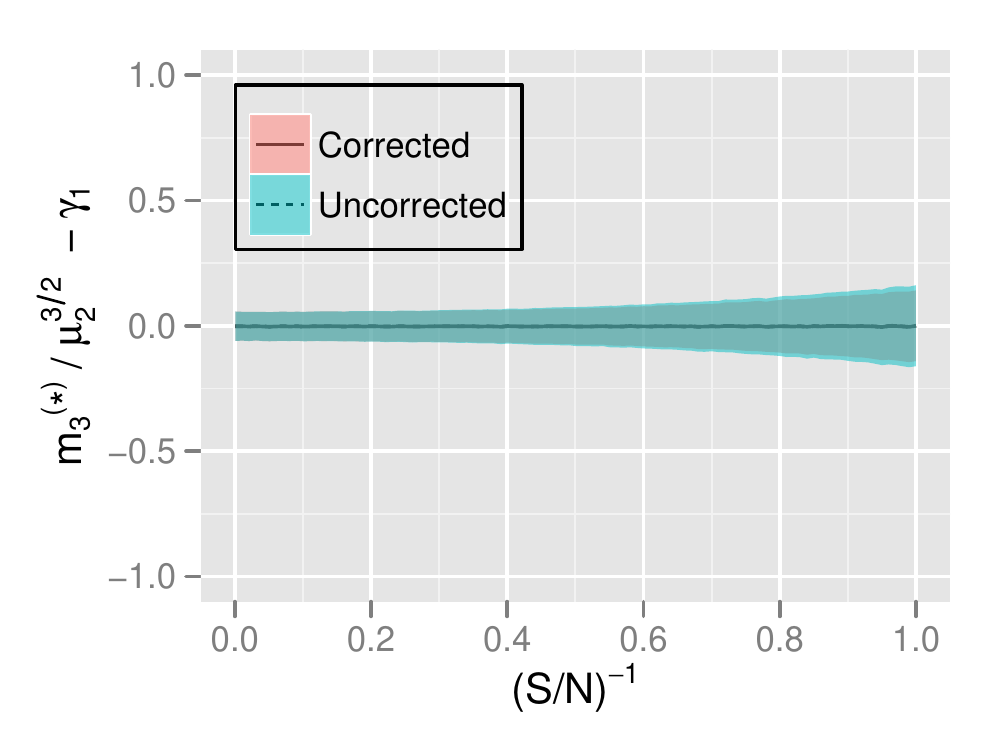}
\end{minipage}
\begin{minipage}{0.5\columnwidth}
\centering
~~~~~~Unweighted \\
\includegraphics[width=\columnwidth]{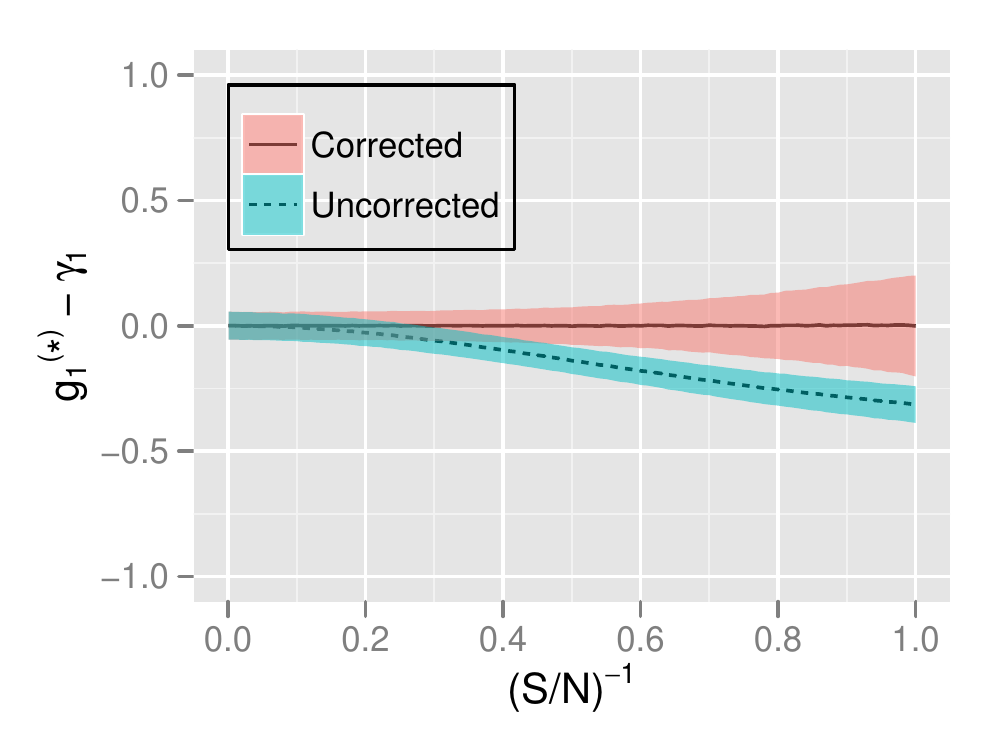}\\
~~~~~~~~Error Weighted\\
\includegraphics[width=\columnwidth]{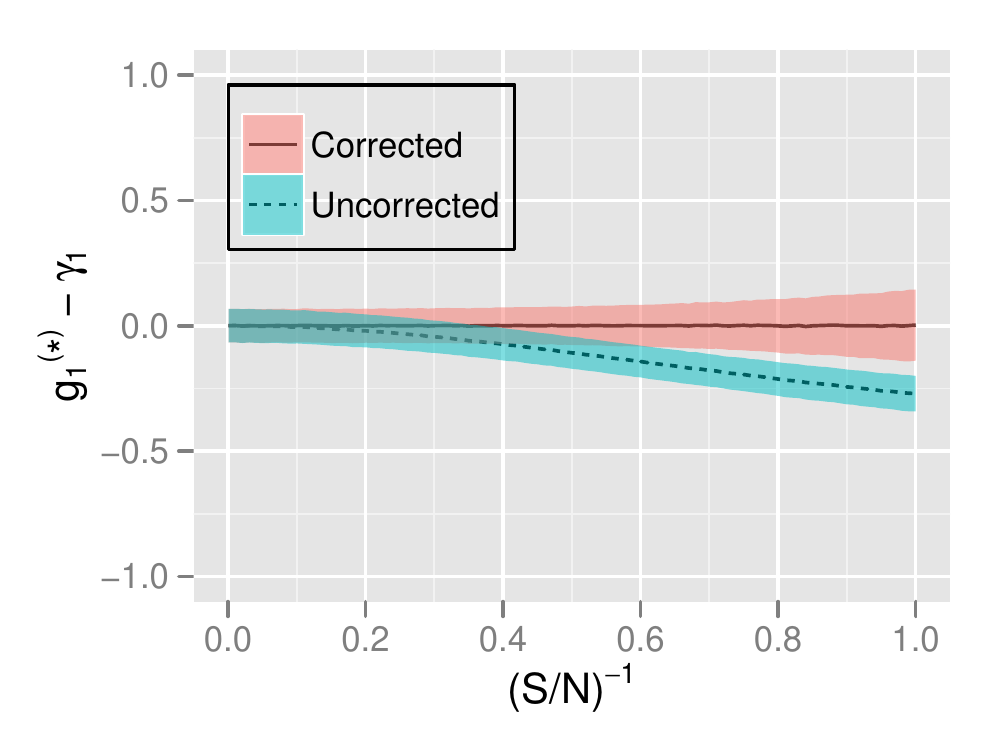}
\end{minipage}
\caption{Noise-biased  ({\it `uncorrected'}\,) versus noise-unbiased ({\it `corrected'}\,) sample skewness for $S/N>1$ and $n=1000$: unweighted in the upper panels and weighted by the inverse of squared measurement errors in the lower panels. Shaded areas encompass one standard deviation from the mean of the distribution of the skewness employing simulations defined by Eqs~(\ref{eq:simuStart})--(\ref{eq:simuCoreEnd}). 
}
\label{fig:M3_1000}
\end{figure}

\begin{figure}
\begin{center}
~~~~~~~~{\bf\fbox{\parbox{0.15\textwidth}{\centering Skewness \\ $(n=1000)$}}}\\
\end{center}
\begin{minipage}{0.5\columnwidth}
\centering
~~~~~~~~Phase Weighted ($a,b\rightarrow 0$)\\
\includegraphics[width=\columnwidth]{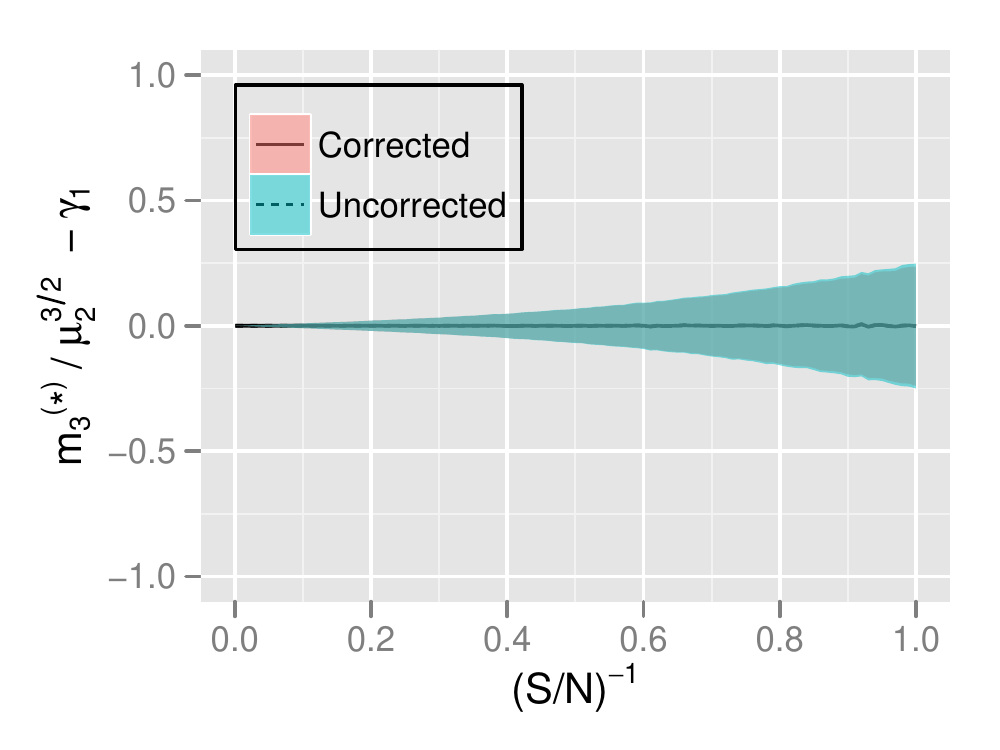}\\
~~~~~~~~Error-Phase Weighted ($a=2,b=0.3$)\\
\includegraphics[width=\columnwidth]{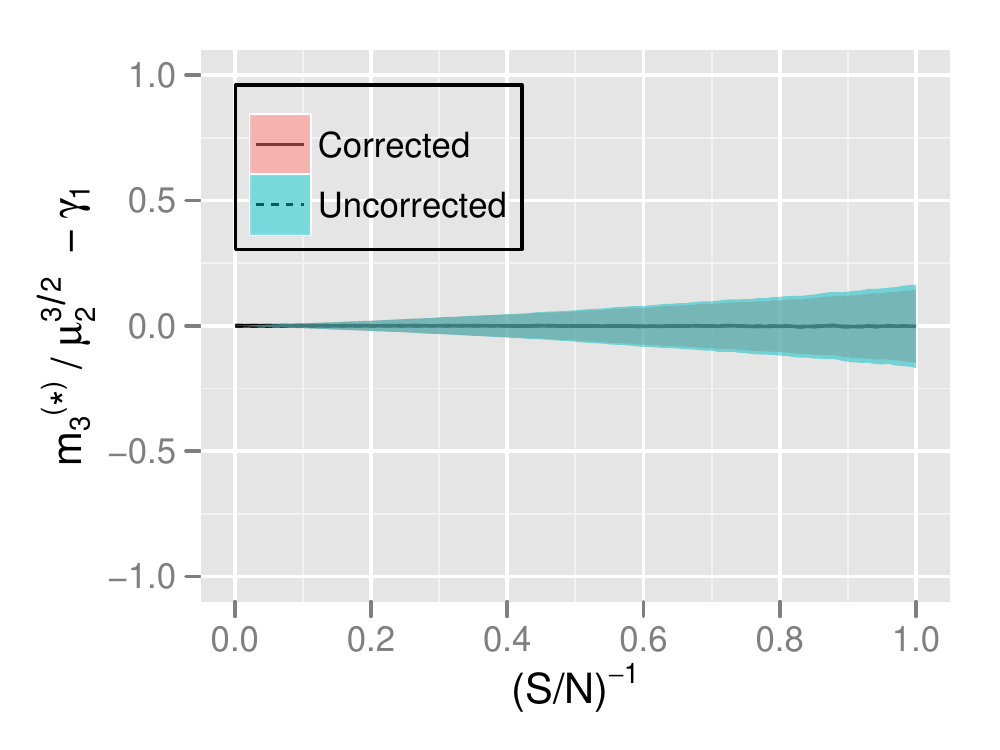}
\end{minipage}
\begin{minipage}{0.5\columnwidth}
\centering
~~~~~~~~Phase Weighted ($a,b\rightarrow 0$)\\
\includegraphics[width=\columnwidth]{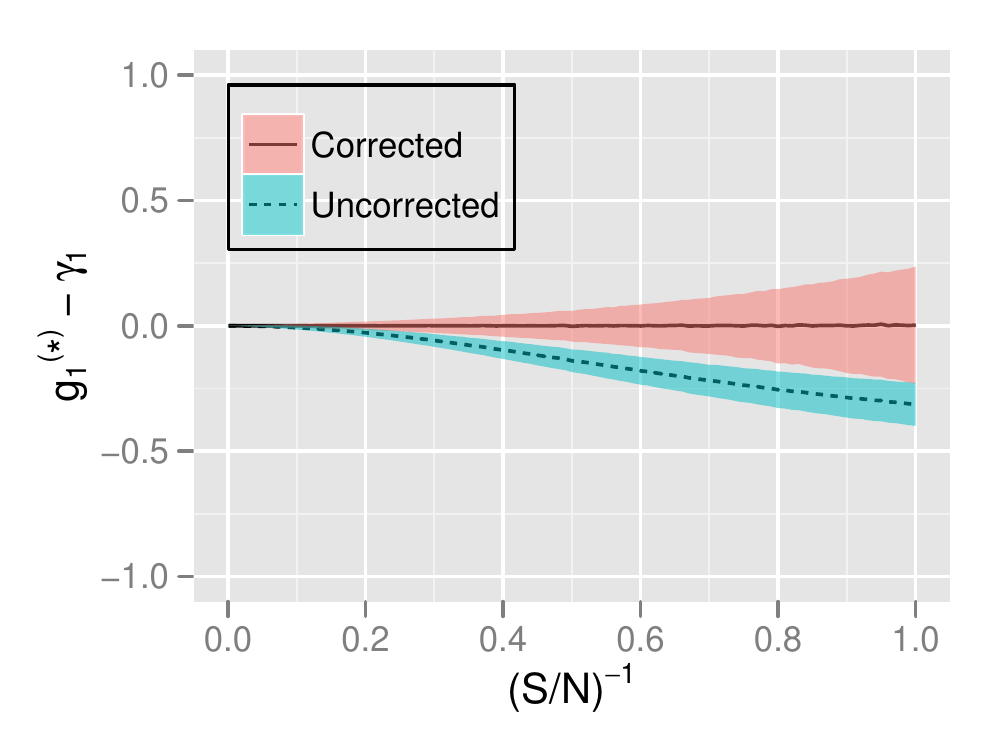}\\
~~~~~~~~Error-Phase Weighted ($a=2,b=0.3$)\\
\includegraphics[width=\columnwidth]{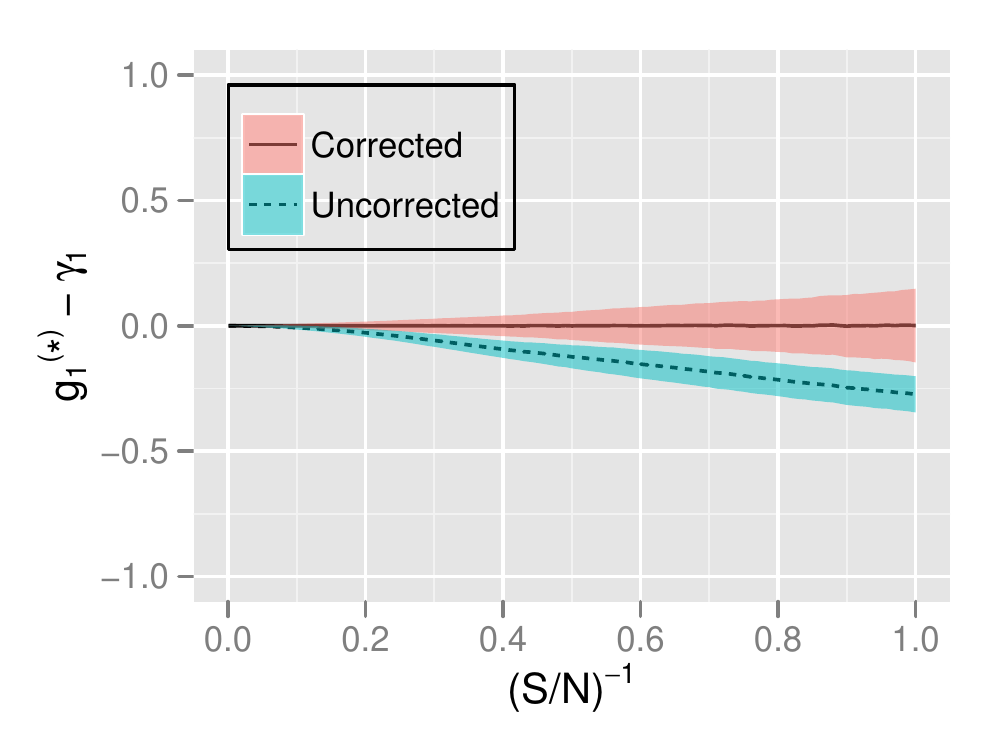}
\end{minipage}
\caption{Noise-biased  ({\it `uncorrected'}\,) versus noise-unbiased ({\it `corrected'}\,) sample skewness for $S/N>1$ and $n=1000$, weighted by phases and errors, according to Eq.~(\ref{eq:SNweights}), with different parameter values, as specified above each panel. 
Shaded areas encompass one standard deviation from the mean of the distribution of the skewness employing simulations defined by Eqs~(\ref{eq:simuStart})--(\ref{eq:simuCoreEnd}). }
\label{fig:M3_1000ph}
\end{figure}

\begin{figure}
\begin{center}
~~~~~~~~{\bf\fbox{\parbox{0.15\textwidth}{\centering Kurtosis \\ $(n=100)$}}}\\
\end{center}
\begin{minipage}{0.5\columnwidth}
\centering
~~~~~~Unweighted \\
\includegraphics[width=\columnwidth]{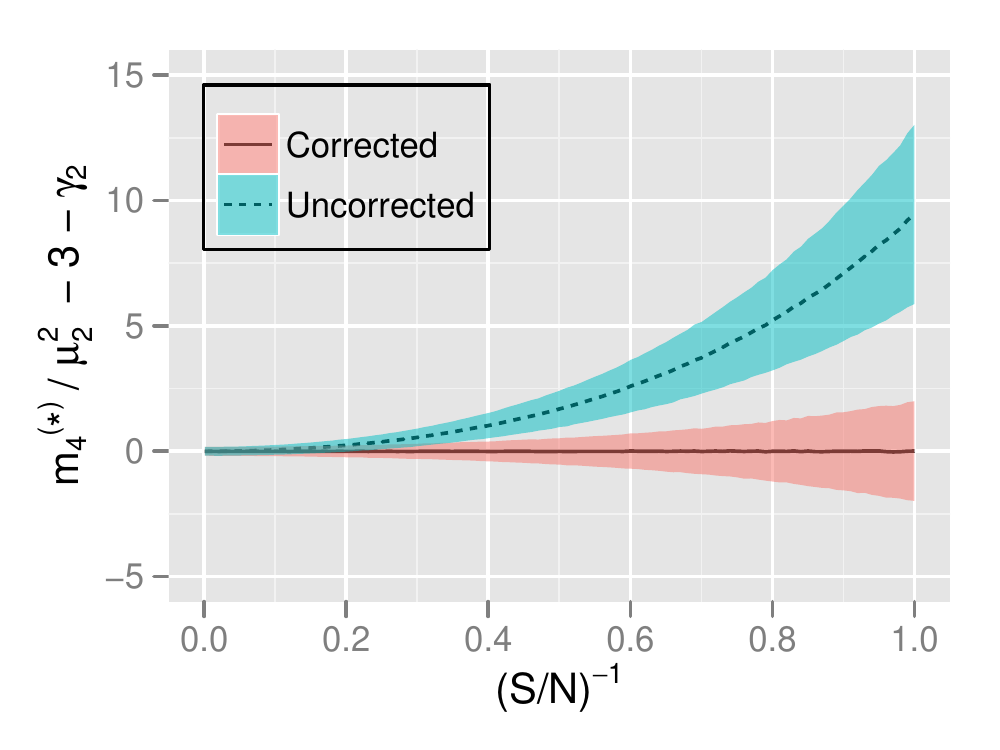}\\
~~~~~~~~Error Weighted\\
\includegraphics[width=\columnwidth]{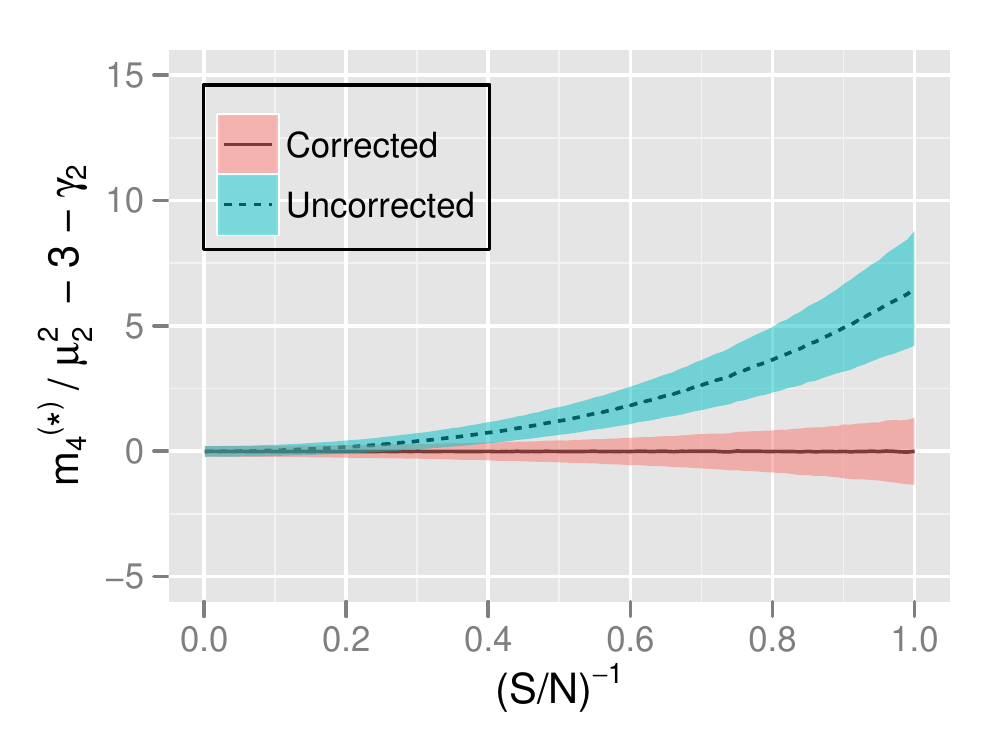}
\end{minipage}
\begin{minipage}{0.5\columnwidth}
\centering
~~~~~~Unweighted \\
\includegraphics[width=\columnwidth]{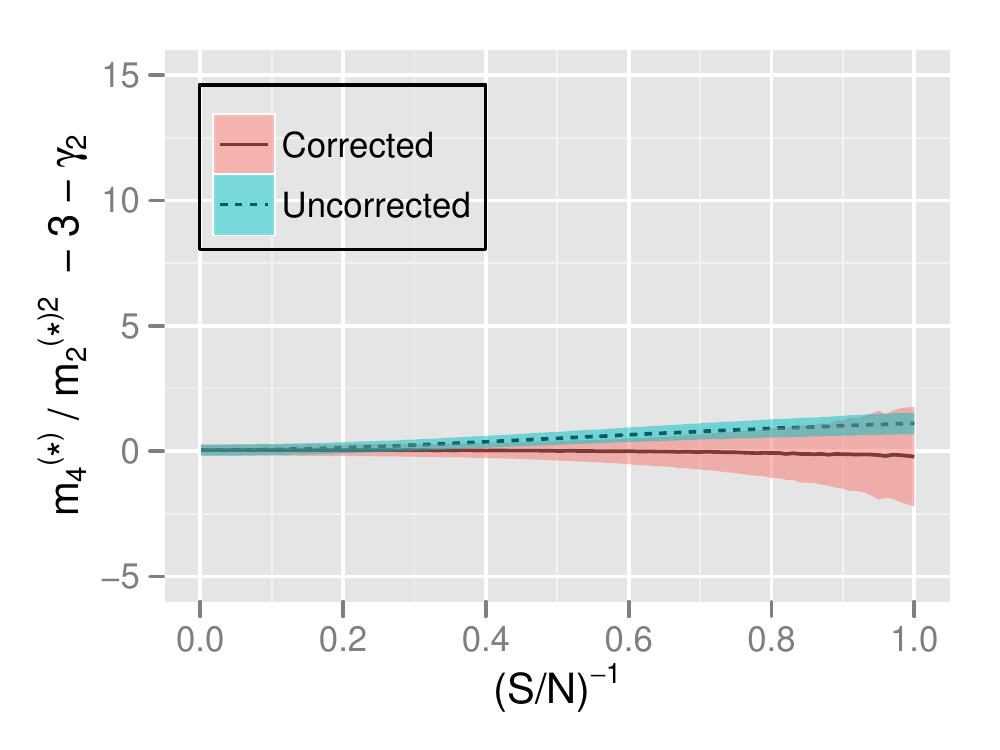}\\
~~~~~~~~Error Weighted\\
\includegraphics[width=\columnwidth]{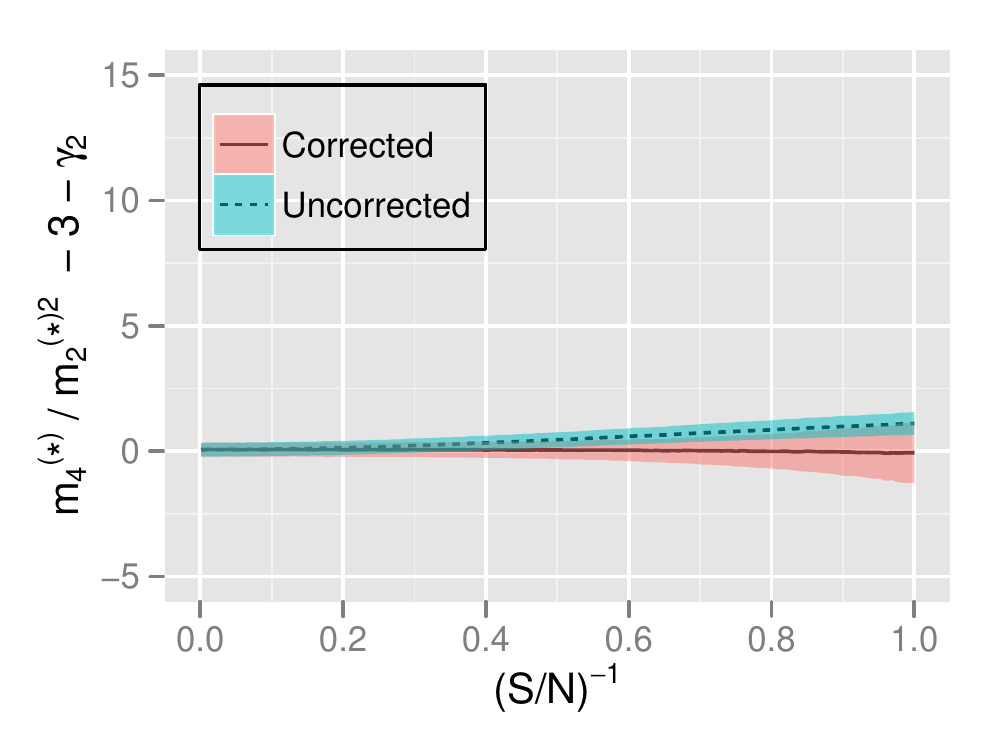}
\end{minipage}
\caption{Noise-biased  ({\it `uncorrected'}\,) versus noise-unbiased ({\it `corrected'}\,) sample kurtosis moment for $S/N>1$ and $n=100$: unweighted in the upper panels and weighted by the inverse of squared measurement errors in the lower panels. Shaded areas encompass one standard deviation from the mean of the distribution of the kurtosis employing simulations defined by Eqs~(\ref{eq:simuStart})--(\ref{eq:simuCoreEnd}). 
}
\label{fig:M4_100}
\end{figure}

\begin{figure}
\begin{center}
~~~~~~~~{\bf\fbox{\parbox{0.15\textwidth}{\centering Kurtosis \\ $(n=100)$}}}\\
\end{center}
\begin{minipage}{0.5\columnwidth}
\centering
~~~~~~~~Phase Weighted ($a,b\rightarrow 0$)\\
\includegraphics[width=\columnwidth]{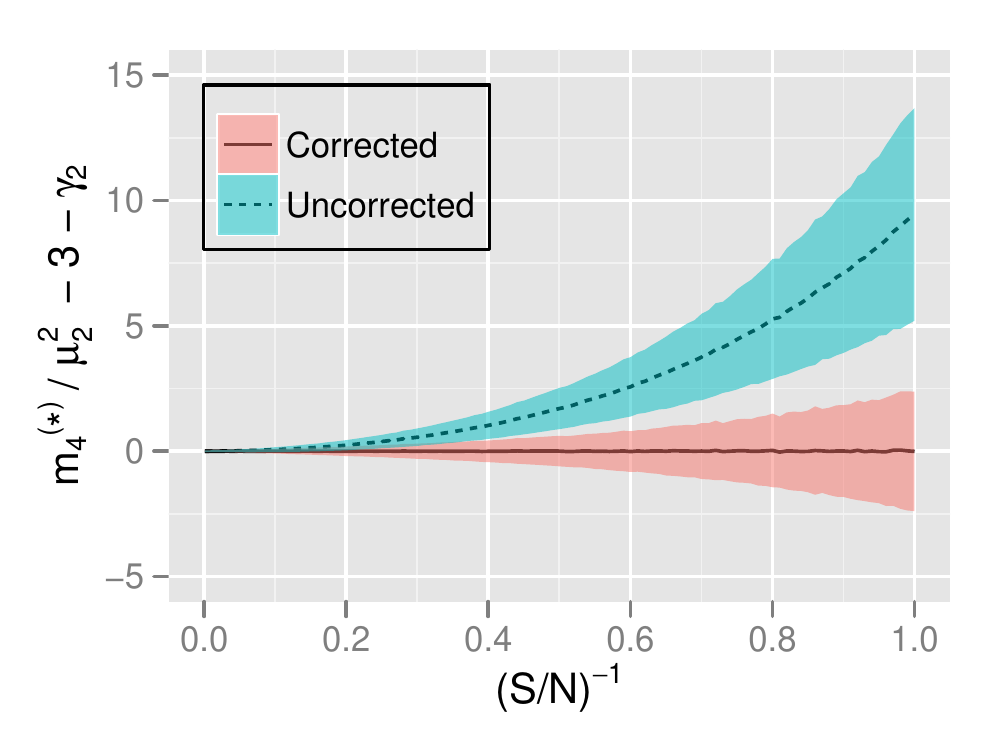}\\
~~~~~~~~Error-Phase Weighted ($a=2,b=0.3$)\\
\includegraphics[width=\columnwidth]{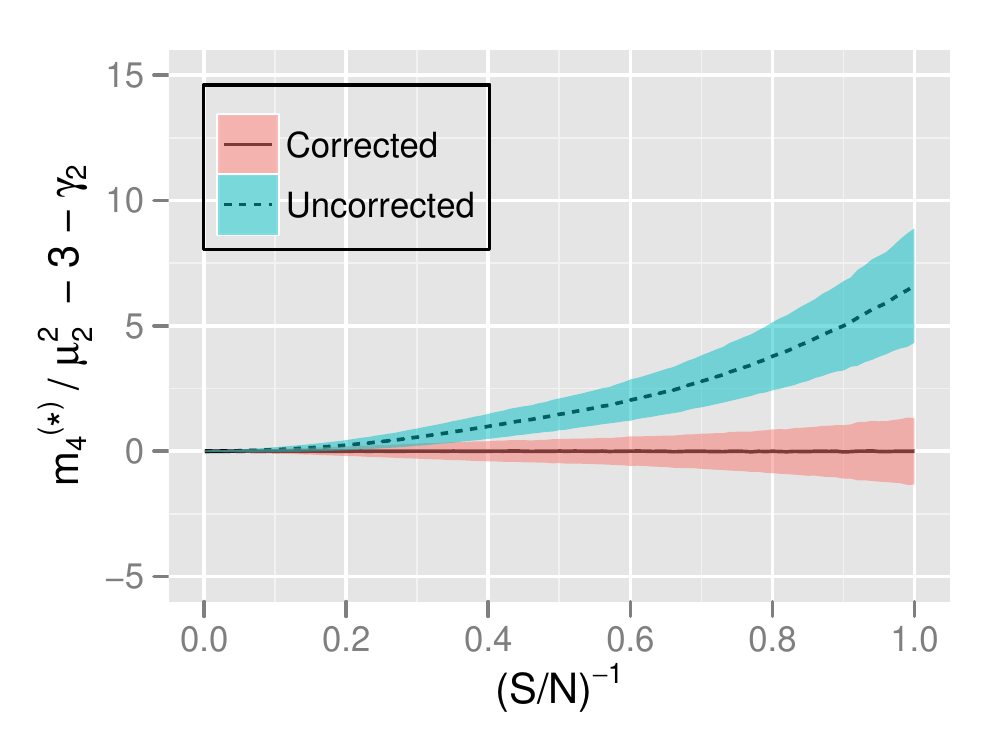}
\end{minipage}
\begin{minipage}{0.5\columnwidth}
\centering
~~~~~~~~Phase Weighted ($a,b\rightarrow 0$)\\
\includegraphics[width=\columnwidth]{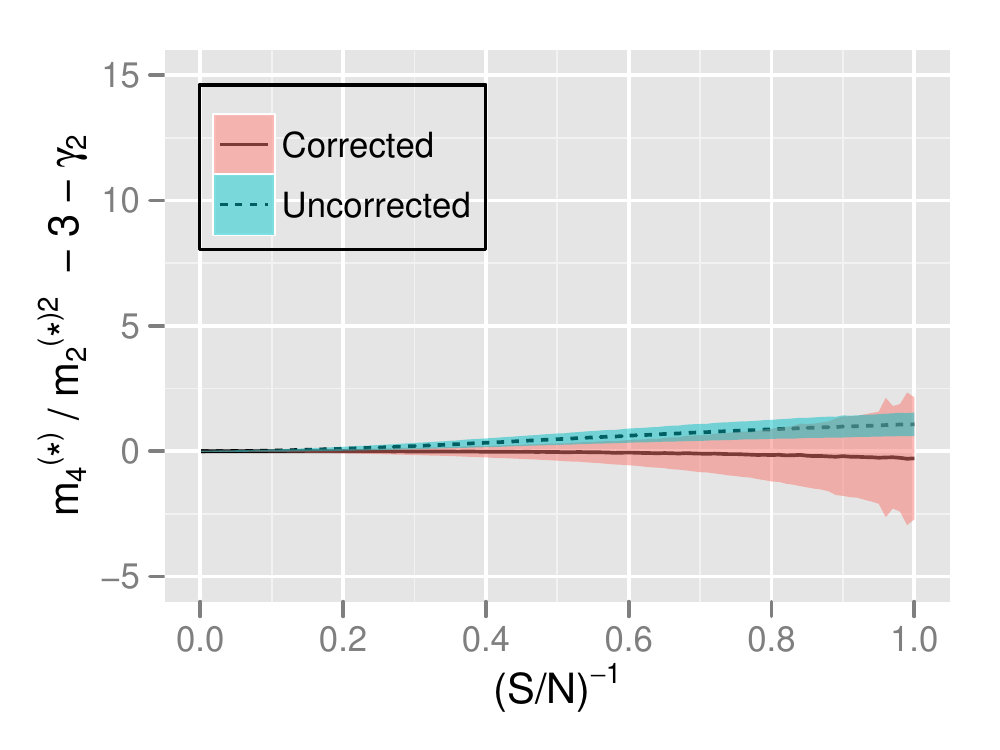}\\
~~~~~~~~Error-Phase Weighted ($a=2,b=0.3$)\\
\includegraphics[width=\columnwidth]{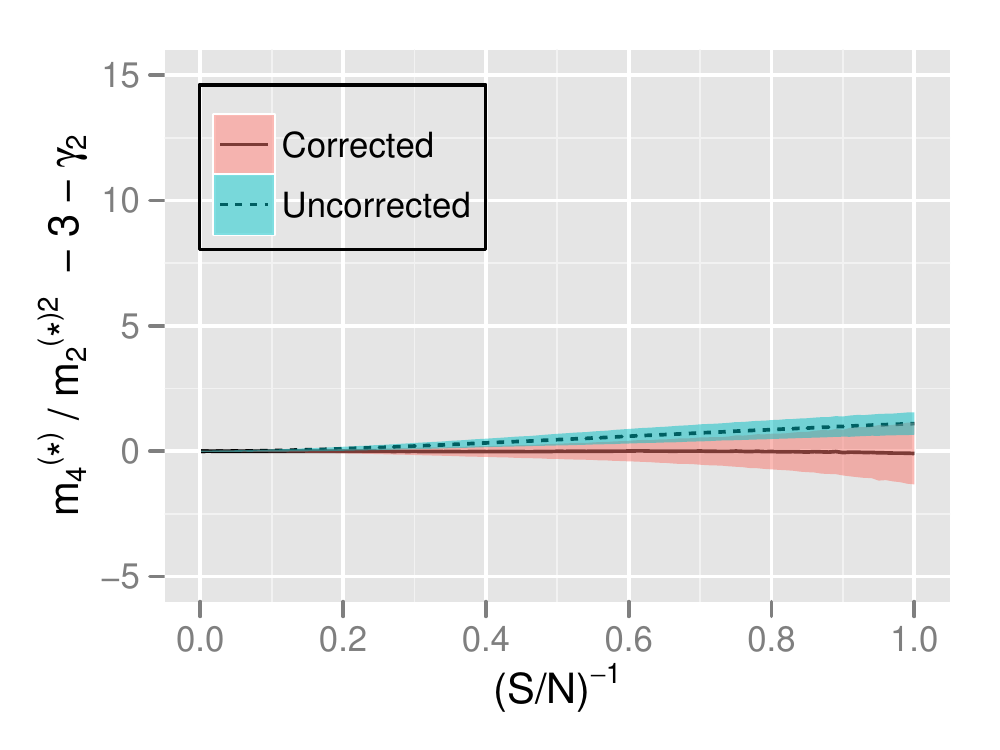}
\end{minipage}
\caption{Noise-biased  ({\it `uncorrected'}\,) versus noise-unbiased ({\it `corrected'}\,) sample kurtosis moment for $S/N>1$ and $n=100$, weighted by phases and errors, according to Eq.~(\ref{eq:SNweights}), with different parameter values, as specified above each panel.  
Shaded areas encompass one standard deviation from the mean of the distribution of the kurtosis employing simulations defined by Eqs~(\ref{eq:simuStart})--(\ref{eq:simuCoreEnd}). }
\label{fig:M4_100ph}
\end{figure}

\begin{figure}
\begin{center}
~~~~~~~~{\bf\fbox{\parbox{0.15\textwidth}{\centering Kurtosis \\ $(n=1000)$}}}\\
\end{center}
\begin{minipage}{0.5\columnwidth}
\centering
~~~~~~Unweighted \\
\includegraphics[width=\columnwidth]{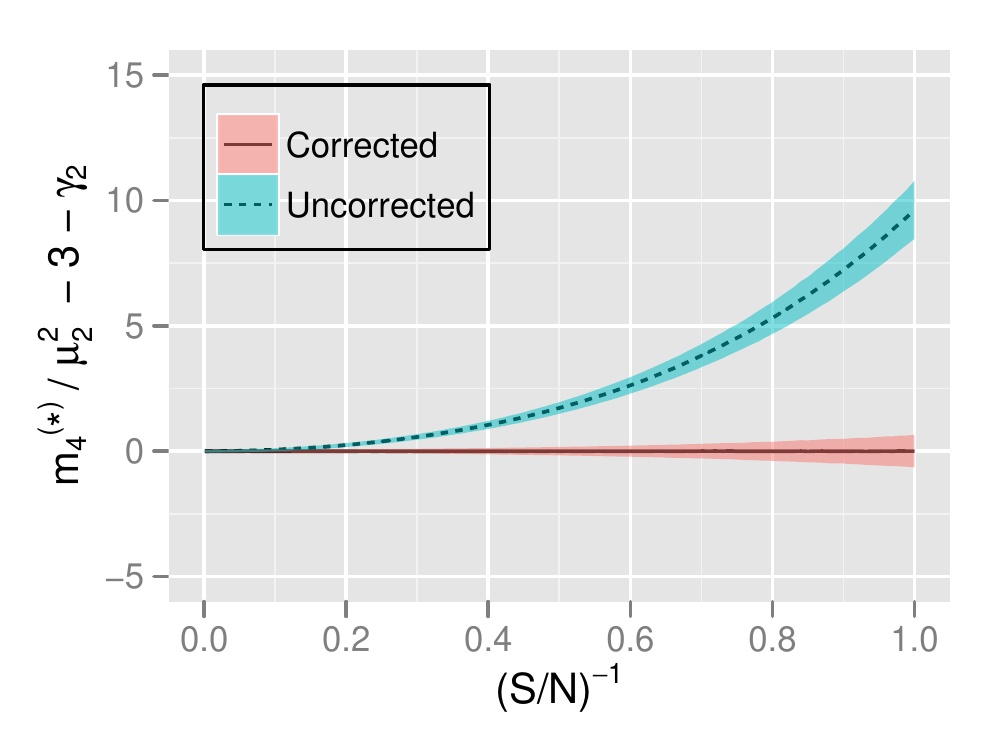}\\
~~~~~~~~Error Weighted\\
\includegraphics[width=\columnwidth]{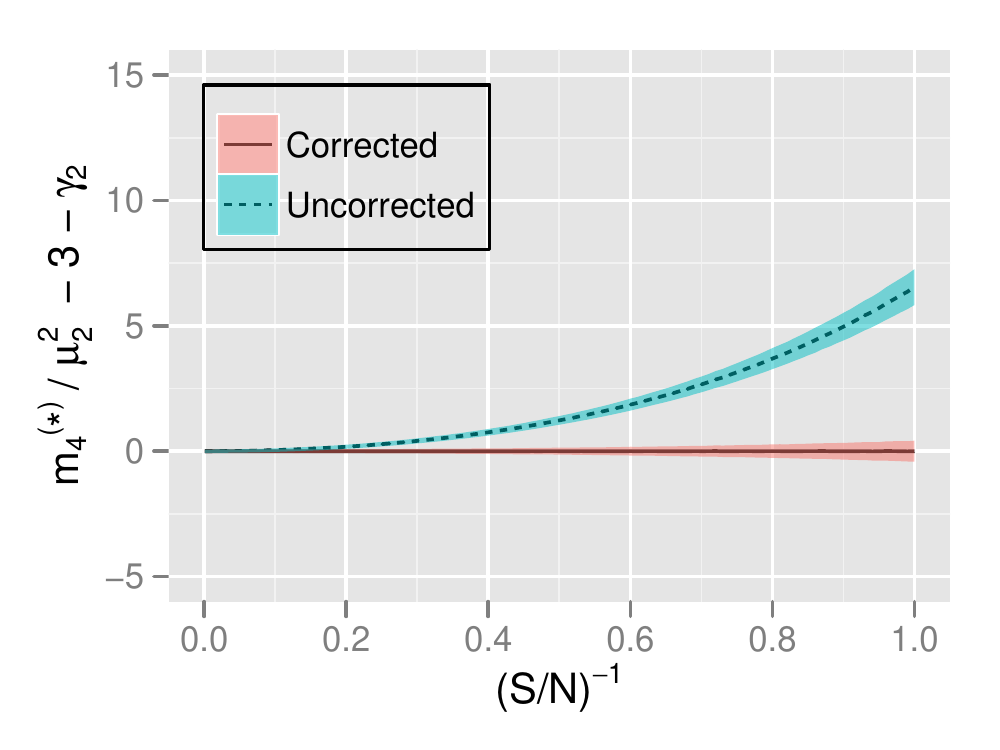}
\end{minipage}
\begin{minipage}{0.5\columnwidth}
\centering
~~~~~~Unweighted \\
\includegraphics[width=\columnwidth]{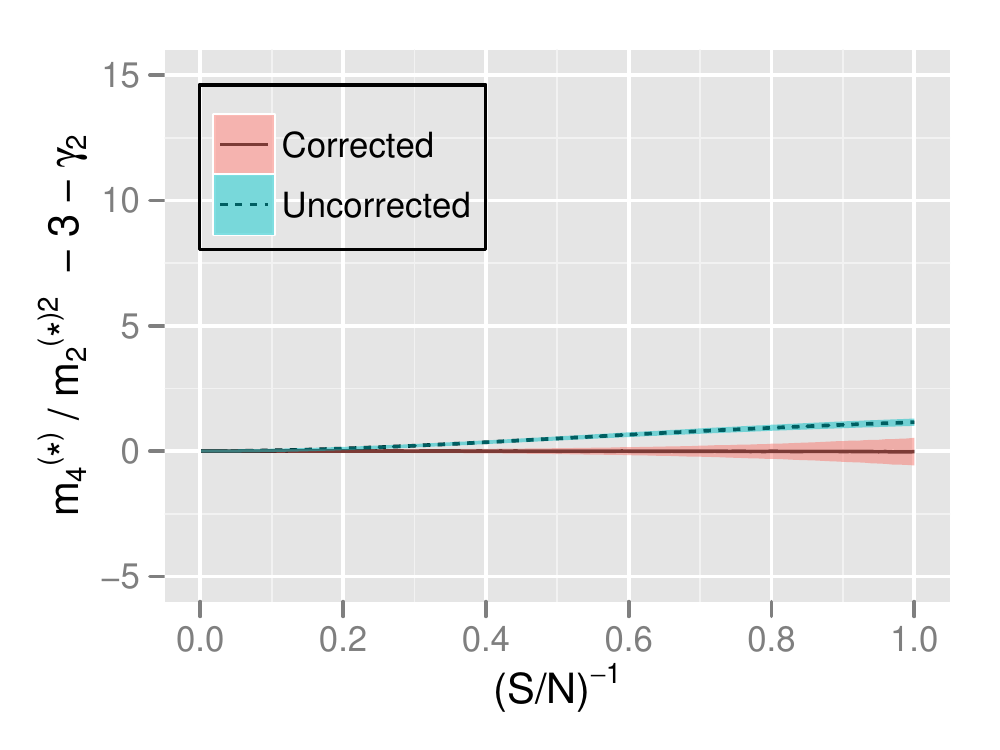}\\
~~~~~~~~Error Weighted\\
\includegraphics[width=\columnwidth]{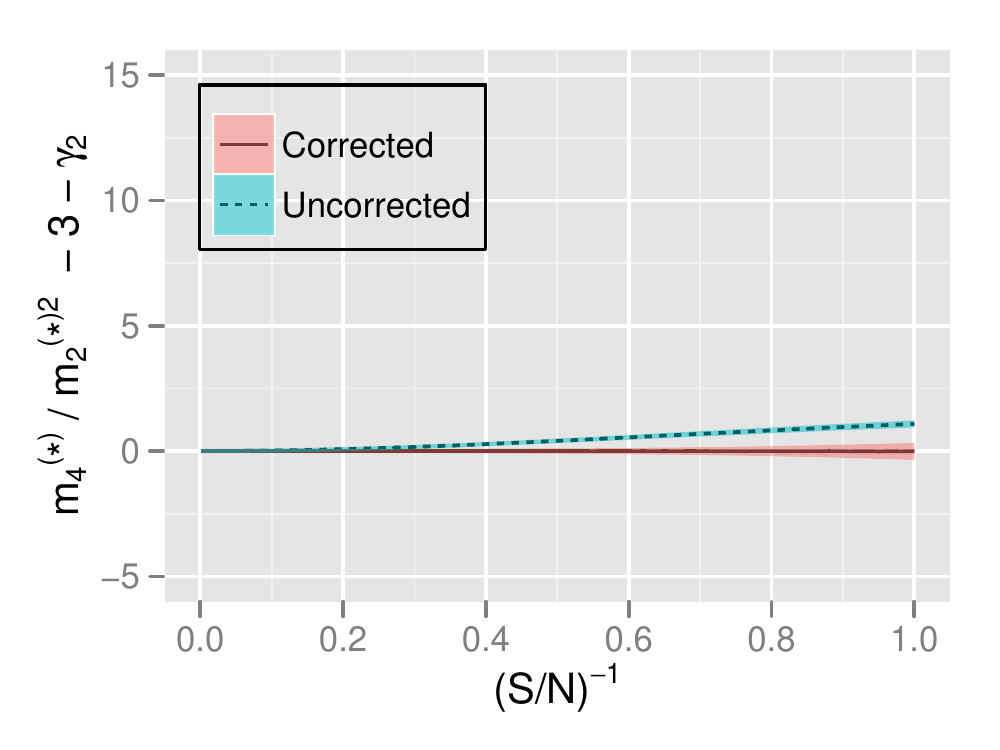}
\end{minipage}
\caption{Noise-biased  ({\it `uncorrected'}\,) versus noise-unbiased ({\it `corrected'}\,) sample kurtosis moment  for $S/N>1$ and $n=1000$: unweighted in the upper panels and weighted by the inverse of squared measurement errors in the lower panels. Shaded areas encompass one standard deviation from the mean of the distribution of the kurtosis employing simulations defined by Eqs~(\ref{eq:simuStart})--(\ref{eq:simuCoreEnd}). 
}
\label{fig:M4_1000}
\end{figure}

\begin{figure}
\begin{center}
~~~~~~~~{\bf\fbox{\parbox{0.15\textwidth}{\centering Kurtosis \\ $(n=1000)$}}}\\
\end{center}
\begin{minipage}{0.5\columnwidth}
\centering
~~~~~~~~Phase Weighted ($a,b\rightarrow 0$)\\
\includegraphics[width=\columnwidth]{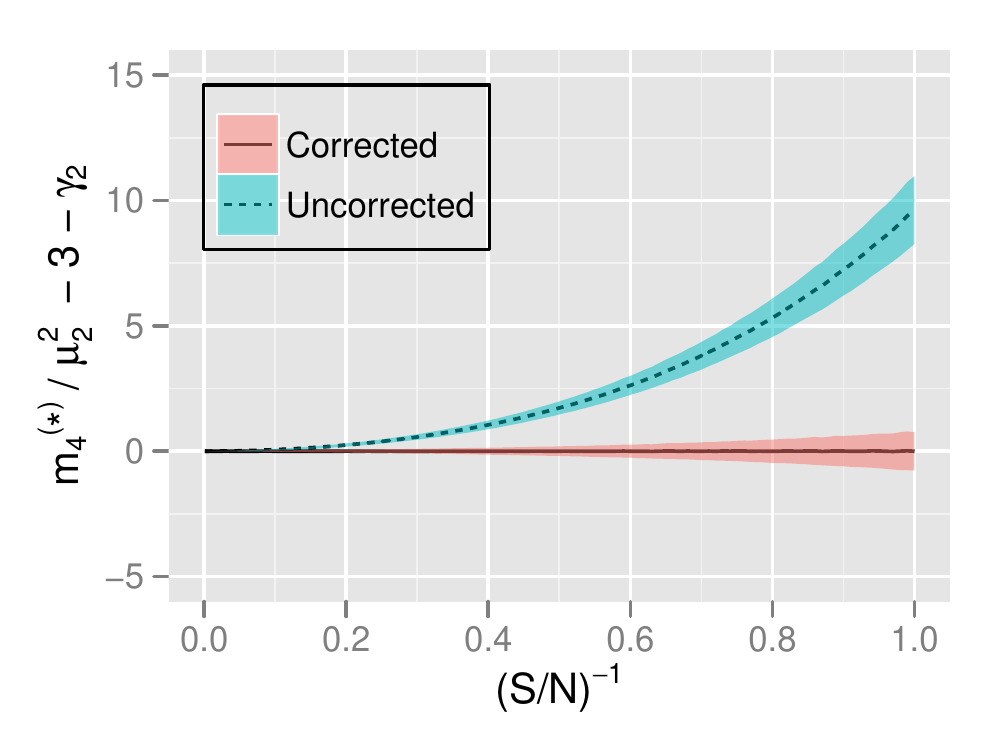}\\
~~~~~~~~Error-Phase Weighted ($a=2,b=0.3$)\\
\includegraphics[width=\columnwidth]{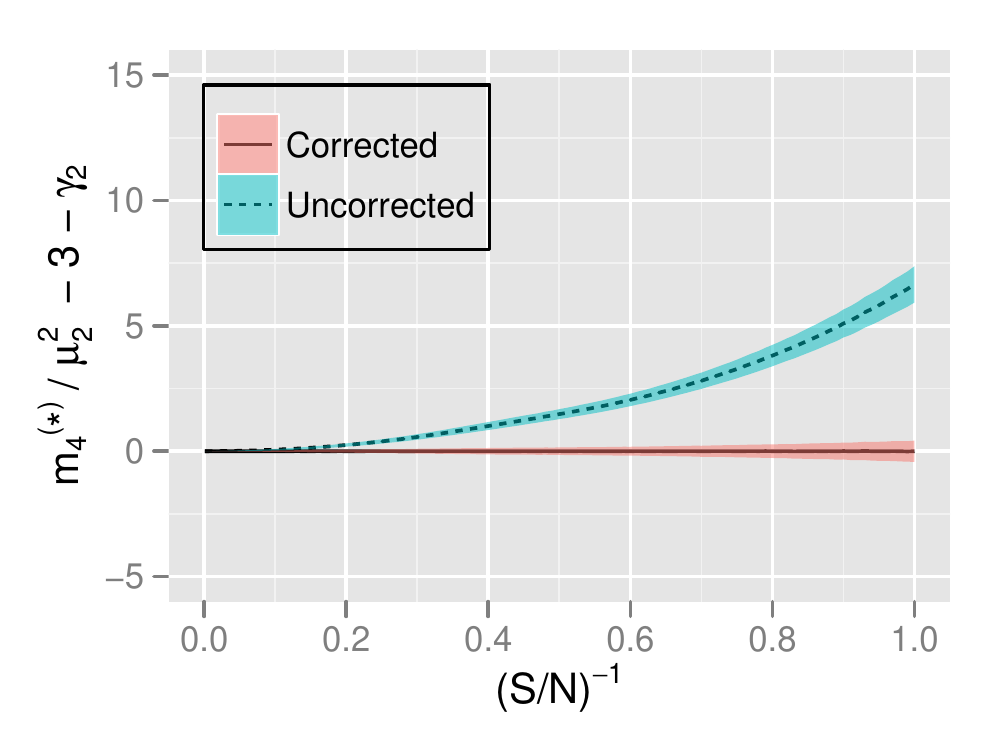}
\end{minipage}
\begin{minipage}{0.5\columnwidth}
\centering
~~~~~~~~Phase Weighted ($a,b\rightarrow 0$)\\
\includegraphics[width=\columnwidth]{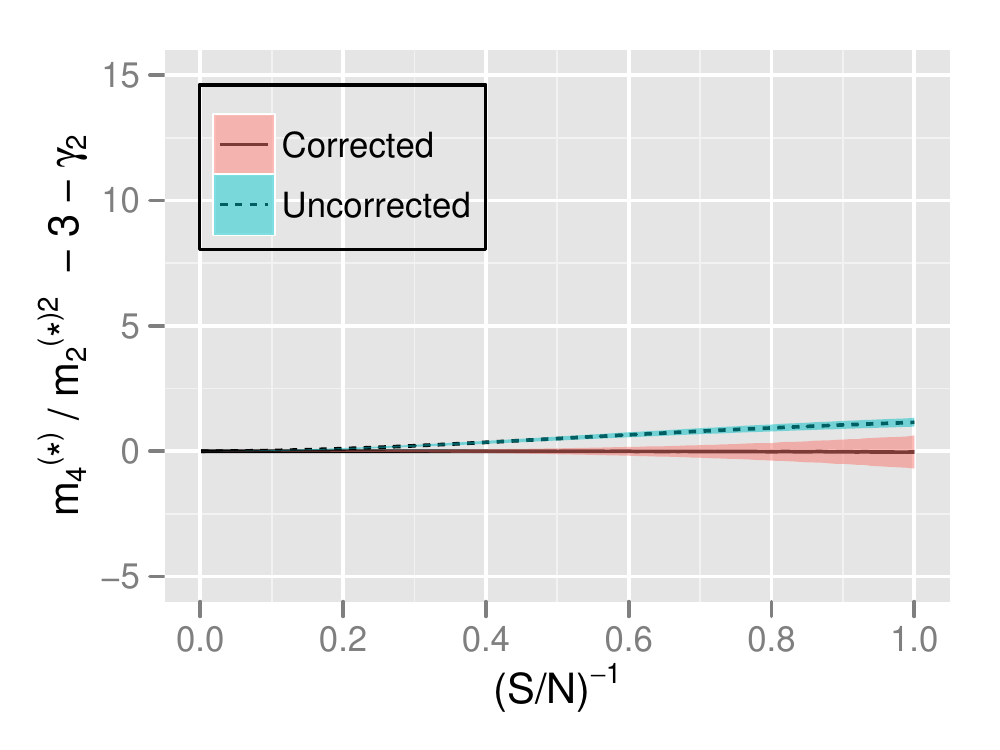}\\
~~~~~~~~Error-Phase Weighted ($a=2,b=0.3$)\\
\includegraphics[width=\columnwidth]{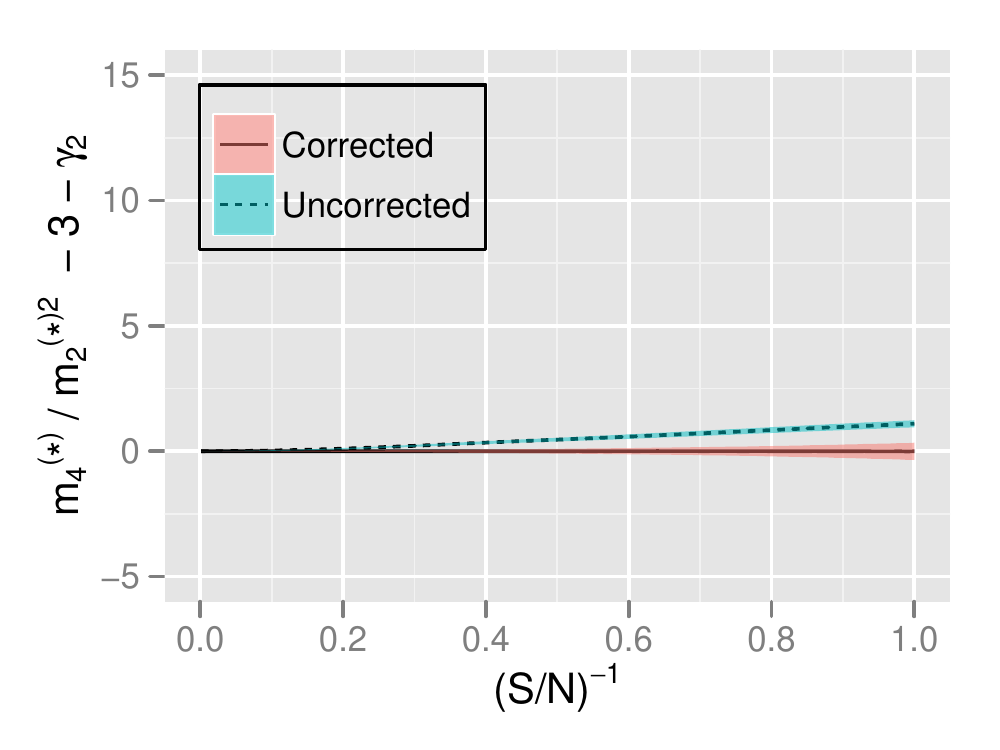}
\end{minipage}
\caption{Noise-biased  ({\it `uncorrected'}\,) versus noise-unbiased ({\it `corrected'}\,) sample kurtosis moment  for $S/N>1$ and $n=1000$, weighted by phases and errors, according to Eq.~(\ref{eq:SNweights}), with different parameter values, as specified above each panel.  
Shaded areas encompass one standard deviation from the mean of the distribution of the kurtosis employing simulations defined by Eqs~(\ref{eq:simuStart})--(\ref{eq:simuCoreEnd}). }
\label{fig:M4_1000ph}
\end{figure}

\begin{figure}
\begin{center}
~~~~~~~~{\bf\fbox{\parbox{0.15\textwidth}{\centering {\em k-}Kurtosis \\ $(n=100)$}}}\\
\end{center}
\begin{minipage}{0.5\columnwidth}
\centering
~~~~~~Unweighted \\
\includegraphics[width=\columnwidth]{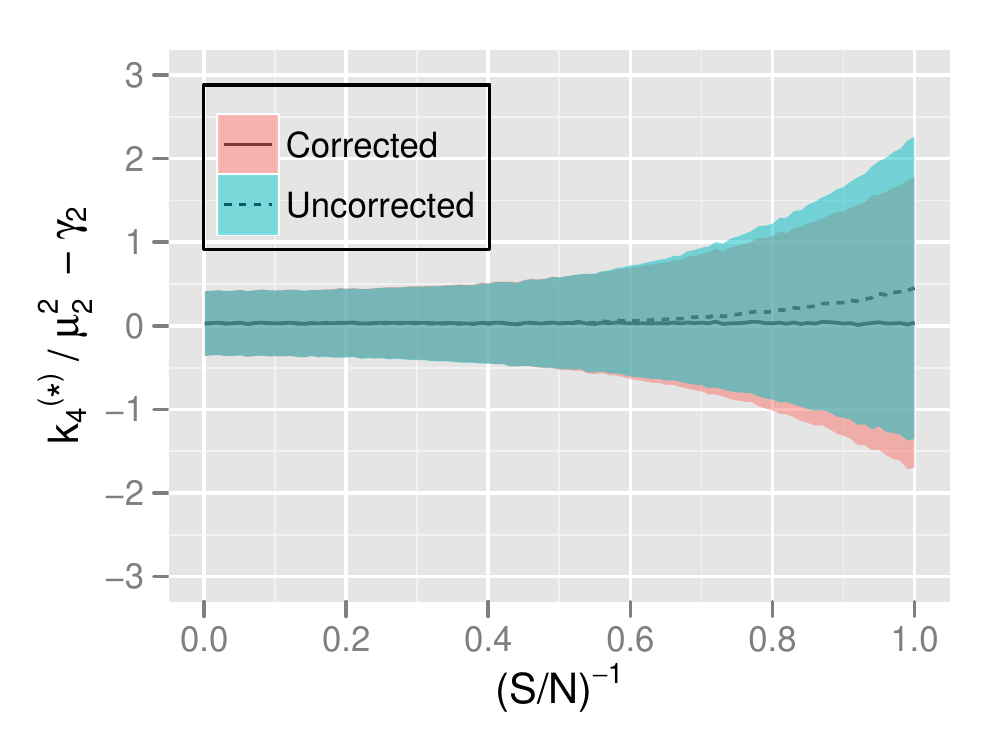}\\
~~~~~~~~Error Weighted\\
\includegraphics[width=\columnwidth]{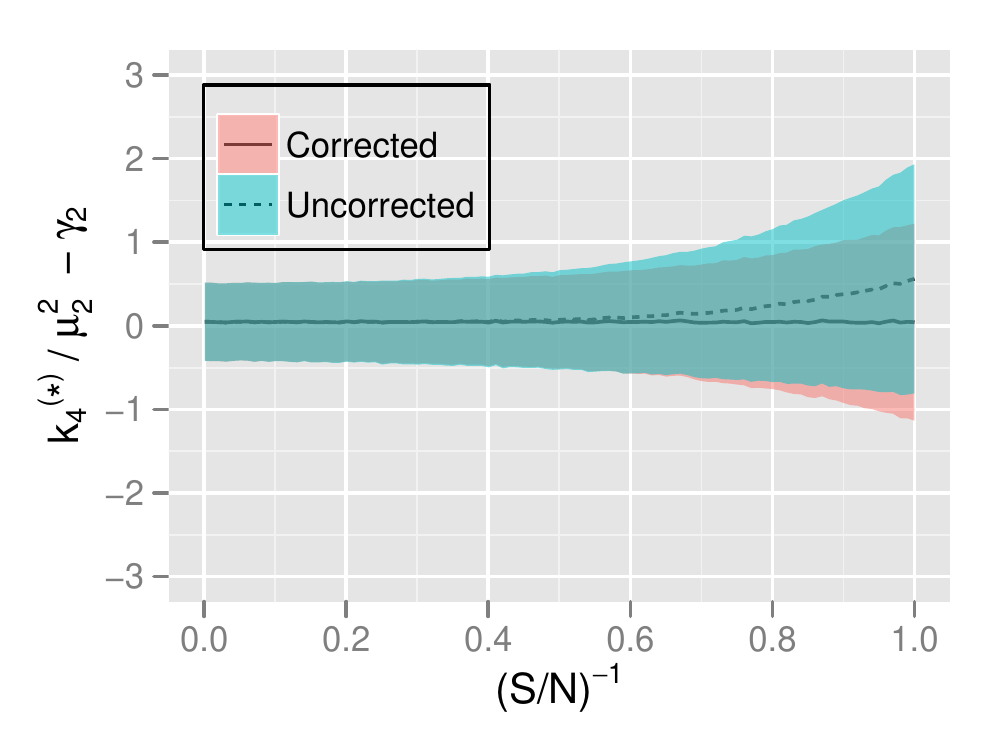}
\end{minipage}
\begin{minipage}{0.5\columnwidth}
\centering
~~~~~~Unweighted \\
\includegraphics[width=\columnwidth]{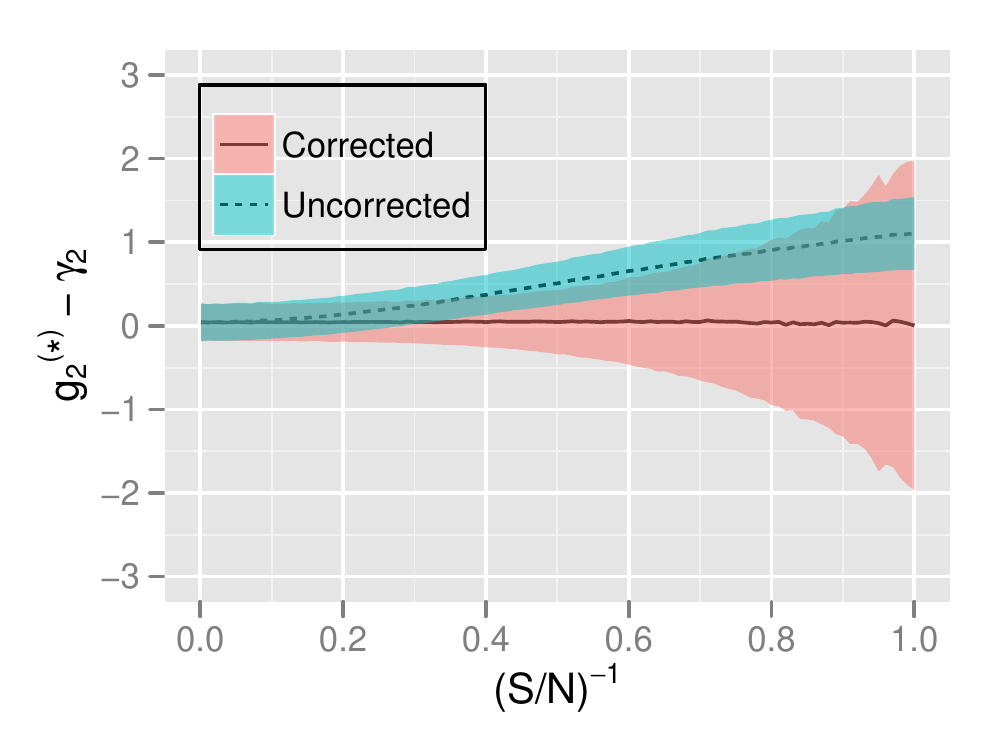}\\
~~~~~~~~Error Weighted\\
\includegraphics[width=\columnwidth]{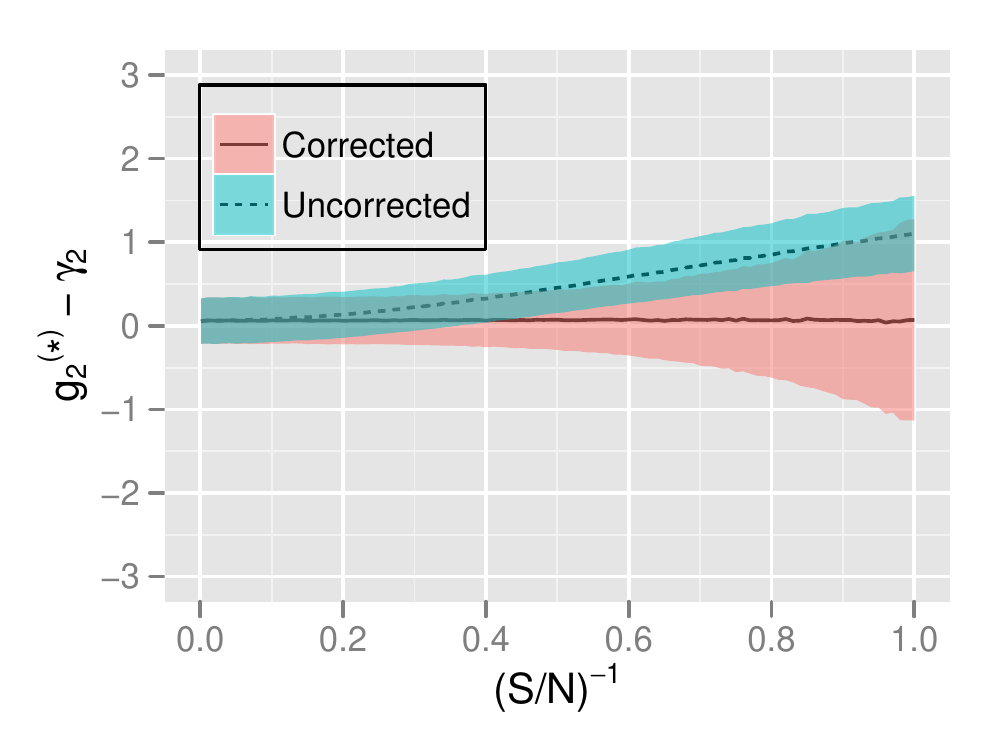}
\end{minipage}
\caption{Noise-biased  ({\it `uncorrected'}\,) versus noise-unbiased ({\it `corrected'}\,) sample kurtosis cumulant for $S/N>1$ and $n=100$: unweighted in the upper panels and weighted by the inverse of squared measurement errors in the lower panels. Shaded areas encompass one standard deviation from the mean of the distribution of the kurtosis employing simulations defined by Eqs~(\ref{eq:simuStart})--(\ref{eq:simuCoreEnd}). 
}
\label{fig:K4_100}
\end{figure}

\begin{figure}
\begin{center}
~~~~~~~~{\bf\fbox{\parbox{0.15\textwidth}{\centering {\em k-}Kurtosis \\ $(n=100)$}}}\\
\end{center}
\begin{minipage}{0.5\columnwidth}
\centering
~~~~~~~~Phase Weighted ($a,b\rightarrow 0$)\\
\includegraphics[width=\columnwidth]{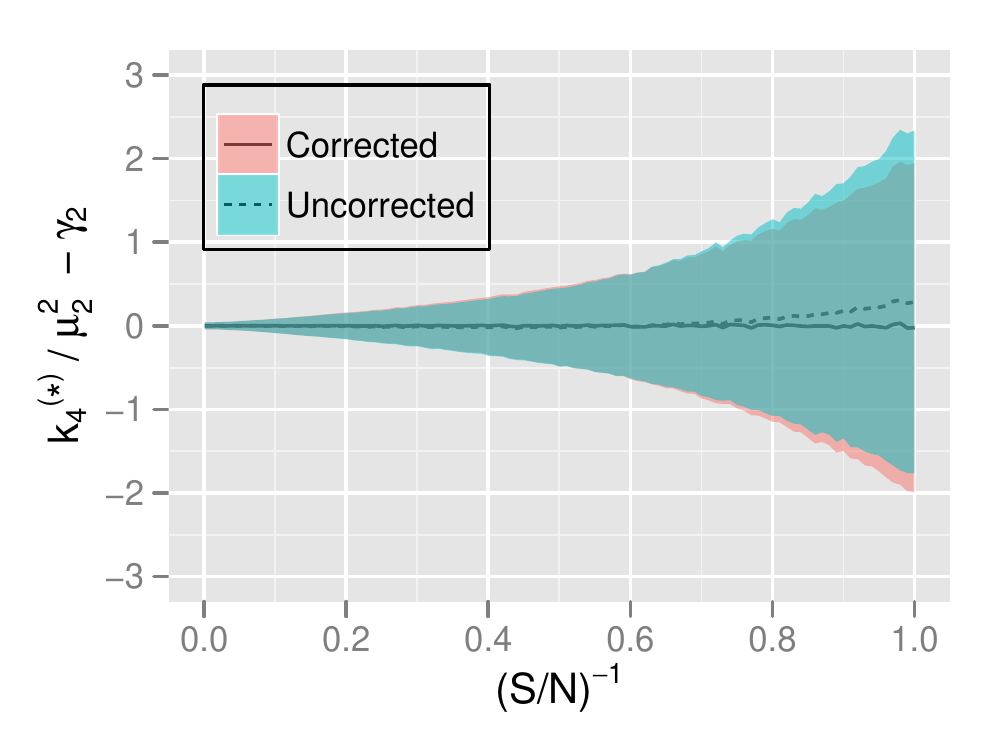}\\
~~~~~~~~Error-Phase Weighted ($a=2,b=0.3$)\\
\includegraphics[width=\columnwidth]{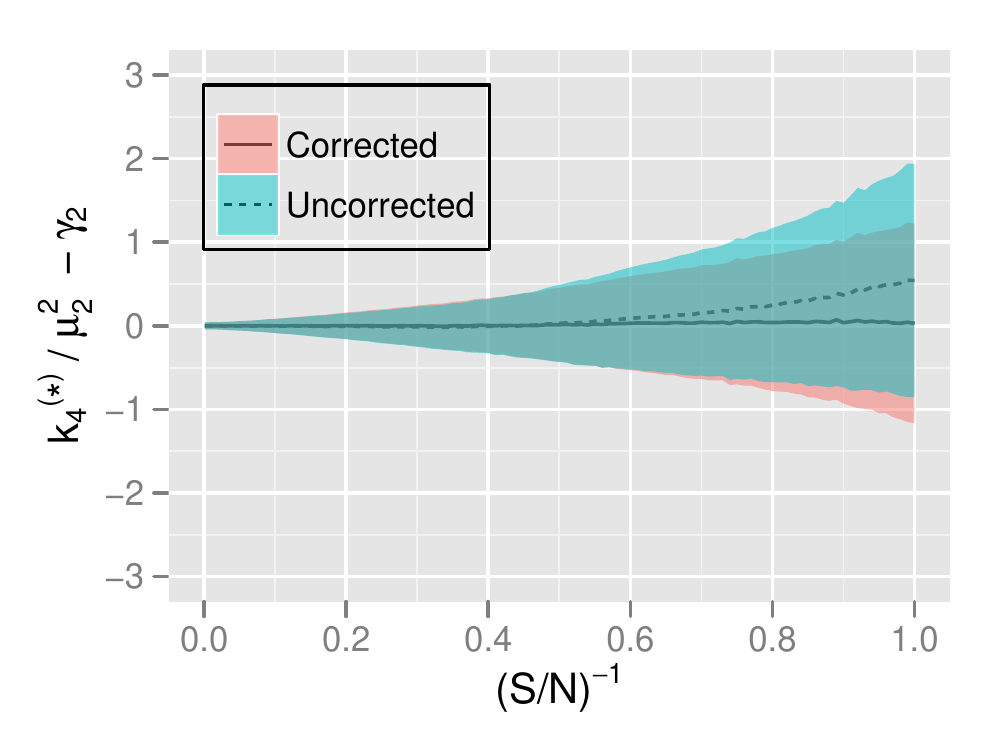}
\end{minipage}
\begin{minipage}{0.5\columnwidth}
\centering
~~~~~~~~Phase Weighted ($a,b\rightarrow 0$)\\
\includegraphics[width=\columnwidth]{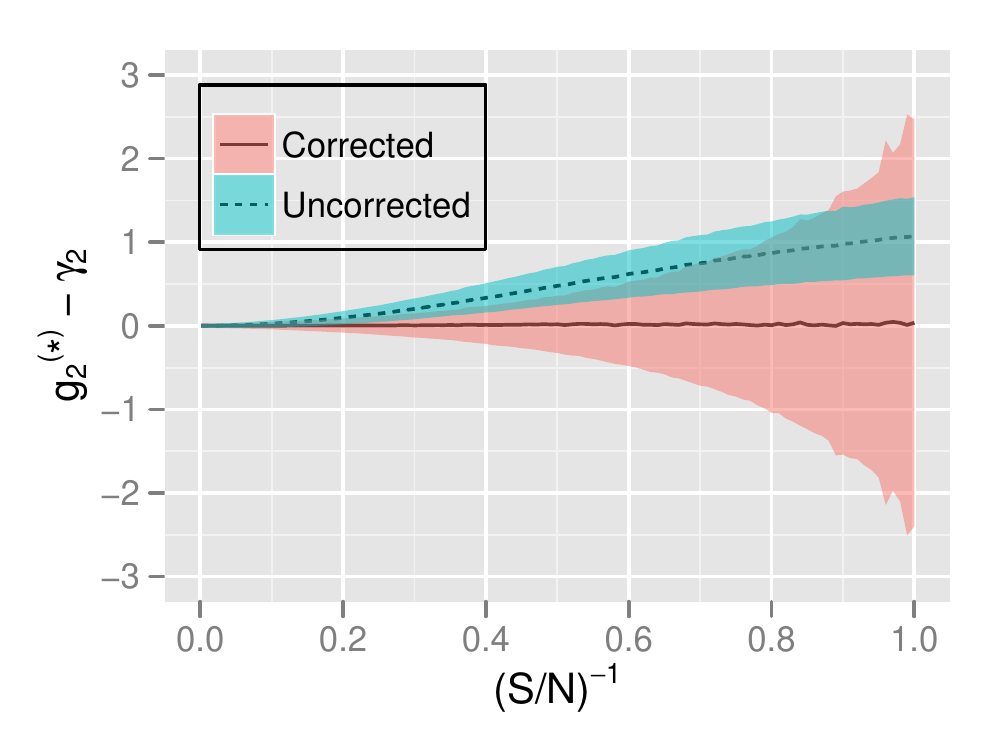}\\
~~~~~~~~Error-Phase Weighted ($a=2,b=0.3$)\\
\includegraphics[width=\columnwidth]{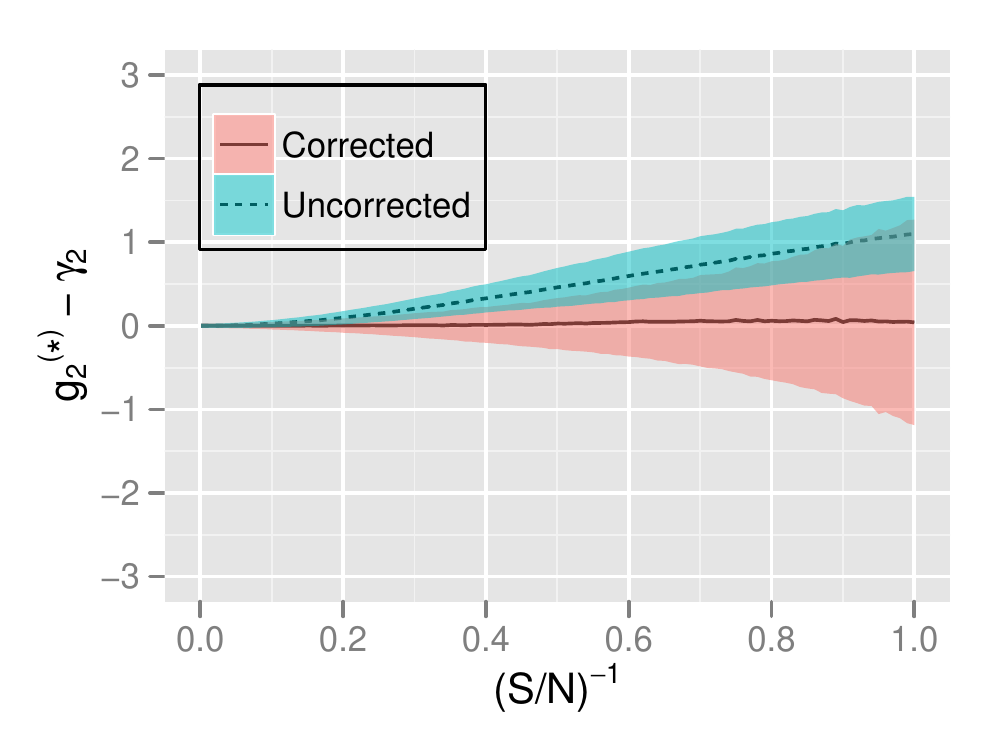}
\end{minipage}
\caption{Noise-biased  ({\it `uncorrected'}\,) versus noise-unbiased ({\it `corrected'}\,) sample kurtosis cumulant for $S/N>1$ and $n=100$, weighted by phases and errors, according to Eq.~(\ref{eq:SNweights}), with different parameter values, as specified above each panel.
Shaded areas encompass one standard deviation from the mean of the distribution of the kurtosis employing simulations defined by Eqs~(\ref{eq:simuStart})--(\ref{eq:simuCoreEnd}).  }
\label{fig:K4_100ph}
\end{figure}

\begin{figure}
\begin{center}
~~~~~~~~{\bf\fbox{\parbox{0.15\textwidth}{\centering {\em k-}Kurtosis \\ $(n=1000)$}}}\\
\end{center}
\begin{minipage}{0.5\columnwidth}
\centering
~~~~~~Unweighted \\
\includegraphics[width=\columnwidth]{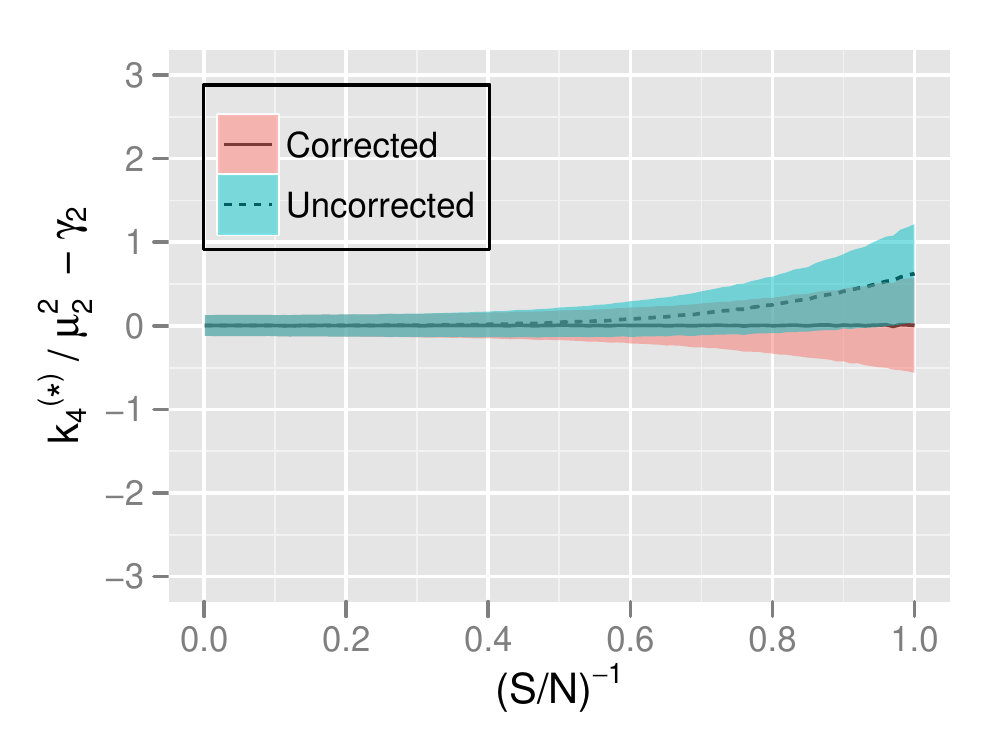}\\
~~~~~~~~Error Weighted\\
\includegraphics[width=\columnwidth]{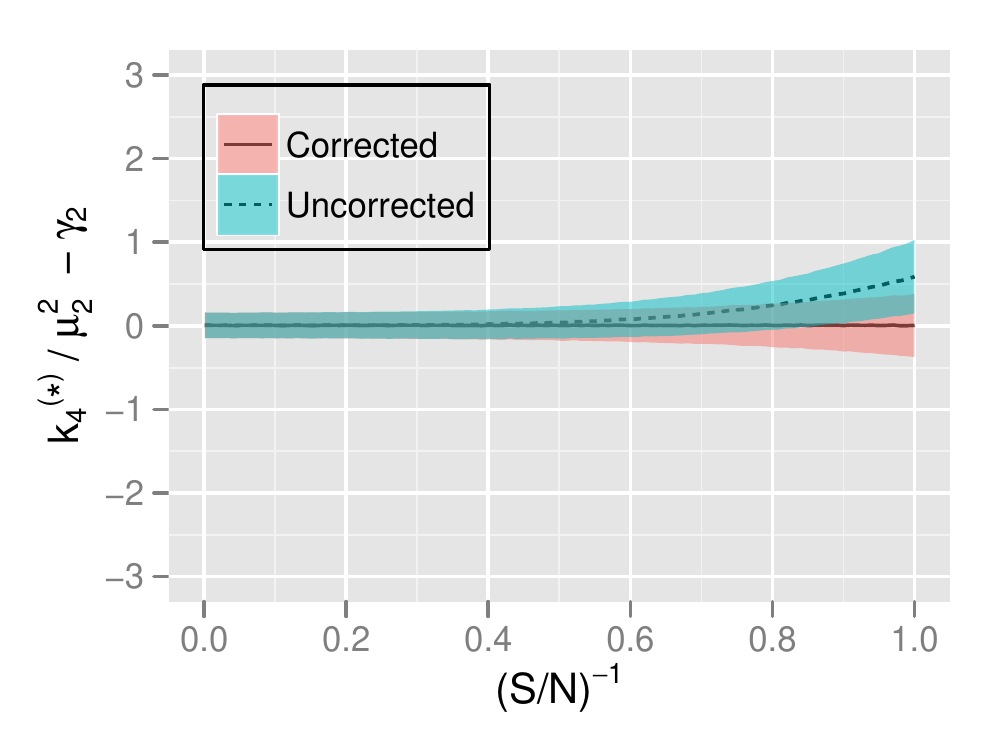}
\end{minipage}
\begin{minipage}{0.5\columnwidth}
\centering
~~~~~~Unweighted \\
\includegraphics[width=\columnwidth]{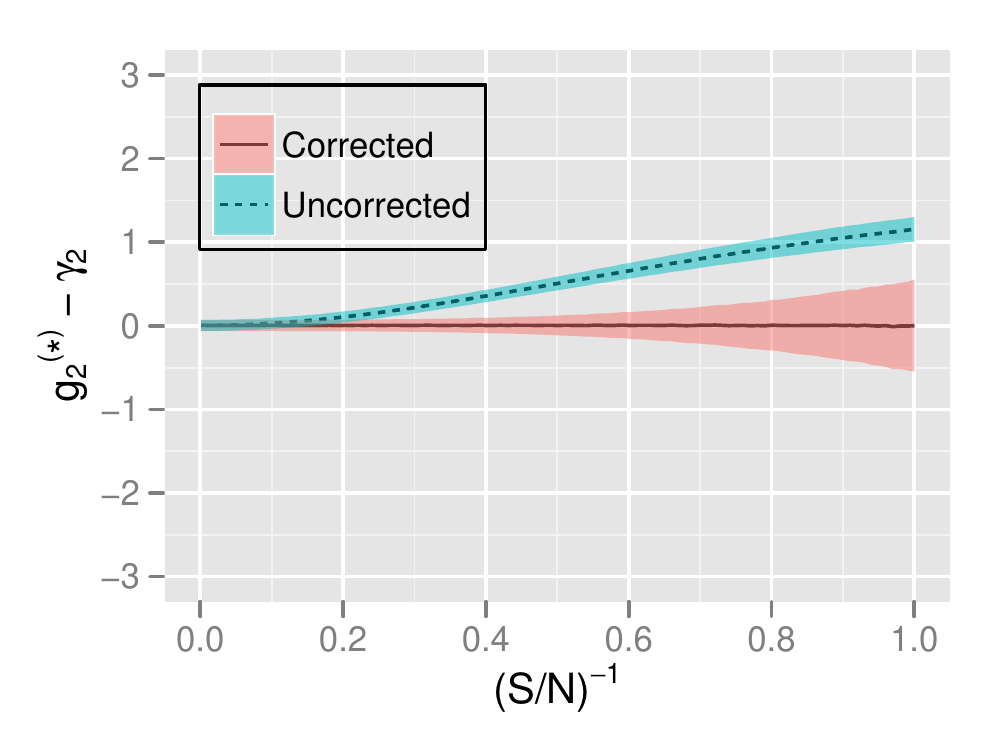}\\
~~~~~~~~Error Weighted\\
\includegraphics[width=\columnwidth]{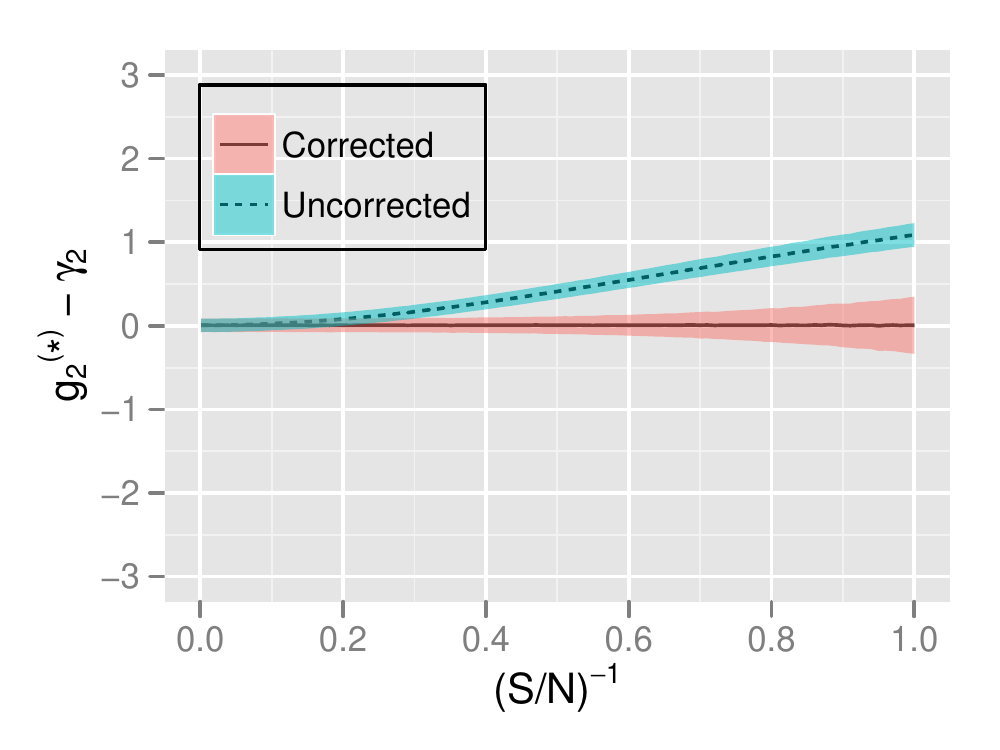}
\end{minipage}
\caption{Noise-biased  ({\it `uncorrected'}\,) versus noise-unbiased ({\it `corrected'}\,) sample kurtosis cumulant for $S/N>1$ and $n=1000$: unweighted in the upper panels and weighted by the inverse of squared measurement errors in the lower panels. Shaded areas encompass one standard deviation from the mean of the distribution of the kurtosis employing simulations defined by Eqs~(\ref{eq:simuStart})--(\ref{eq:simuCoreEnd}). 
}
\label{fig:K4_1000}
\end{figure}

\begin{figure}
\begin{center}
~~~~~~~~{\bf\fbox{\parbox{0.15\textwidth}{\centering {\em k-}Kurtosis \\ $(n=1000)$}}}\\
\end{center}
\begin{minipage}{0.5\columnwidth}
\centering
~~~~~~~~Phase Weighted ($a,b\rightarrow 0$)\\
\includegraphics[width=\columnwidth]{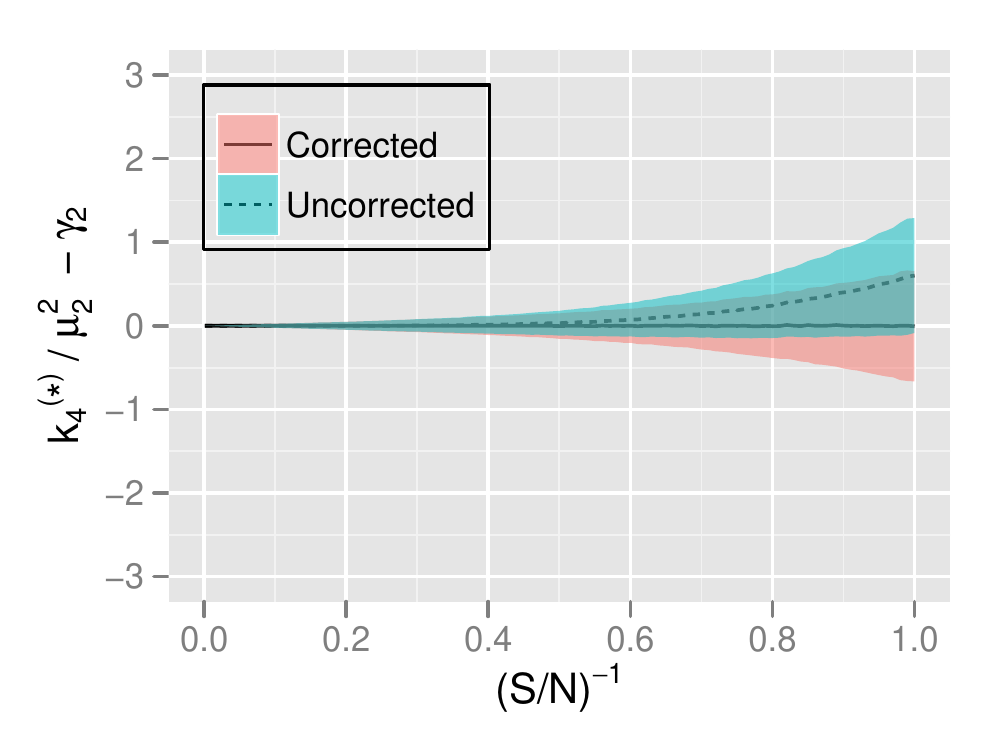}\\
~~~~~~~~Error-Phase Weighted ($a=2,b=0.3$)\\
\includegraphics[width=\columnwidth]{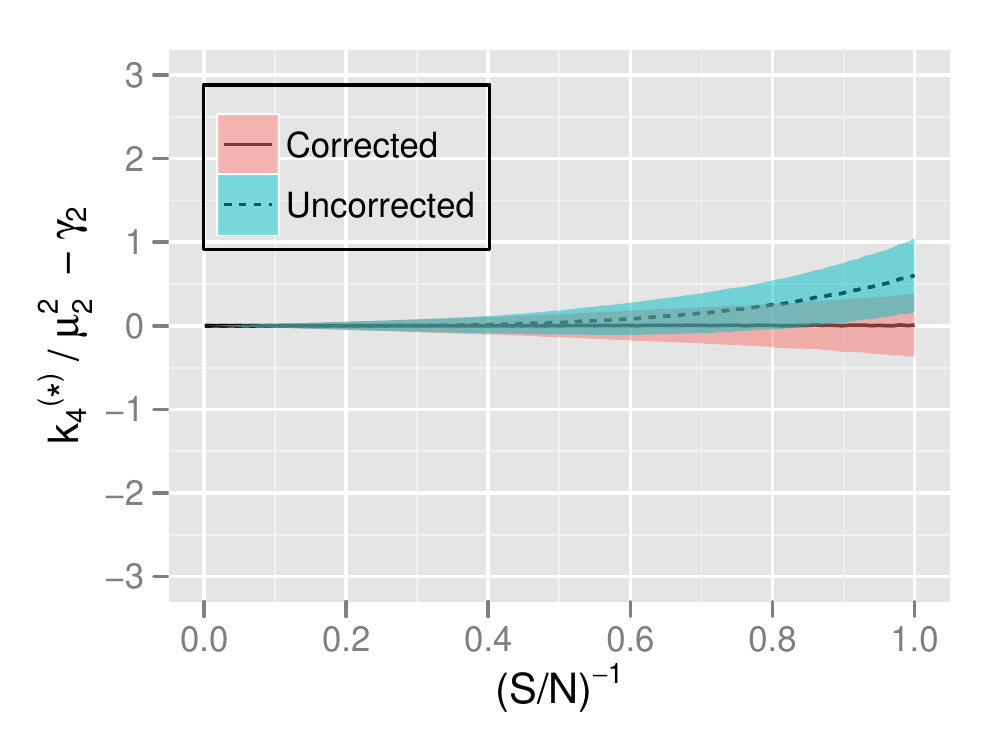}
\end{minipage}
\begin{minipage}{0.5\columnwidth}
\centering
~~~~~~~~Phase Weighted ($a,b\rightarrow 0$)\\
\includegraphics[width=\columnwidth]{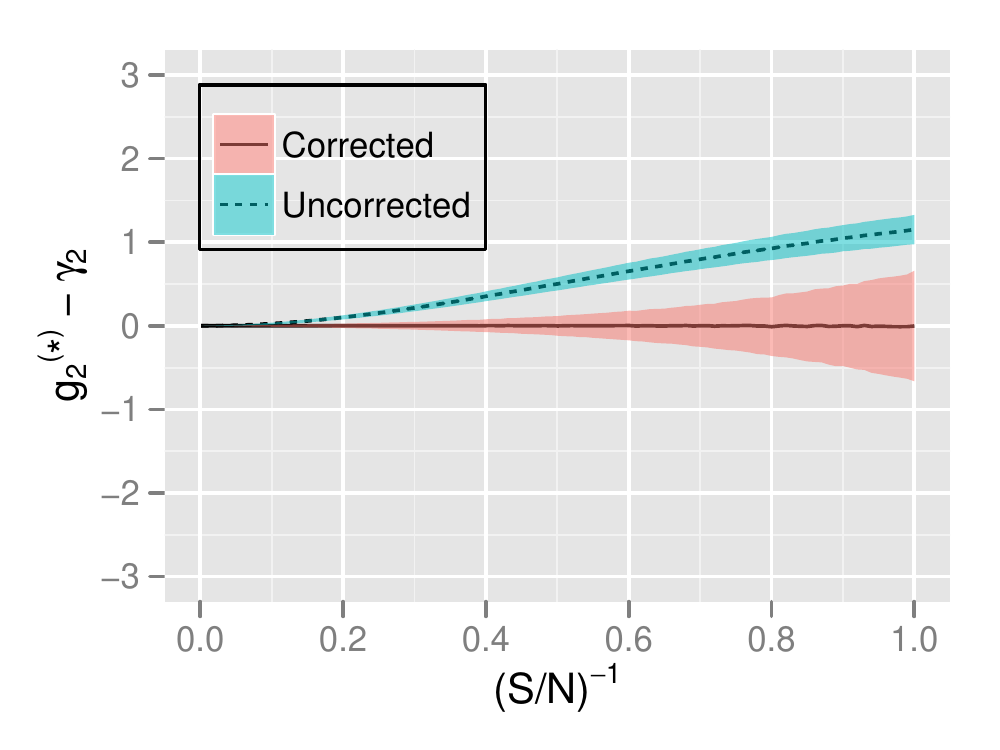}\\
~~~~~~~~Error-Phase Weighted ($a=2,b=0.3$)\\
\includegraphics[width=\columnwidth]{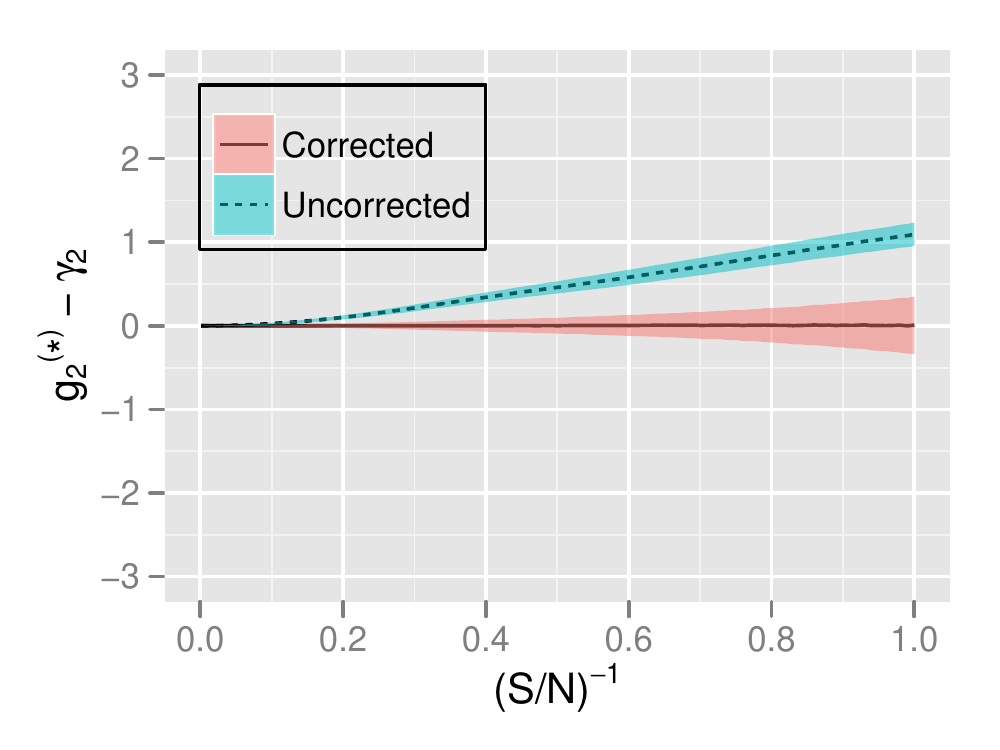}
\end{minipage}
\caption{Noise-biased  ({\it `uncorrected'}\,) versus noise-unbiased ({\it `corrected'}\,) sample kurtosis cumulant for $S/N>1$ and $n=1000$, weighted by phases and errors, according to Eq.~(\ref{eq:SNweights}), with different parameter values, as specified above each panel.  
Shaded areas encompass one standard deviation from the mean of the distribution of the kurtosis employing simulations defined by Eqs~(\ref{eq:simuStart})--(\ref{eq:simuCoreEnd}).  }
\label{fig:K4_1000ph}
\end{figure}

Figures~\ref{fig:M2_100}--\ref{fig:K4_1000ph} compare noise-biased ({\it `uncorrected'}\,) and noise-unbiased ({\it `corrected'}\,) estimators as a function of $S/N$, evaluating the following deviations from the population values:
\begin{align}
&m_2/\mu_2-1~~~~~~~~~\mbox{vs}~~m^*_2/\mu_2-1, \\
&m_3/\mu_2^{3/2}-\gamma_1~~~~~\mbox{vs}~~m^*_3/\mu_2^{3/2}-\gamma_1,~~~~~~g_1-\gamma_1~~~~~~~~~~~~~\mbox{vs}~~g^*_1-\gamma_1,\\
&m_4/\mu_2^2-3-\gamma_2~~\mbox{vs}~~m^*_4/\mu_2^2-3-\gamma_2,~~~m_4/m_2^2-3-\gamma_2~~\mbox{vs}~~m^*_4/(m^*_2)^2-3-\gamma_2,\\
&k_4/\mu_2^2-\gamma_2~~~~~~~~\,\mbox{vs}~~k^*_4/\mu_2^2-\gamma_2,~~~~~~~~~g_2-\gamma_2~~~~~~~~~~~~~~\mbox{vs}~~g^*_2-\gamma_2,
\end{align}
in both weighted and unweighted cases, for $n=100$ and 1000. The dependence on $n$ is described in more details in \citet{RimoldiniUnbiased}. Estimators standardized by both true and estimated variance are presented to help interpret the behaviour of the ratios from their components.

All figures confirm that `corrected' and `uncorrected' estimators have similar precision and accuracy at high $S/N$ levels (typically for $S/N>10$). Noise-unbiased estimators are found to be the most accurate in all cases and over the whole $S/N$ range tested. Their precision is generally similar to the noise-uncorrected counterparts,  apart from estimators standardized by the estimated variance, such as $g_1$, $g_2$ and $m_4/m_2^2$, for which the uncorrected version can be much more precise (although biased) for $S/N<2$, typically.
As expected, the precision of estimators employing $n=1000$ measurements per sample was greater than the one obtained with sample sizes of $n=100$. 

Weighting by the inverse of squared measurement errors made the estimators slightly less precise at high $S/N$ ratios, but  more precise and accurate at low $S/N$ levels (except for the mean).

Weighting by phase intervals led to a significant improvement in precision of all estimators in the limit of large $S/N$ ratios and a reduction of precision at low $S/N$ (apart from the case of the mean).
Tuning parameters such as $a=2$ and $b=0.3$ in Eq.~(\ref{eq:SNweights}) were able to mitigate the imprecision at low $S/N$  reducing to the error-weighted results, which appeared to be the most accurate and precise in the limit of low $S/N$ ratios (in these simulations). This  solution might provide a reasonable compromise between precision and accuracy of all estimators, at least for $S/N>1$. 

Figures~\ref{fig:M3_100}--\ref{fig:M3_1000ph} show that the skewness moment $m_3$ is quite unbiased by noise, while the standardized version $g_1$ is underestimated at high $S/N$ because of the overestimated variance $m_2$ (as shown in Figs~\ref{fig:M2_100}--\ref{fig:M2_1000}). While the accuracy of $g_1$ deteriorates at low $S/N$, its precision is much less affected by noise.

The kurtosis moment $m_4$ (Figs~\ref{fig:M4_100}--\ref{fig:M4_1000ph}) is  less precise and accurate than the noise-unbiased equivalent, and  its normalization by the squared variance reduces dramatically its inaccuracy and imprecision (since $m_2$ and $m_4$ exhibit a similar trend as a function of $S/N$).
The kurtosis cumulant $k_4$, instead, is much closer to its noise-unbiased counterpart, as shown in Figs~\ref{fig:K4_100}--\ref{fig:K4_1000ph}.
The normalization of $k_4$ by the squared variance improves its precision  
at the cost of lower accuracy for $S/N<10$:
the bias of $g_2$ is similar to (greater than) the precision of $g_2^*$ for $n=100$ ($n=1000$).

The lower the $S/N$ level is, the less precise estimators are and the noise-unbiased variance can be underestimated (and even become non-positive). 
Thus, 
the skewness and kurtosis estimators standardized by $k^*_2$ or $K^*_2$, as in Eqs~(\ref{eq:gi1})--(\ref{eq:gi2}), should be avoided in circumstances that combine small sample sizes (up to a few dozens of elements) and low S/N ratios (of the order of a few or less).

Figures related to moments and cumulants of irregularly sampled sinusoidal signals are very similar to the ones presented herein, with the exception of $g_1$, which would have a similar precision but with no bias, as a consequence of the null skewness of a sinusoidal signal (since the mean of $k_3$ estimates is zero, they are not biased by the standardization with an overestimated noise-biased variance).

From the comparison of noise-biased and unbiased estimators with different weighting schemes, it appears that,  for large sample sizes, noise-unbiased phase-weighted  estimators are usually the most accurate for $S/N>2$ (apart from  the special cases of standardized skewness and kurtosis when their true value is zero). For noisy signals (e.g., $S/N<2$),  error weighting seems the most appropriate, at least with Gaussian uncertainties, thus {\em noise-unbiased error-phase weighted estimators can provide a satisfactory  compromise in general}.
Further improvements might be achieved by tuning parameters better fitted to  estimators and signals of interest, in view of specific requirements of precision and accuracy.

\section{Conclusions}
\label{sec:concl}
Exact expressions of noise-unbiased skewness and kurtosis were provided in the unweighted and weighted formulations, under the assumption of independent data and Gaussian uncertainties.  Such estimators can be particularly useful in the processing, interpretation and comparison of data characterized by low $S/N$ regimes.

Simulations of an irregularly sampled skewed periodic signal were employed to compare noise-biased and unbiased estimators as a function of $S/N$ in the unweighted, inverse-squared error weighted and phase-weighted schemes.
While noise-unbiased estimators were found more accurate in general, they were less precise than the uncorrected counterparts at low $S/N$ ratios. 
The application of a mixed weighting scheme involving phase intervals and uncertainties 
was able to balance precision and accuracy on a wide range of $S/N$ levels.
The effect of noise-unbiased estimators and different weighting schemes on the characterization and classification of astronomical time series is described in \citet{RimoldiniWeighted}.

\section*{Acknowledgments}
The author thanks M. S\"uveges for many discussions and valuable comments on the original manuscript.

\newpage

\appendix
\section{Derivation of noise-unbiased moments}
\label{app:derivations}
The derivations presented in this Appendix involve weighted estimators under the assumption of independent measurements, uncertainties and weights. 
Definitions and some of the relations often employed herein are listed below.
\begin{itemize}
\item For brevity, $m_r=m_r(\mathbf{x})$, and $\sum_i$ and $\prod_i$ are implied to involve all (from the 1-st to the $n$-th) terms, unless explicitly stated otherwise.
\item The following integral solutions are often employed:
\begin{align}
\langle x^s_i \rangle = \int_{-\infty}^{\infty}\frac{{x'_i}^s}{\sqrt{2 \pi}\, \epsilon_i} \exp\left[-\frac{(x'_i-\xi_i)^2}{2\epsilon_i^2}\right] {\mathrm d}x'_i=
\begin{cases}
\xi_i &\text{ for } s=1 \\
\xi_i^2+\epsilon_i^2 &\text{ for } s=2 \\
\xi_i^3+3\xi_i\epsilon_i^2 &\text{ for } s=3 \\
\xi_i^4+6\xi_i^2\epsilon_i^2+3\epsilon_i^4 &\text{ for } s=4.
\end{cases}
\end{align}
\item The expected value $\langle m\rangle$ of a generic estimator $m(\mathbf{x})=\sum_i a_i x_i^s\sum_{j\neq i} b_j x_j^t\sum_{k\neq i,j} c_k x_k^u\sum_{l\neq i,j,k} d_l x_l^v$ of independent data with Gaussian uncertainties is computed as follows 
\begin{align}
\langle m\rangle &=\int_{\field{R}^n}m(\mathbf{x}') \prod_{i=1}^n \frac{1}{\sqrt{2 \pi}\, \epsilon_i} \exp\left[-\frac{(x'_i-\xi_i)^2}{2\epsilon_i^2}\right]\, {\mathrm d}^n\!x'\\
&=\prod_{h}\int_{-\infty}^{\infty} m(\mathbf{x}')\,\frac{1}{\sqrt{2 \pi}\, \epsilon_h} \exp\left[-\frac{(x'_h-\xi_h)^2}{2\epsilon_h^2}\right] {\mathrm d}x'_h \\
&=\sum_i a_i \int_{-\infty}^{\infty} \frac{{x'_i}^s}{\sqrt{2 \pi}\, \epsilon_i} \exp\left[-\frac{(x'_i-\xi_i)^2}{2\epsilon_i^2}\right] {\mathrm d}x'_i
~\sum_{j\neq i} b_j \int_{-\infty}^{\infty} \frac{{x'_j}^t}{\sqrt{2 \pi}\, \epsilon_j} \exp\left[-\frac{(x'_j-\xi_j)^2}{2\epsilon_j^2}\right] {\mathrm d}x'_j~\times\notag\\
&~~~~ \times\sum_{k\neq i,j} c_k \int_{-\infty}^{\infty} \frac{{x'_k}^{\!u}}{\sqrt{2 \pi}\, \epsilon_k} \exp\left[-\frac{(x'_k-\xi_k)^2}{2\epsilon_k^2}\right] {\mathrm d}x'_k
\sum_{l\neq i,j,k} d_l \int_{-\infty}^{\infty} \frac{{x'_l}^v}{\sqrt{2 \pi}\, \epsilon_l} \exp\left[-\frac{(x'_l-\xi_l)^2}{2\epsilon_l^2}\right] {\mathrm d}x'_l.
\end{align}
\item The results of the following expressions are employed:\\
$(\sum_iw_ix_i)^{3}= \\
~~~~ = \sum_iw_ix_i \,(\sum_jw_jx_j)^2\\
~~~~ = \sum_iw_ix_i \, (\sum_jw_j^2x_j^2 + \sum_jw_jx_j\sum_{k\neq j}w_kx_k)\\
~~~~ =\sum_iw_i^3x_i^3+3\sum_iw_i^2x_i^2\sum_{j\neq i}w_jx_j+\sum_iw_ix_i\sum_{j\neq i}w_jx_j\sum_{k\neq i,j}w_kx_k.\\
\\
(\sum_iw_ix_i)^{4}=\\
~~~~ = \sum_iw_ix_i \sum_jw_jx_j \, (\sum_kw_k^2x_k^2 + \sum_kw_kx_k\sum_{l\neq k}w_lx_l)\\
~~~~ = \sum_iw_ix_i \,(\sum_jw_j^3x_j^3 + 3 \sum_jw_j^2x_j^2\sum_{k\neq j}w_kx_k + \sum_jw_jx_j \sum_{k\neq j}w_kx_k\sum_{l\neq j,k}w_lx_l)\\
~~~~ =\sum_iw_i^4x_i^4+4\sum_iw_i^3x_i^3\sum_{j\neq i}w_jx_j+6\sum_iw_i^2x_i^2\sum_{j\neq i}w_jx_j\sum_{k\neq i,j}w_kx_k+\\
~~~~~~~  +3\sum_iw_i^2x_i^2\sum_{j\neq i}w_j^2x_j^2+\sum_iw_ix_i\sum_{j\neq i}w_jx_j\sum_{k\neq i,j}w_kx_k\sum_{l\neq i,j,k}w_lx_l.\\
\\
\sum_iw_ix_i^2\,(\sum_jw_jx_j)^{2}=\\
~~~~ = \sum_iw_ix_i^2 \, (\sum_jw_j^2x_j^2 + \sum_jw_jx_j\sum_{k\neq j}w_kx_k)\\
~~~~ = \sum_iw_i^3x_i^4+\sum_iw_ix_i^2\sum_{j\neq i}w_j^2x_j^2+2\sum_iw_i^2x_i^3\sum_{j\neq i}w_jx_j+\sum_iw_ix_i^2\sum_{j\neq i}w_jx_j\sum_{k\neq i,j}w_kx_k.\\
\\
\sum_iw_i\epsilon_i^2\sum_{j\neq i}w_j\xi_j\sum_{k\neq i,j}w_k\xi_k=\\
~~~~ = \sum_iw_i\epsilon_i^2\sum_{j}w_j\xi_j\sum_{k}w_k\xi_k-\sum_iw_i\epsilon_i^2\sum_{j\neq i}w_j^2\xi_j^2-2\sum_iw_i^2\xi_i\epsilon_i^2\sum_{j\neq i}w_j\xi_j-\sum_iw_i^3\xi_i^2\epsilon_i^2\\
~~~~ = V_1^2\mbox{$\bar{\xi}$}^{\,2}\sum_iw_i\epsilon_i^2-\sum_iw_i\epsilon_i^2\sum_{j}w_j^2\xi_j^2+\sum_iw_i^3\xi_i^2\epsilon_i^2
-2V_1\mbox{$\bar{\xi}$}\sum_iw_i^2\xi_i\epsilon_i^2+2\sum_iw_i^3\xi_i^2\epsilon_i^2-\sum_iw_i^3\xi_i^2\epsilon_i^2\\
~~~~ = V_1^2\mbox{$\bar{\xi}$}^{\,2}\sum_iw_i\epsilon_i^2-\sum_iw_i\epsilon_i^2\sum_{j}w_j^2\xi_j^2
-2V_1\mbox{$\bar{\xi}$}\sum_iw_i^2\xi_i\epsilon_i^2+2\sum_iw_i^3\xi_i^2\epsilon_i^2.\\
$\newpage$
\sum_iw_i^2\epsilon_i^2\sum_{j\neq i}w_j\xi_j\sum_{k\neq i,j}w_k\xi_k=\\
~~~~ = \sum_iw_i^2\epsilon_i^2\sum_{j}w_j\xi_j\sum_{k}w_k\xi_k-\sum_iw_i^2\epsilon_i^2\sum_{j\neq i}w_j^2\xi_j^2-2\sum_iw_i^3\xi_i\epsilon_i^2\sum_{j\neq i}w_j\xi_j-\sum_iw_i^4\xi_i^2\epsilon_i^2\\
~~~~ = V_1^2\mbox{$\bar{\xi}$}^{\,2}\sum_iw_i^2\epsilon_i^2-\sum_iw_i^2\epsilon_i^2\sum_{j}w_j^2\xi_j^2+\sum_iw_i^4\xi_i^2\epsilon_i^2
-2V_1\mbox{$\bar{\xi}$}\sum_iw_i^3\xi_i\epsilon_i^2+2\sum_iw_i^4\xi_i^2\epsilon_i^2-\sum_iw_i^4\xi_i^2\epsilon_i^2\\
~~~~ = V_1^2\mbox{$\bar{\xi}$}^{\,2}\sum_iw_i^2\epsilon_i^2-\sum_iw_i^2\epsilon_i^2\sum_{j}w_j^2\xi_j^2
-2V_1\mbox{$\bar{\xi}$}\sum_iw_i^3\xi_i\epsilon_i^2+2\sum_iw_i^4\xi_i^2\epsilon_i^2.
$
\end{itemize}

\subsection{Outline of results}
The expressions of the elements pursued along the derivation of noise-unbiased estimators (detailed in Sec.~\ref{app:details}) are summarized below, following the notation introduced in Sections~\ref{sec:notation} and \ref{sec:method}.
\begin{align}
&\langle m_2\rangle =\frac{1}{V_1}\sum_{i} w_i \left(\xi_i-\mbox{$\bar{\xi}$}\,\right)^2 +\frac{1}{V_1}\sum_{i} w_i \epsilon_i^2 \left(1-\frac{w_i}{V_1}\right)   = \langle k_2\rangle \\
&\langle m_3\rangle =\frac{1}{V_1}\sum_i w_i \left(\xi_i-\mbox{$\bar{\xi}$}\,\right)^3+\frac{3}{V_1}\sum_{i}w_i \epsilon_i^2  \left(\xi_i-\mbox{$\bar{\xi}$} \,\right)\left(1-\frac{2 w_i}{V_1} \right)= \langle k_3\rangle \\
&\langle m_4\rangle  =\frac{1}{V_1}\sum_i w_i \left(\xi_i-\mbox{$\bar{\xi}$}\,\right)^4+  \frac{6}{V_1}\sum_{i}w_i  \epsilon_i^2 \left[\left(\xi_i-\mbox{$\bar{\xi}$}\,\right)^2\left(1-\frac{2w_i}{V_1} \right)+\epsilon_i^2\left(\frac{1}{2}-\frac{2w_i}{V_1}+\frac{2w_i^2}{V_1^2} \right)\right]+ \notag \\
&~~~~~~~~~~ +\frac{6}{V_1^3} \sum_{i}w_i^2\epsilon_i^2 \left[ \sum_{j} w_j \left( \xi_j-\mbox{$\bar{\xi}$}\,\right)^2
+\sum_{j} w_j \epsilon_j^2 \left(1-\frac{3w_j}{2V_1}\right) \right] \\
& \langle m_2^2\rangle =\left[ \frac{1}{V_1}\sum_i w_i \left(\xi_i-\mbox{$\bar{\xi}$}\,\right)^2\right]^2+\frac{4}{V_1^2}\sum_{i} w_i^2\epsilon_i^2 \left(\xi_i-\mbox{$\bar{\xi}$}\,\right)^2 
 +\frac{2}{V_1^2} \sum_{i} w_i \left(\xi_i-\mbox{$\bar{\xi}$}\, \right)^2\sum_{j} w_j \epsilon_j^2\left(1-\frac{w_j}{V_1} \right)+  \notag\\
&~~~~~~~~~~  +\frac{2}{V_1^2}\sum_{i} w_i^2\epsilon_i^4\left(1-\frac{2w_i}{V_1} \right)
+\frac{1}{V_1^2}\left[\sum_{i} w_i \epsilon_i^2\left(1-\frac{w_i}{V_1} \right) \right]^2
+\frac{2}{V_1^4}\left(\sum_{i} w_i^2 \epsilon_i^2 \right)^2 \\
&\langle k_4\rangle  = \langle m_4\rangle-3\,\langle m_2^2\rangle \\
&\notag\\
&\mbox{If~~}f(\mbox{{\boldmath $\xi$}},\mbox{{\boldmath $\epsilon$}})= \sum_{i} c_i \left(\xi_i-\mbox{$\bar{\xi}$}\, \right),
\mbox{~~~then~~}f^*\!(\mathbf{x},\mbox{{\boldmath $\epsilon$}})=\sum_{i} c_i \left(x_i-\bar{x} \right).\\
&\mbox{If~~}f(\mbox{{\boldmath $\xi$}},\mbox{{\boldmath $\epsilon$}})= \sum_{i} c_i \left(\xi_i-\mbox{$\bar{\xi}$}\, \right)^2{\!,}\,
\mbox{~~then~~}f^*\!(\mathbf{x},\mbox{{\boldmath $\epsilon$}})=\sum_{i} c_i \left[\left(x_i-\bar{x} \right)^2-\epsilon_i^2\left(1-\frac{2w_i}{V_1} \right)
-\frac{1}{V_1^2}\sum_{j} w_j^2\epsilon_j^2\right].\\
&\notag\\
&m^*_2 = m_2-\frac{1}{V_1}\sum_{i}w_i  \epsilon_i^2\left(1-\frac{w_i}{V_1} \right) = k^*_2 \\
&m^*_3 = m_3-\frac{3}{V_1}\sum_{i}w_i \epsilon_i^2  \left(x_i-\bar{x}\right)\left(1-\frac{2 w_i}{V_1} \right)= k^*_3 \\
&m^*_4  = m_4- \frac{6}{V_1}\sum_{i}w_i  \epsilon_i^2 \left[\left(x_i-\bar{x}\right)^2\left(1-\frac{2w_i}{V_1} \right)-\frac{\epsilon_i^2}{2}\left(1-\frac{2w_i}{V_1}\right)^2
+\frac{m^*_2 w_i}{V_1}\right] -\frac{3}{V_1^4}\left(\sum_{i} w_i^2\epsilon_i^2\right)^2\\
&(m_2^2)^* =\left(m^*_2\right)^2 -\frac{4}{V_1^2}\sum_{i} w_i^2\epsilon_i^2 \left[ \left(x_i-\bar{x}\right)^2 -\frac{\epsilon_i^2}{2}\left(1-\frac{2w_i}{V_1} \right)\right]
+\frac{2}{V_1^4}\left(\sum_{i} w_i^2\epsilon_i^2\right)^2\\
&k^*_4  = m^*_4-3\,(m_2^2)^*
\end{align}

\subsection{Detailed computations}
\label{app:details}
\begin{align}
m_2&=\frac{1}{V_1}\sum_{i}w_i \left(x_i-\bar{x}\right)^2\\
&=\frac{1}{V_1}\sum_{i} w_i x_i^2-\frac{2}{V_1}\bar{x}\sum_iw_ix_i+\bar{x}^2\\
&=\frac{1}{V_1}\sum_{i} w_i x_i^2-\bar{x}^2\\
&=\frac{1}{V_1}\sum_{i} w_i x_i^2-\left(\frac{1}{V_1}\sum_{i} w_i x_i\right)^{\!2}\\
&=\frac{1}{V_1}\sum_{i} w_i x_i^2-\frac{1}{V_1^2}\sum_{i} w_i^2 x_i^2-\frac{1}{V_1^2}\sum_{i} w_i x_i\sum_{j\neq i} w_j x_j\\
&\notag\\
m_3&=\frac{1}{V_1}\sum_{i}w_i \left(x_i-\bar{x}\right)^3\\
&=\frac{1}{V_1}\sum_{i} w_i x_i^3-\frac{3}{V_1}\bar{x}\sum_iw_ix_i^2+\frac{3}{V_1}\bar{x}^2\sum_iw_ix_i-\bar{x}^3\\
&=\frac{1}{V_1}\sum_{i} w_i x_i^3-\frac{3}{V_1}\bar{x}\sum_iw_ix_i^2+2\bar{x}^3\\
&=\frac{1}{V_1}\sum_{i} w_i x_i^3-\frac{3}{V_1^2}\sum_iw_ix_i^2\sum_jw_jx_j+2\left(\frac{1}{V_1}\sum_iw_ix_i\right)^{\!3}\\
&=\frac{1}{V_1}\sum_{i} w_i x_i^3-\frac{3}{V_1^2}\left(\sum_iw_i^2x_i^3+\sum_iw_ix_i^2\sum_{j\neq i}w_jx_j\right)+\notag\\
&~~~~ +\frac{2}{V_1^3}\left(\sum_iw_i^3x_i^3+3\sum_iw_i^2x_i^2\sum_{j\neq i}w_jx_j+\sum_iw_ix_i\sum_{j\neq i}w_jx_j\sum_{k\neq i,j}w_kx_k\right)\\
&\notag\\
m_4&=\frac{1}{V_1}\sum_{i}w_i \left(x_i-\bar{x}\right)^4\\
&=\frac{1}{V_1}\sum_{i} w_i x_i^4-\frac{4}{V_1}\bar{x}\sum_iw_ix_i^3+\frac{6}{V_1}\bar{x}^2\sum_iw_ix_i^2-\frac{4}{V_1}\bar{x}^3\sum_iw_ix_i-\bar{x}^4\\
&=\frac{1}{V_1}\sum_{i} w_i x_i^4-\frac{4}{V_1}\bar{x}\sum_iw_ix_i^3+\frac{6}{V_1}\bar{x}^2\sum_iw_ix_i^2-3\bar{x}^4\\
&=\frac{1}{V_1}\sum_{i} w_i x_i^4-\frac{4}{V_1^2}\sum_iw_ix_i^3\sum_jw_jx_j+\frac{6}{V_1^3}\sum_iw_ix_i^2\left(\sum_jw_jx_j\right)^{\!2}-\frac{3}{V_1^4}\left(\sum_iw_ix_i\right)^{\!4}\\
&=\frac{1}{V_1}\sum_{i} w_i x_i^4-\frac{4}{V_1^2}\left(\sum_iw_i^2x_i^4+\sum_iw_ix_i^3\sum_{j\neq i}w_jx_j\right)+\frac{6}{V_1^3}\left(\sum_iw_i^3x_i^4+\right.\notag\\
&~~~~ \left.+\sum_iw_ix_i^2\sum_{j\neq i}w_j^2x_j^2+2\sum_iw_i^2x_i^3\sum_{j\neq i}w_jx_j+\sum_iw_ix_i^2\sum_{j\neq i}w_jx_j\sum_{k\neq i,j}w_kx_k\right)+\notag
\end{align}
\begin{align}
&~~~~ -\frac{3}{V_1^4}\left(\sum_iw_i^4x_i^4+4\sum_iw_i^3x_i^3\sum_{j\neq i}w_jx_j+6\sum_iw_i^2x_i^2\sum_{j\neq i}w_jx_j\sum_{k\neq i,j}w_kx_k+ \right. \notag\\
&~~~~ \left. +3\sum_iw_i^2x_i^2\sum_{j\neq i}w_j^2x_j^2+\sum_iw_ix_i\sum_{j\neq i}w_jx_j\sum_{k\neq i,j}w_kx_k\sum_{l\neq i,j,k}w_lx_l\right)\\
m_2^2&=\left[\frac{1}{V_1}\sum_{i}w_i \left(x_i-\bar{x}\right)^2\right]^2\\
&=\left[\frac{1}{V_1}\sum_{i}w_ix_i^2- \left(\frac{1}{V_1}\sum_{i}w_ix_i\right)^{\!2}\right]^2\\
&=\frac{1}{V_1^2}\left(\sum_{i}w_ix_i^2\right)^2-\frac{2}{V_1^3} \sum_{i}w_ix_i^2\left(\sum_{j}w_jx_j\right)^{\!2}+\frac{1}{V_1^4}\left(\sum_{i}w_ix_i\right)^{\!4}\\
&=\frac{1}{V_1^2}\left(\sum_{i}w_ix_i^2\right)^2-\frac{2}{V_1^3} \sum_{i}w_ix_i^2\left(\sum_{j}w_jx_j\right)^{\!2}+\frac{1}{V_1^4}\left(\sum_{i}w_ix_i\right)^{\!4}\\
&=\frac{1}{V_1^2}\left(\sum_{i}w_i^2x_i^4+\sum_{i}w_ix_i^2\sum_{j\neq i}w_jx_j^2\right)
-\frac{2}{V_1^3}\left(\sum_iw_i^3x_i^4+\sum_iw_ix_i^2\sum_{j\neq i}w_j^2x_j^2+\right.\notag\\
&~~~~ \left.+2\sum_iw_i^2x_i^3\sum_{j\neq i}w_jx_j+\sum_iw_ix_i^2\sum_{j\neq i}w_jx_j\sum_{k\neq i,j}w_kx_k\right)+\notag\\
&~~~~ +\frac{1}{V_1^4}\left(\sum_iw_i^4x_i^4+4\sum_iw_i^3x_i^3\sum_{j\neq i}w_jx_j+6\sum_iw_i^2x_i^2\sum_{j\neq i}w_jx_j\sum_{k\neq i,j}w_kx_k+ \right. \notag\\
&~~~~ \left. +3\sum_iw_i^2x_i^2\sum_{j\neq i}w_j^2x_j^2+\sum_iw_ix_i\sum_{j\neq i}w_jx_j\sum_{k\neq i,j}w_kx_k\sum_{l\neq i,j,k}w_lx_l\right)\\
&\notag\\
\langle m_2\rangle &=\prod_{i}\int_{-\infty}^{\infty} m_2(\mathbf{x}')\,\frac{1}{\sqrt{2 \pi}\, \epsilon_i} \exp\left[-\frac{(x'_i-\xi_i)^2}{2\epsilon_i^2}\right] {\mathrm d}x'_i\\
&=\frac{1}{V_1}\sum_{i} w_i \left(\xi_i^2+\epsilon_i^2\right) - \frac{1}{V_1^2}\sum_{i} w_i^2 \left(\xi_i^2+\epsilon_i^2\right) -  \frac{1}{V_1^2}\sum_{i} w_i \xi_i \sum_{j\neq i} w_j \xi_j \\
&=\frac{1}{V_1}\sum_{i} w_i \xi_i^2 +\frac{1}{V_1}\sum_{i} w_i \epsilon_i^2 - \frac{1}{V_1^2}\sum_{i} w_i^2 \epsilon_i^2 - \left(\frac{1}{V_1^2}\sum_{i} w_i^2 \xi_i^2 +  \frac{1}{V_1^2}\sum_{i} w_i \xi_i \sum_{j\neq i} w_j \xi_j \right)\\
&=\frac{1}{V_1}\sum_{i} w_i \xi_i^2 +\frac{1}{V_1}\sum_{i} w_i \epsilon_i^2 \left(1-\frac{w_i}{V_1}\right) - \mbox{$\bar{\xi}$}^{\,2}\\
&=\frac{1}{V_1}\sum_{i} w_i \left(\xi_i-\mbox{$\bar{\xi}$}\,\right)^2 +\frac{1}{V_1}\sum_{i} w_i \epsilon_i^2 \left(1-\frac{w_i}{V_1}\right) = \langle k_2\rangle
\end{align}

\begin{align}
\langle m_3\rangle &=\prod_{i}\int_{-\infty}^{\infty} m_3(\mathbf{x}')\,\frac{1}{\sqrt{2 \pi}\, \epsilon_i} \exp\left[-\frac{(x'_i-\xi_i)^2}{2\epsilon_i^2}\right] {\mathrm d}x'_i\\
&=\frac{1}{V_1}\sum_{i} w_i \left(\xi_i^3+3\xi_i\epsilon_i^2\right)-\frac{3}{V_1^2}\left[\sum_iw_i^2 \left(\xi_i^3+3\xi_i\epsilon_i^2\right)+\sum_iw_i \left(\xi_i^2+\epsilon_i^2\right)\sum_{j\neq i}w_j\xi_j\right]+\notag\\
&~~~~ +\frac{2}{V_1^3}\left[\sum_iw_i^3 \left(\xi_i^3+3\xi_i\epsilon_i^2\right)+3\sum_iw_i^2 \left(\xi_i^2+\epsilon_i^2\right)\sum_{j\neq i}w_j\xi_j+\sum_iw_i\xi_i\sum_{j\neq i}w_j\xi_j\sum_{k\neq i,j}w_k\xi_k\right]\\
&=\frac{1}{V_1}\sum_{i} w_i\xi_i^3+\frac{3}{V_1}\sum_{i} w_i \xi_i\epsilon_i^2
-\frac{3}{V_1^2}\sum_iw_i^2 \xi_i^3 -\frac{9}{V_1^2}\sum_iw_i^2 \xi_i\epsilon_i^2
-\frac{3}{V_1^2}\sum_iw_i \xi_i^2\sum_{j\neq i}w_j\xi_j+\notag\\
&~~~~ -\frac{3}{V_1^2}\sum_iw_i \epsilon_i^2\sum_{j\neq i}w_j\xi_j
+\frac{2}{V_1^3}\sum_iw_i^3 \xi_i^3 
+\frac{6}{V_1^3}\sum_iw_i^3 \xi_i\epsilon_i^2 
+\frac{6}{V_1^3}\sum_iw_i^2 \xi_i^2\sum_{j\neq i}w_j\xi_j +\notag\\
&~~~~ +\frac{6}{V_1^3}\sum_iw_i^2 \epsilon_i^2\sum_{j\neq i}w_j\xi_j 
+\frac{2}{V_1^3}\sum_iw_i\xi_i\sum_{j\neq i}w_j\xi_j\sum_{k\neq i,j}w_k\xi_k\\
&=\frac{1}{V_1}\sum_{i} w_i\xi_i^3
+\frac{3}{V_1}\sum_{i} w_i \xi_i\epsilon_i^2\left(1-\frac{3 w_i}{V_1} \right)
-\frac{3}{V_1}\,\mbox{$\bar{\xi}$}\,\sum_iw_i \xi_i^2 
-\frac{3}{V_1^2}\sum_iw_i \epsilon_i^2\sum_{j\neq i}w_j\xi_j+\notag\\
&~~~~ +2\,\mbox{$\bar{\xi}$}^{\,3}
+\frac{6}{V_1^2}\,\mbox{$\bar{\xi}$}\,\sum_iw_i^2 \epsilon_i^2\\
&=\frac{1}{V_1}\sum_i w_i \left(\xi_i-\mbox{$\bar{\xi}$}\,\right)^3
+\frac{3}{V_1}\sum_{i} w_i \epsilon_i^2\left(\xi_i-\frac{3 w_i}{V_1}\,\xi_i + \frac{2 w_i}{V_1}\,\mbox{$\bar{\xi}$}\right)
-\frac{3}{V_1}\sum_iw_i \epsilon_i^2 \left(\mbox{$\bar{\xi}$}- \frac{w_i}{V_1}\,\xi_i\right) \\
&=\frac{1}{V_1}\sum_i w_i \left(\xi_i-\mbox{$\bar{\xi}$}\,\right)^3
+\frac{3}{V_1}\sum_{i} w_i \epsilon_i^2\left[\xi_i- \mbox{$\bar{\xi}$} -\frac{2 w_i}{V_1}\left(\xi_i-\mbox{$\bar{\xi}$}\,\right)\right]\\
&=\frac{1}{V_1}\sum_i w_i \left(\xi_i-\mbox{$\bar{\xi}$}\,\right)^3+\frac{3}{V_1}\sum_{i}w_i \epsilon_i^2  \left(\xi_i-\mbox{$\bar{\xi}$} \,\right)\left(1-\frac{2 w_i}{V_1} \right)= \langle k_3\rangle
\end{align}

\begin{align}
\langle m_4\rangle &=\prod_{i}\int_{-\infty}^{\infty} m_4(\mathbf{x}')\,\frac{1}{\sqrt{2 \pi}\, \epsilon_i} \exp\left[-\frac{(x'_i-\xi_i)^2}{2\epsilon_i^2}\right] {\mathrm d}x'_i\\
&=\frac{1}{V_1}\sum_{i} w_i \left( \xi_i^4+6\xi_i^2\epsilon_i^2+3\epsilon_i^4\right)-\frac{4}{V_1^2}\left[\sum_iw_i^2\left( \xi_i^4+6\xi_i^2\epsilon_i^2+3\epsilon_i^4\right)+\sum_iw_i\left( \xi_i^3+3\xi_i\epsilon_i^2\right)\sum_{j\neq i}w_j\xi_j\right]+\notag\\
&~~~~ +\frac{6}{V_1^3}\left[\sum_iw_i^3 \left( \xi_i^4+6\xi_i^2\epsilon_i^2+3\epsilon_i^4\right)+\sum_iw_i\left( \xi_i^2+\epsilon_i^2\right)\sum_{j\neq i}w_j^2\left(\xi_j^2+\epsilon_j^2\right)+\right. \notag\\
&~~~~ \left.+2\sum_iw_i^2\left(\xi_i^3+3\xi_i\epsilon_i^2\right)\sum_{j\neq i}w_j\xi_j+\sum_iw_i\left(\xi_i^2+\epsilon_i^2\right)\sum_{j\neq i}w_j\xi_j\sum_{k\neq i,j}w_k\xi_k\right]+\notag\\
&~~~~ -\frac{3}{V_1^4}\left[\sum_iw_i^4 \left( \xi_i^4+6\xi_i^2\epsilon_i^2+3\epsilon_i^4\right)+4\sum_iw_i^3\left(\xi_i^3+3\xi_i\epsilon_i^2\right)\sum_{j\neq i}w_j\xi_j+\right. \notag\\
&~~~~ +6\sum_iw_i^2\left(\xi_i^2+\epsilon_i^2\right)\sum_{j\neq i}w_j\xi_j\sum_{k\neq i,j}w_k\xi_k+3\sum_iw_i^2\left(\xi_i^2+\epsilon_i^2\right)\sum_{j\neq i}w_j^2\left(\xi_j^2+\epsilon_j^2\right)+  \notag\\
&~~~~ \left. +\sum_iw_i\xi_i\sum_{j\neq i}w_j\xi_j\sum_{k\neq i,j}w_k\xi_k\sum_{l\neq i,j,k}w_l\xi_l\right]\\
&=\frac{1}{V_1}\sum_{i} w_i \xi_i^4
+\frac{6}{V_1}\sum_{i} w_i\xi_i^2\epsilon_i^2
+\frac{3}{V_1}\sum_{i} w_i \epsilon_i^4
-\frac{4}{V_1^2}\sum_iw_i^2 \xi_i^4
-\frac{24}{V_1^2}\sum_iw_i^2\xi_i^2\epsilon_i^2
-\frac{12}{V_1^2}\sum_iw_i^2\epsilon_i^4 +\notag\\
&~~~~ -\frac{4}{V_1^2}\sum_iw_i \xi_i^3\sum_{j\neq i}w_j\xi_j
-\frac{12}{V_1^2}\sum_iw_i \xi_i\epsilon_i^2\sum_{j\neq i}w_j\xi_j
+\frac{6}{V_1^3}\sum_iw_i^3  \xi_i^4
+\frac{36}{V_1^3}\sum_iw_i^3\xi_i^2\epsilon_i^2+\notag\\
&~~~~ +\frac{18}{V_1^3}\sum_iw_i^3 \epsilon_i^4
+\frac{6}{V_1^3}\sum_iw_i\xi_i^2\sum_{j\neq i}w_j^2\xi_j^2
+\frac{6}{V_1^3}\sum_iw_i\xi_i^2\sum_{j\neq i}w_j^2\epsilon_j^2
+\frac{6}{V_1^3}\sum_iw_i\epsilon_i^2\sum_{j\neq i}w_j^2\xi_j^2+\notag\\
&~~~~ +\frac{6}{V_1^3}\sum_iw_i\epsilon_i^2\sum_{j\neq i}w_j^2\epsilon_j^2
+\frac{12}{V_1^3}\sum_iw_i^2\xi_i^3\sum_{j\neq i}w_j\xi_j
+\frac{36}{V_1^3}\sum_iw_i^2\xi_i\epsilon_i^2\sum_{j\neq i}w_j\xi_j+\notag\\
&~~~~ +\frac{6}{V_1^3}\sum_iw_i\xi_i^2\sum_{j\neq i}w_j\xi_j\sum_{k\neq i,j}w_k\xi_k
+\frac{6}{V_1^3}\sum_iw_i\epsilon_i^2\sum_{j\neq i}w_j\xi_j\sum_{k\neq i,j}w_k\xi_k
-\frac{3}{V_1^4}\sum_iw_i^4 \xi_i^4+\notag\\
&~~~~ -\frac{18}{V_1^4}\sum_iw_i^4 \xi_i^2\epsilon_i^2
-\frac{9}{V_1^4}\sum_iw_i^4\epsilon_i^4
-\frac{12}{V_1^4}\sum_iw_i^3\xi_i^3\sum_{j\neq i}w_j\xi_j
-\frac{36}{V_1^4}\sum_iw_i^3\xi_i\epsilon_i^2\sum_{j\neq i}w_j\xi_j+\notag\\
&~~~~ -\frac{18}{V_1^4}\sum_iw_i^2\xi_i^2\sum_{j\neq i}w_j\xi_j\sum_{k\neq i,j}w_k\xi_k
-\frac{18}{V_1^4}\sum_iw_i^2\epsilon_i^2\sum_{j\neq i}w_j\xi_j\sum_{k\neq i,j}w_k\xi_k+\notag\\
&~~~~ -\frac{9}{V_1^4}\sum_iw_i^2\xi_i^2\sum_{j\neq i}w_j^2\xi_j^2
-\frac{9}{V_1^4}\sum_iw_i^2\xi_i^2\sum_{j\neq i}w_j^2\epsilon_j^2
-\frac{9}{V_1^4}\sum_iw_i^2\epsilon_i^2\sum_{j\neq i}w_j^2\xi_j^2+\notag\\
&~~~~ -\frac{9}{V_1^4}\sum_iw_i^2\epsilon_i^2\sum_{j\neq i}w_j^2\epsilon_j^2
-\frac{3}{V_1^4}\sum_iw_i\xi_i\sum_{j\neq i}w_j\xi_j\sum_{k\neq i,j}w_k\xi_k\sum_{l\neq i,j,k}w_l\xi_l
\end{align}
\begin{align}
&=\frac{1}{V_1}\sum_i w_i \left(\xi_i-\mbox{$\bar{\xi}$}\,\right)^4
+\frac{6}{V_1}\sum_{i} w_i\epsilon_i^2\left(\xi_i-\mbox{$\bar{\xi}$}\,\right)^2
-\frac{6}{V_1}\,\mbox{$\bar{\xi}$}^{\,2}\sum_i w_i \epsilon_i^2
-\frac{12}{V_1^2}\sum_iw_i^2\xi_i^2\epsilon_i^2+ \notag\\
&~~~~ +\frac{3}{V_1}\sum_{i} w_i \epsilon_i^4
-\frac{12}{V_1^2}\sum_iw_i^2\epsilon_i^4 
+\frac{36}{V_1^2}\,\mbox{$\bar{\xi}$}\,\sum_iw_i^2\xi_i\epsilon_i^2
+\frac{12}{V_1^3}\sum_iw_i^3 \epsilon_i^4
+\frac{6}{V_1^3}\sum_iw_i\epsilon_i^2\sum_{j}w_j^2\epsilon_j^2+\notag\\
&~~~~ +\frac{6}{V_1^3}\sum_iw_i\xi_i^2\sum_{j}w_j^2\epsilon_j^2
+\frac{6}{V_1^3}\sum_iw_i\epsilon_i^2\sum_{j}w_j^2\xi_j^2
-\frac{12}{V_1^3}\sum_iw_i^3\xi_i^2\epsilon_i^2 +\notag\\
&~~~~ 
+\frac{6}{V_1^3}\left(V_1^2\mbox{$\bar{\xi}$}^{\,2}\sum_iw_i\epsilon_i^2-\sum_iw_i\epsilon_i^2\sum_{j}w_j^2\xi_j^2
-2V_1\mbox{$\bar{\xi}$}\,\sum_iw_i^2\xi_i\epsilon_i^2+2\sum_iw_i^3\xi_i^2\epsilon_i^2\right)+\notag\\
&~~~~ 
-\frac{18}{V_1^4}\sum_iw_i^2\epsilon_i^2\sum_{j}w_j^2\xi_j^2
-\frac{9}{V_1^4}\left(\sum_iw_i^2\epsilon_i^2\right)^2
-\frac{36}{V_1^3}\,\mbox{$\bar{\xi}$}\,\sum_iw_i^3\xi_i\epsilon_i^2
+\frac{36}{V_1^4}\sum_iw_i^4\xi_i^2\epsilon_i^2+\notag\\
&~~~~ 
-\frac{18}{V_1^4}\left(V_1^2\mbox{$\bar{\xi}$}^{\,2}\sum_iw_i^2\epsilon_i^2-\sum_iw_i^2\epsilon_i^2\sum_{j}w_j^2\xi_j^2
-2V_1\mbox{$\bar{\xi}$}\,\sum_iw_i^3\xi_i\epsilon_i^2+2\sum_iw_i^4\xi_i^2\epsilon_i^2\right)\\
&=\frac{1}{V_1}\sum_i w_i \left(\xi_i-\mbox{$\bar{\xi}$}\,\right)^4
+\frac{6}{V_1}\sum_{i} w_i\epsilon_i^2\left(\xi_i-\mbox{$\bar{\xi}$}\,\right)^2
-\frac{12}{V_1^2}\sum_iw_i^2\xi_i^2\epsilon_i^2+ \notag\\
&~~~~ +\frac{3}{V_1}\sum_{i} w_i \epsilon_i^4\left(1-\frac{4w_i}{V_1}+\frac{4w_i^2}{V_1^2} \right)
+\frac{24}{V_1^2}\,\mbox{$\bar{\xi}$}\,\sum_iw_i^2\xi_i\epsilon_i^2
+\frac{6}{V_1^3}\sum_iw_i\epsilon_i^2\sum_{j}w_j^2\epsilon_j^2+\notag\\
&~~~~ +\frac{6}{V_1^3}\sum_iw_i\xi_i^2\sum_{j}w_j^2\epsilon_j^2
-\frac{9}{V_1^4}\left(\sum_iw_i^2\epsilon_i^2\right)^2
-\frac{18}{V_1^2}\,\mbox{$\bar{\xi}$}^{\,2}\sum_iw_i^2\epsilon_i^2\\
&=\frac{1}{V_1}\sum_i w_i \left(\xi_i-\mbox{$\bar{\xi}$}\,\right)^4
+\frac{6}{V_1}\sum_{i} w_i\epsilon_i^2\left(\xi_i-\mbox{$\bar{\xi}$}\,\right)^2
-\frac{12}{V_1^2}\sum_iw_i^2\epsilon_i^2\left(\xi_i^2-2\xi_i\mbox{$\bar{\xi}$}+\mbox{$\bar{\xi}$}^{\,2} \right)+ \notag\\
&~~~~ +\frac{3}{V_1}\sum_{i} w_i \epsilon_i^4\left(1-\frac{2w_i}{V_1} \right)^2
+\frac{6}{V_1^3}\sum_{i}w_i^2\epsilon_i^2\sum_jw_j\epsilon_j^2+\notag\\
&~~~~ +\frac{6}{V_1^3}\sum_{i}w_i^2\epsilon_i^2\left(\sum_jw_j\xi_j^2-V_1\mbox{$\bar{\xi}$}^{\,2}\right)
-\frac{9}{V_1^4}\left(\sum_iw_i^2\epsilon_i^2\right)^2\\
&=\frac{1}{V_1}\sum_i w_i \left(\xi_i-\mbox{$\bar{\xi}$}\,\right)^4+  \frac{6}{V_1}\sum_{i}w_i  \epsilon_i^2 \left(\xi_i-\mbox{$\bar{\xi}$}\,\right)^2\left(1-\frac{2w_i}{V_1} \right)
+\frac{3}{V_1}\sum_{i} w_i \epsilon_i^4\left(1-\frac{2w_i}{V_1} \right)^2+ \notag \\
&~~~~ +\frac{6}{V_1^3} \sum_{i}w_i^2\epsilon_i^2 \left[ \sum_{j} w_j \left( \xi_j-\mbox{$\bar{\xi}$}\,\right)^2
+\sum_{j} w_j \epsilon_j^2 \left(1-\frac{3w_j}{2V_1}\right) \right] 
\end{align}

\begin{align}
\langle m_2^2\rangle &=\prod_{i}\int_{-\infty}^{\infty} m_2^2(\mathbf{x}')\,\frac{1}{\sqrt{2 \pi}\, \epsilon_i} \exp\left[-\frac{(x'_i-\xi_i)^2}{2\epsilon_i^2}\right] {\mathrm d}x'_i\\
&=\frac{1}{V_1^2}\left[\sum_{i}w_i^2\left(\xi_i^4+6\xi_i^2\epsilon_i^2+3\epsilon_i^4\right)+\sum_{i}w_i\left(\xi_i^2+\epsilon_i^2\right)\sum_{j\neq i}w_j\left(\xi_j^2+\epsilon_j^2\right)\right]+\notag\\
&~~~~ -\frac{2}{V_1^3}\left[\sum_iw_i^3\left(\xi_i^4+6\xi_i^2\epsilon_i^2+3\epsilon_i^4\right)+\sum_iw_i\left(\xi_i^2+\epsilon_i^2\right)\sum_{j\neq i}w_j^2\left(\xi_j^2+\epsilon_j^2\right)+\right.\notag\\
&~~~~ \left.+2\sum_iw_i^2\left(\xi_i^3+3\xi_i\epsilon_i^2\right)\sum_{j\neq i}w_j\xi_j+\sum_iw_i\left(\xi_i^2+\epsilon_i^2\right)\sum_{j\neq i}w_j\xi_j\sum_{k\neq i,j}w_k\xi_k\right]+\notag\\
&~~~~ +\frac{1}{V_1^4}\left[\sum_iw_i^4\left(\xi_i^4+6\xi_i^2\epsilon_i^2+3\epsilon_i^4\right)+4\sum_iw_i^3\left(\xi_i^3+3\xi_i\epsilon_i^2\right)\sum_{j\neq i}w_j\xi_j+\right.\notag\\
&~~~~ +6\sum_iw_i^2\left(\xi_i^2+\epsilon_i^2\right)\sum_{j\neq i}w_j\xi_j\sum_{k\neq i,j}w_k\xi_k+3\sum_iw_i^2\left(\xi_i^2+\epsilon_i^2\right)\sum_{j\neq i}w_j^2\left(\xi_j^2+\epsilon_j^2\right)+ \notag\\
&~~~~ \left. +\sum_iw_i\xi_i\sum_{j\neq i}w_j\xi_j\sum_{k\neq i,j}w_k\xi_k\sum_{l\neq i,j,k}w_l\xi_l\right]\\
&=\frac{1}{V_1^2}\sum_{i}w_i^2\xi_i^4
+\frac{6}{V_1^2}\sum_{i}w_i^2\xi_i^2\epsilon_i^2
+\frac{3}{V_1^2}\sum_{i}w_i^2\epsilon_i^4
+\frac{1}{V_1^2}\sum_{i}w_i\xi_i^2\sum_{j\neq i}w_j\xi_j^2
+\frac{1}{V_1^2}\sum_{i}w_i\xi_i^2\sum_{j\neq i}w_j\epsilon_j^2+\notag\\
&~~~~ +\frac{1}{V_1^2}\sum_{i}w_i\epsilon_i^2\sum_{j\neq i}w_j\xi_j^2
+\frac{1}{V_1^2}\sum_{i}w_i\epsilon_i^2\sum_{j\neq i}w_j\epsilon_j^2
-\frac{2}{V_1^3}\sum_iw_i^3\xi_i^4
-\frac{12}{V_1^3}\sum_iw_i^3\xi_i^2\epsilon_i^2
-\frac{6}{V_1^3}\sum_iw_i^3\epsilon_i^4+\notag\\
&~~~ -\frac{2}{V_1^3}\sum_iw_i\xi_i^2\sum_{j\neq i}w_j^2\xi_j^2
-\frac{2}{V_1^3}\sum_iw_i\xi_i^2\sum_{j\neq i}w_j^2\epsilon_j^2
-\frac{2}{V_1^3}\sum_iw_i\epsilon_i^2\sum_{j\neq i}w_j^2\xi_j^2
-\frac{2}{V_1^3}\sum_iw_i\epsilon_i^2\sum_{j\neq i}w_j^2\epsilon_j^2+\notag\\
&~~~~ -\frac{4}{V_1^3}\sum_iw_i^2\xi_i^3\sum_{j\neq i}w_j\xi_j
-\frac{12}{V_1^3}\sum_iw_i^2\xi_i\epsilon_i^2\sum_{j\neq i}w_j\xi_j
-\frac{2}{V_1^3}\sum_iw_i\xi_i^2\sum_{j\neq i}w_j\xi_j\sum_{k\neq i,j}w_k\xi_k+\notag\\
&~~~~ -\frac{2}{V_1^3}\sum_iw_i\epsilon_i^2\sum_{j\neq i}w_j\xi_j\sum_{k\neq i,j}w_k\xi_k
+\frac{1}{V_1^4}\sum_iw_i^4\xi_i^4
+\frac{6}{V_1^4}\sum_iw_i^4\xi_i^2\epsilon_i^2
+\frac{3}{V_1^4}\sum_iw_i^4\epsilon_i^4+\notag\\
&~~~~ +\frac{4}{V_1^4}\sum_iw_i^3\xi_i^3\sum_{j\neq i}w_j\xi_j
+\frac{12}{V_1^4}\sum_iw_i^3\xi_i\epsilon_i^2\sum_{j\neq i}w_j\xi_j
+\frac{6}{V_1^4}\sum_iw_i^2\xi_i^2\sum_{j\neq i}w_j\xi_j\sum_{k\neq i,j}w_k\xi_k+\notag\\
&~~~~ +\frac{6}{V_1^4}\sum_iw_i^2\epsilon_i^2\sum_{j\neq i}w_j\xi_j\sum_{k\neq i,j}w_k\xi_k
+\frac{3}{V_1^4}\sum_iw_i^2\xi_i^2\sum_{j\neq i}w_j^2\xi_j^2
+\frac{3}{V_1^4}\sum_iw_i^2\xi_i^2\sum_{j\neq i}w_j^2\epsilon_j^2+\notag\\
&~~~~ +\frac{3}{V_1^4}\sum_iw_i^2\epsilon_i^2\sum_{j\neq i}w_j^2\xi_j^2
+\frac{3}{V_1^4}\sum_iw_i^2\epsilon_i^2\sum_{j\neq i}w_j^2\epsilon_j^2
+\frac{1}{V_1^4}\sum_iw_i\xi_i\sum_{j\neq i}w_j\xi_j\sum_{k\neq i,j}w_k\xi_k\sum_{l\neq i,j,k}w_l\xi_l
\end{align}
\begin{align}
&=\left[ \frac{1}{V_1}\sum_i w_i \left(\xi_i-\mbox{$\bar{\xi}$}\,\right)^2\right]^2
+\frac{6}{V_1^2}\sum_{i}w_i^2\xi_i^2\epsilon_i^2
+\frac{3}{V_1^2}\sum_{i}w_i^2\epsilon_i^4
+\frac{2}{V_1^2}\sum_{i}w_i\xi_i^2\sum_{j\neq i}w_j\epsilon_j^2+\notag\\
&~~~~ +\frac{1}{V_1^2}\sum_{i}w_i\epsilon_i^2\sum_{j\neq i}w_j\epsilon_j^2
-\frac{12}{V_1^3}\sum_iw_i^3\xi_i^2\epsilon_i^2
-\frac{6}{V_1^3}\sum_iw_i^3\epsilon_i^4
-\frac{2}{V_1^3}\sum_iw_i\xi_i^2\sum_{j\neq i}w_j^2\epsilon_j^2+\notag\\
&~~~~ -\frac{2}{V_1^3}\sum_iw_i\epsilon_i^2\sum_{j\neq i}w_j^2\xi_j^2
-\frac{2}{V_1^3}\sum_iw_i\epsilon_i^2\sum_{j\neq i}w_j^2\epsilon_j^2
 -\frac{12}{V_1^3}\sum_iw_i^2\xi_i\epsilon_i^2\sum_{j\neq i}w_j\xi_j+\notag\\
&~~~~ -\frac{2}{V_1^3}\left(V_1^2\mbox{$\bar{\xi}$}^{\,2}\sum_iw_i\epsilon_i^2-\sum_iw_i\epsilon_i^2\sum_{j}w_j^2\xi_j^2 -2V_1\mbox{$\bar{\xi}$}\,\sum_iw_i^2\xi_i\epsilon_i^2+2\sum_iw_i^3\xi_i^2\epsilon_i^2\right)+\notag\\
&~~~~ +\frac{6}{V_1^4}\sum_iw_i^4\xi_i^2\epsilon_i^2
+\frac{3}{V_1^4}\sum_iw_i^4\epsilon_i^4
+\frac{12}{V_1^4}\sum_iw_i^3\xi_i\epsilon_i^2\sum_{j\neq i}w_j\xi_j +\notag\\
&~~~~ +\frac{6}{V_1^4}\left(V_1^2\mbox{$\bar{\xi}$}^{\,2}\sum_iw_i^2\epsilon_i^2-\sum_iw_i^2\epsilon_i^2\sum_{j}w_j^2\xi_j^2-2V_1\mbox{$\bar{\xi}$}\,\sum_iw_i^3\xi_i\epsilon_i^2+2\sum_iw_i^4\xi_i^2\epsilon_i^2\right)+\notag\\
&~~~~ +\frac{6}{V_1^4}\sum_iw_i^2\xi_i^2\sum_{j\neq i}w_j^2\epsilon_j^2
+\frac{3}{V_1^4}\sum_iw_i^2\epsilon_i^2\sum_{j\neq i}w_j^2\epsilon_j^2\\
&=\left[ \frac{1}{V_1}\sum_i w_i \left(\xi_i-\mbox{$\bar{\xi}$}\,\right)^2\right]^2
+\frac{4}{V_1^2}\sum_{i}w_i^2\xi_i^2\epsilon_i^2
+\frac{2}{V_1^2}\sum_{i}w_i^2\epsilon_i^4
+\frac{2}{V_1^2}\sum_{i}w_i\xi_i^2\sum_{j}w_j\epsilon_j^2+\notag\\
&~~~~ +\frac{1}{V_1^2}\left(\sum_{i}w_i\epsilon_i^2\right)^2
-\frac{4}{V_1^3}\sum_iw_i^3\epsilon_i^4
-\frac{2}{V_1^3}\sum_iw_i\xi_i^2\sum_{j}w_j^2\epsilon_j^2
-\frac{2}{V_1^3}\sum_iw_i\epsilon_i^2\sum_{j}w_j^2\epsilon_j^2 + \notag\\
&~~~~ -\frac{8}{V_1^2}\,\mbox{$\bar{\xi}$}\,\sum_iw_i^2\xi_i\epsilon_i^2
-\frac{2}{V_1}\,\mbox{$\bar{\xi}$}^{\,2}\sum_iw_i\epsilon_i^2
+\frac{6}{V_1^2}\,\mbox{$\bar{\xi}$}^{\,2}\sum_iw_i^2\epsilon_i^2
+\frac{3}{V_1^4}\left(\sum_iw_i^2\epsilon_i^2\right)^2\\
&=\left[ \frac{1}{V_1}\sum_i w_i \left(\xi_i-\mbox{$\bar{\xi}$}\,\right)^2\right]^2
+\frac{4}{V_1^2}\sum_{i}w_i^2\epsilon_i^2\left(\xi_i^2 -2\xi_i \mbox{$\bar{\xi}$}+\mbox{$\bar{\xi}$}^{\,2}\right)
+\frac{2}{V_1^2}\sum_{i}w_i^2\epsilon_i^4+\notag\\
&~~~~ +\frac{2}{V_1^2}\sum_{i}w_i\left(\xi_i^2- \mbox{$\bar{\xi}$}^{\,2}\right)\sum_{j}w_j\epsilon_j^2
 +\frac{1}{V_1^2}\left(\sum_{i}w_i\epsilon_i^2\right)^2
-\frac{4}{V_1^3}\sum_iw_i^3\epsilon_i^4 + \notag\\
&~~~~ -\frac{2}{V_1^3}\sum_iw_i\left(\xi_i^2- \mbox{$\bar{\xi}$}^{\,2}\right)\sum_{j}w_j^2\epsilon_j^2
-\frac{2}{V_1^3}\sum_iw_i\epsilon_i^2\sum_{j}w_j^2\epsilon_j^2 
+\frac{3}{V_1^4}\left(\sum_iw_i^2\epsilon_i^2\right)^2\\
&=\left[ \frac{1}{V_1}\sum_i w_i \left(\xi_i-\mbox{$\bar{\xi}$}\,\right)^2\right]^2+\frac{4}{V_1^2}\sum_{i} w_i^2\epsilon_i^2 \left(\xi_i-\mbox{$\bar{\xi}$}\,\right)^2 
 +\frac{2}{V_1^2} \sum_{i} w_i \left(\xi_i-\mbox{$\bar{\xi}$}\, \right)^2\sum_{j} w_j \epsilon_j^2\left(1-\frac{w_j}{V_1} \right)+  \notag\\
&~~~~  +\frac{2}{V_1^2}\sum_{i} w_i^2\epsilon_i^4\left(1-\frac{2w_i}{V_1} \right)
+\frac{1}{V_1^2}\left[\sum_{i} w_i \epsilon_i^2\left(1-\frac{w_i}{V_1} \right) \right]^2
+\frac{2}{V_1^4}\left(\sum_{i} w_i^2 \epsilon_i^2 \right)^2 \\
&\notag\\
\langle k_4\rangle  &= \langle m_4\rangle-3\,\langle m_2^2\rangle
\end{align}

\begin{align}
\langle m_2(\mathbf{x})\rangle - m_2(\mbox{{\boldmath $\xi$}})=&\,\frac{1}{V_1}\sum_{i} w_i \epsilon_i^2 \left(1-\frac{w_i}{V_1}\right) \label{eq:f_m2} \\
\langle m_3(\mathbf{x})\rangle - m_3(\mbox{{\boldmath $\xi$}})=&\,\frac{3}{V_1}\sum_{i}w_i \epsilon_i^2  \left(\xi_i-\mbox{$\bar{\xi}$} \,\right)\left(1-\frac{2 w_i}{V_1} \right) \label{eq:f_m3}\\
\langle m_4(\mathbf{x})\rangle - m_4(\mbox{{\boldmath $\xi$}}) =&\, \frac{6}{V_1}\sum_{i}w_i  \epsilon_i^2 \left(\xi_i-\mbox{$\bar{\xi}$}\,\right)^2\left(1-\frac{2w_i}{V_1} \right)
+\frac{3}{V_1}\sum_{i} w_i \epsilon_i^4\left(1-\frac{2w_i}{V_1} \right)^2+ \notag \\
&\,+\frac{6}{V_1^3} \sum_{i}w_i^2\epsilon_i^2 \left[ \sum_{j} w_j \left( \xi_j-\mbox{$\bar{\xi}$}\,\right)^2
+\sum_{j} w_j \epsilon_j^2 \left(1-\frac{3w_j}{2V_1}\right) \right] \label{eq:f_m4}\\
\langle m_2^2(\mathbf{x})\rangle - m_2^2(\mbox{{\boldmath $\xi$}})=&\,\frac{4}{V_1^2}\sum_{i} w_i^2\epsilon_i^2 \left(\xi_i-\mbox{$\bar{\xi}$}\,\right)^2 
 +\frac{2}{V_1^2} \sum_{i} w_i \left(\xi_i-\mbox{$\bar{\xi}$}\, \right)^2\sum_{j} w_j \epsilon_j^2\left(1-\frac{w_j}{V_1} \right)+  \notag\\
& +\frac{2}{V_1^2}\sum_{i} w_i^2\epsilon_i^4\left(1-\frac{2w_i}{V_1} \right)
+\frac{1}{V_1^2}\left[\sum_{i} w_i \epsilon_i^2\left(1-\frac{w_i}{V_1} \right) \right]^2
+\frac{2}{V_1^4}\left(\sum_{i} w_i^2 \epsilon_i^2 \right)^2 \label{eq:f_m22}
\end{align}

Since the right-hand side of Eq.~(\ref{eq:f_m2}) does not depend on $\mbox{{\boldmath $\xi$}}$, $f^*\!(\mathbf{x},\mbox{{\boldmath $\epsilon$}})=f(\mbox{{\boldmath $\xi$}},\mbox{{\boldmath $\epsilon$}})=\langle m_2(\mathbf{x})\rangle-m_2(\mbox{{\boldmath $\xi$}})$ and the expression of the noise-unbiased sample variance $m_2^*=m_2(\mathbf{x})-f^*\!(\mathbf{x},\mbox{{\boldmath $\epsilon$}})$ is  found immediately:
\begin{align}
m^*_2 = m_2-\frac{1}{V_1}\sum_{i}w_i  \epsilon_i^2\left(1-\frac{w_i}{V_1} \right) = k^*_2.
\end{align}

In order to remove the dependence on $\mbox{{\boldmath $\xi$}}$ in Eqs~(\ref{eq:f_m3})--(\ref{eq:f_m22}), $f^*\!(\mathbf{x},\mbox{{\boldmath $\epsilon$}})$ is derived from $f(\mbox{{\boldmath $\xi$}},\mbox{{\boldmath $\epsilon$}})$ such that $\langle f^*\!(\mathbf{x},\mbox{{\boldmath $\epsilon$}})\rangle=f(\mbox{{\boldmath $\xi$}},\mbox{{\boldmath $\epsilon$}})$.
In the case of skewness, $f(\mbox{{\boldmath $\xi$}},\mbox{{\boldmath $\epsilon$}})$ has the following form:
\begin{align}
f(\mbox{{\boldmath $\xi$}},\mbox{{\boldmath $\epsilon$}})= \sum_{i} c_i \left(\xi_i-\mbox{$\bar{\xi}$}\, \right), \label{eq:f_skew}
\end{align}
where $c_i$ denotes the coefficient of the $i$-th term.
The computation of $\langle\,\sum_i c_i (x_i-\bar{x})\, \rangle$ leads to:
\begin{align}
\langle\,\sum_i c_i (x_i-\bar{x}) \,\rangle
&=\sum_i c_i \int_{-\infty}^{\infty} \frac{x'_i}{\sqrt{2 \pi}\, \epsilon_i} \exp\left[-\frac{(x'_i-\xi_i)^2}{2\epsilon_i^2}\right] {\mathrm d}x'_i\,+\notag\\
&~~~~ -\frac{2}{V_1}\sum_i c_i \sum_jw_j\int_{-\infty}^{\infty} \frac{x'_j}{\sqrt{2 \pi}\, \epsilon_j} \exp\left[-\frac{(x'_j-\xi_j)^2}{2\epsilon_j^2}\right] {\mathrm d}x'_j\\
&=\sum_{i} c_i \xi_i-\frac{1}{V_1}\sum_i c_i \sum_jw_j\xi_j\\
&=\sum_{i} c_i \left(\xi_i-\mbox{$\bar{\xi}$}\,\right). \label{eq:fstar_skew}
\end{align}
Since Eq.~(\ref{eq:fstar_skew}) equals Eq.~(\ref{eq:f_skew}), it follows
\begin{align}
f^*\!(\mathbf{x},\mbox{{\boldmath $\epsilon$}})=\sum_{i} c_i \left(x_i-\bar{x} \right),
\end{align}
and the noise-unbiased sample skewness $m_3^*=m_3(\mathbf{x})-f^*\!(\mathbf{x},\mbox{{\boldmath $\epsilon$}})$ is
\begin{align}
m^*_3 =m_3-\frac{3}{V_1}\sum_{i}w_i \epsilon_i^2  \left(x_i-\bar{x}\right)\left(1-\frac{2 w_i}{V_1} \right)= k^*_3.
\end{align}

For the kurtosis moment and cumulant, $f(\mbox{{\boldmath $\xi$}},\mbox{{\boldmath $\epsilon$}})$ involves $\mbox{{\boldmath $\xi$}}$-dependent terms of the  form
$\sum_{i} c_i \left(\xi_i-\mbox{$\bar{\xi}$}\, \right)^2$.
The computation of $\langle\,\sum_i c_i (x_i-\bar{x})^2 \rangle$ leads to:
\begin{align}
\langle\,\sum_i c_i (x_i-\bar{x})^2 \rangle
&=\sum_i c_i \int_{-\infty}^{\infty} \frac{{x'_i}^2}{\sqrt{2 \pi}\, \epsilon_i} \exp\left[-\frac{(x'_i-\xi_i)^2}{2\epsilon_i^2}\right] {\mathrm d}x'_i\,+\notag\\
&~~~~ -\frac{2}{V_1}\sum_i c_i w_i\int_{-\infty}^{\infty} \frac{{x'_i}^2}{\sqrt{2 \pi}\, \epsilon_i} \exp\left[-\frac{(x'_i-\xi_i)^2}{2\epsilon_i^2}\right] {\mathrm d}x'_i+\notag\\
&~~~~ -\frac{2}{V_1}\sum_i c_i \sum_{j\neq i}w_j\int_{-\infty}^{\infty} \frac{x'_i\,x'_j}{2 \pi\, \epsilon_i\epsilon_j} \exp\left[-\frac{(x'_i-\xi_i)^2}{2\epsilon_i^2}-\frac{(x'_j-\xi_j)^2}{2\epsilon_j^2}\right] {\mathrm d}x'_i\, {\mathrm d}x'_j+\notag\\
&~~~~ +\frac{1}{V_1^2}\sum_i c_i \sum_jw^2_j\int_{-\infty}^{\infty} \frac{{x'_j}^2}{\sqrt{2 \pi}\, \epsilon_j} \exp\left[-\frac{(x'_j-\xi_j)^2}{2\epsilon_j^2}\right] {\mathrm d}x'_j+\notag\\
&~~~~ +\frac{1}{V_1^2}\sum_i c_i \sum_jw_j\sum_{k\neq j}w_k\int_{-\infty}^{\infty} \frac{x'_j\,x'_k}{2 \pi\, \epsilon_j\epsilon_k} \exp\left[-\frac{(x'_j-\xi_j)^2}{2\epsilon_j^2}-\frac{(x'_k-\xi_k)^2}{2\epsilon_k^2}\right] {\mathrm d}x'_j\,{\mathrm d}x'_k \\
&=\sum_i c_i\left(\xi_i^2+\epsilon_i^2\right)-\frac{2}{V_1}\sum_i c_i w_i \left(\xi_i^2+\epsilon_i^2\right)-\frac{2}{V_1}\sum_i c_i \xi_i \sum_{j\neq i} w_j\xi_j +\notag\\
&~~~~ + \frac{1}{V_1^2}\sum_i c_i \sum_j w_j^2\left(\xi_j^2+\epsilon_j^2\right) + \frac{1}{V_1^2}\sum_i c_i \sum_j w_j \xi_j \sum_{k\neq j} w_k \xi_k\\
&=\sum_i c_i\xi_i^2+\sum_i c_i\epsilon_i^2-2\,\mbox{$\bar{\xi}$}\,\sum_i c_i \xi_i-\frac{2}{V_1}\sum_i c_i w_i\epsilon_i^2 +\notag\\
&~~~~ +  \frac{1}{V_1^2}\sum_i c_i \sum_j w_j^2\epsilon_j^2 +\mbox{$\bar{\xi}$}^{\,2} \sum_i c_i \\
&=\sum_{i} c_i \left(\xi_i-\mbox{$\bar{\xi}$}\, \right)^2+\sum_{i} c_i \epsilon_i^2\left(1-\frac{2w_i}{V_1} \right)+ \frac{1}{V_1^2}\sum_{i} c_i\sum_{j} w_j^2\epsilon_j^2.
\end{align}
Thus,  each of the terms of the form $\sum_{i} c_i \left(\xi_i-\mbox{$\bar{\xi}$}\, \right)^2$ in Eqs~(\ref{eq:f_m4})--(\ref{eq:f_m22}) can be replaced by the  expression
\begin{align}
\sum_{i} c_i \left[\left(x_i-\bar{x} \right)^2-\epsilon_i^2\left(1-\frac{2w_i}{V_1} \right)-\frac{1}{V_1^2}\sum_{j} w_j^2\epsilon_j^2\right],
\end{align}
and the noise-unbiased sample kurtosis moment $m_4^*$ and cumulant $k^*_4$ are found as follows:
\begin{align}
m^*_4  &=m_4- \frac{6}{V_1}\sum_{i}w_i  \epsilon_i^2 \left[\left(x_i-\bar{x} \right)^2\left(1-\frac{2w_i}{V_1} \right)-\epsilon_i^2\left(1-\frac{2w_i}{V_1} \right)^2-\frac{1}{V_1^2}\left(1-\frac{2w_i}{V_1} \right)\sum_{j} w_j^2\epsilon_j^2\right] +\notag\\
&~~~~ -\frac{3}{V_1}\sum_{i} w_i \epsilon_i^4\left(1-\frac{2w_i}{V_1} \right)^2 
-\frac{6}{V_1^3} \sum_{i}w_i^2\epsilon_i^2 \left[ \sum_{j} w_j \left(x_j-\bar{x} \right)^2-\sum_{j} w_j\epsilon_j^2\left(1-\frac{2w_j}{V_1} \right)+\right.\notag\\
&~~~~ \left.-\frac{1}{V_1}\sum_{j} w_j^2\epsilon_j^2  +\sum_{j} w_j \epsilon_j^2 \left(1-\frac{3w_j}{2V_1}\right) \right] \\
&=m_4- \frac{6}{V_1}\sum_{i}w_i  \epsilon_i^2\left(x_i-\bar{x} \right)^2\left(1-\frac{2w_i}{V_1} \right)+\frac{3}{V_1}\sum_{i} w_i \epsilon_i^4\left(1-\frac{2w_i}{V_1} \right)^2 +\notag\\
&~~~~  -\frac{6}{V_1^3} \sum_{i}w_i^2\epsilon_i^2 \left[ \sum_{j} w_j \left(x_j-\bar{x} \right)^2 - \sum_{j}w_j \epsilon_j^2 \left(1-\frac{w_j}{V_1} \right)\right] -\frac{3}{V_1^4}\left(\sum_{i} w_i^2 \epsilon_i^2\right)^2\\
&=m_4- \frac{6}{V_1}\sum_{i}w_i  \epsilon_i^2 \left[\left(x_i-\bar{x}\right)^2\left(1-\frac{2w_i}{V_1} \right)-\frac{\epsilon_i^2}{2}\left(1-\frac{2w_i}{V_1}\right)^2
+\frac{m^*_2 w_i}{V_1}\right] -\frac{3}{V_1^4}\left(\sum_{i} w_i^2\epsilon_i^2\right)^2\\
(m_2^2)^* &=m_2^2
-\frac{4}{V_1^2}\sum_{i} w_i^2\epsilon_i^2 \left[\left(x_i-\bar{x} \right)^2-\epsilon_i^2\left(1-\frac{2w_i}{V_1} \right)-\frac{1}{V_1^2}\sum_{j} w_j^2\epsilon_j^2\right]+\notag\\
&~~~~ -\frac{2}{V_1^2} \sum_{i} w_i\left[\left(x_i-\bar{x} \right)^2-\epsilon_i^2\left(1-\frac{2w_i}{V_1} \right)-\frac{1}{V_1^2}\sum_{j} w_j^2\epsilon_j^2\right]
\sum_{j} w_j \epsilon_j^2\left(1-\frac{w_j}{V_1} \right)+  \notag\\
&~~~~ -\frac{2}{V_1^2}\sum_{i} w_i^2\epsilon_i^4\left(1-\frac{2w_i}{V_1} \right)
-\frac{1}{V_1^2}\left[\sum_{i} w_i \epsilon_i^2\left(1-\frac{w_i}{V_1} \right) \right]^2
-\frac{2}{V_1^4}\left(\sum_{i} w_i^2 \epsilon_i^2 \right)^2\\
&=m_2^2
-\frac{2}{V_1}\, m_2\sum_{i} w_i \epsilon_i^2\left(1-\frac{w_i}{V_1} \right)
+\frac{1}{V_1^2}\left[\sum_{i} w_i \epsilon_i^2\left(1-\frac{w_i}{V_1} \right) \right]^2 + \notag\\
&~~~~ -\frac{4}{V_1^2}\sum_{i} w_i^2\epsilon_i^2 \left[\left(x_i-\bar{x} \right)^2-\frac{\epsilon_i^2}{2}\left(1-\frac{2w_i}{V_1} \right)\right]
-\frac{2}{V_1^2}\left(\sum_{i} w_i \epsilon_i^2 \right)^2+\frac{4}{V_1^3}\sum_{i} w_i \epsilon_i^2\sum_{j} w_j^2 \epsilon_j^2+\notag\\
&~~~~ +\frac{2}{V_1^2} \sum_{i} w_i\left[\epsilon_i^2\left(1-\frac{2w_i}{V_1} \right)+\frac{1}{V_1^2}\sum_{j} w_j^2\epsilon_j^2\right]
\left(\sum_{j} w_j \epsilon_j^2-\frac{1}{V_1}\sum_{j} w_j^2 \epsilon_j^2 \right)\\
&=\left(m^*_2\right)^2 -\frac{4}{V_1^2}\sum_{i} w_i^2\epsilon_i^2 \left[ \left(x_i-\bar{x}\right)^2 -\frac{\epsilon_i^2}{2}\left(1-\frac{2w_i}{V_1} \right)\right]
-\frac{2}{V_1^2}\left(\sum_{i} w_i \epsilon_i^2 \right)^2+\notag\\
&~~~~ +\frac{4}{V_1^3}\sum_{i} w_i \epsilon_i^2\sum_{j} w_j^2 \epsilon_j^2
+\frac{2}{V_1^2} \left(\sum_{i} w_i\epsilon_i^2\right)^2
-\frac{4}{V_1^3} \sum_{i} w_i^2\epsilon_i^2\sum_{j} w_j \epsilon_j^2
-\frac{2}{V_1^3} \sum_{i} w_i \epsilon_i^2\sum_{j} w_j^2 \epsilon_j^2 + \notag\\
&~~~~ +\frac{4}{V_1^4} \left(\sum_{i} w_i^2 \epsilon_i^2\right)^2 
+\frac{2}{V_1^3}\sum_{i} w_i^2\epsilon_i^2 \sum_{j} w_j \epsilon_j^2 
- \frac{2}{V_1^4}\left(\sum_{i} w_i^2\epsilon_i^2\right)^2 \\
&=\left(m^*_2\right)^2 -\frac{4}{V_1^2}\sum_{i} w_i^2\epsilon_i^2 \left[ \left(x_i-\bar{x}\right)^2 -\frac{\epsilon_i^2}{2}\left(1-\frac{2w_i}{V_1} \right)\right]
+\frac{2}{V_1^4}\left(\sum_{i} w_i^2\epsilon_i^2\right)^2\\
k^*_4 &= m^*_4-3\,(m_2^2)^* .
\end{align}

\end{document}